\documentclass[reprint,eqsecnum,floats,aps,amsmath,amssymb,nofootinbib,prd,onecolumn, showpacs]{revtex4-1}

\usepackage{graphicx}
\usepackage{amsmath,amssymb}
\usepackage{hyperref}
\usepackage{graphicx}
\usepackage{subfigure}
\usepackage{arydshln}
\usepackage{inputenc}
\usepackage{graphicx}
\usepackage{amsmath,amssymb}
\usepackage{hyperref}

\begin{document}

\title{Hybrid Loop Quantum Cosmology: An Overview} 

\author{Beatriz Elizaga Navascu\'es}
\email{w.iac20060@kurenai.waseda.jp}
\affiliation{Institute for Quantum Gravity, Friedrich-Alexander University Erlangen-N{\"u}rnberg, Staudstra{\ss}e 7, 91058 Erlangen, Germany}
\author{Guillermo  A. Mena Marug\'an}
\email{mena@iem.cfmac.csic.es}
\affiliation{Instituto de Estructura de la Materia, IEM-CSIC, Serrano 121, 28006 Madrid, Spain}

\begin{abstract}
Loop Quantum Gravity is a nonperturbative and background independent program for the quantization of General Relativity. Its underlying formalism has been applied successfully to the study of cosmological spacetimes, both to test the principles and techniques of the theory and to discuss its physical consequences. These applications have opened a new area of research known as Loop Quantum Cosmology. The hybrid approach addresses the quantization of cosmological systems that include fields. This proposal combines the description of a finite number of degrees of freedom using Loop Quantum Cosmology, typically corresponding to a homogeneous background, and a Fock quantization of the field content of the model. In this review we first present a summary of the foundations of homogeneous Loop Quantum Cosmology and we then revisit the hybrid quantization approach, applying it to the study of Gowdy spacetimes with linearly polarized gravitational waves on toroidal spatial sections, and to the analysis of cosmological perturbations in preinflationary and inflationary stages of the Universe. The main challenge is to extract predictions about quantum geometry effects that eventually might be confronted with cosmological observations. This is the first extensive review of the hybrid approach in the literature on Loop Quantum Cosmology.
\end{abstract}

\maketitle

\section{Introduction}

Modern Physics has two basic pillars in Quantum Mechanics and Einstein's theory of General Relativity (GR). However, the latter is a geometric description of the gravitational field that does not incorporate the principles of Quantum Mechanics. Numerous attempts have been made to construct a quantum theory of the spacetime geometry but, at present, there is still no proposal that the scientific community accepts unanimously as fully satisfactory.  One of the proposals for the quantum description of gravity that has reached more impact with a robust mathematical development is Loop Quantum Gravity (LQG) \cite{ashlqg,ashlewlqg, lqgThiemann}. It is a quantization formalism for globally hyperbolic spacetimes, based on a canonical and non-perturbative formulation of the geometric degrees of freedom. The fundamental novelty with respect to other pre-existing canonical proposals [such as the Wheeler-DeWitt (WdW) quantization, also called quantum geometrodynamics \cite{WdW,Hall}] lies in the use of techniques imported from Yang-Mills gauge theories \cite{yangmills}, known by their success in explaining non-perturbative regimes of the strong and electroweak interactions. In addition, the formulation of LQG is independent of any spacetime background structure and is aimed to respect the general covariance of Einstein's theory (in its canonical formulation). To achieve this goal, LQG adopts the quantization scheme proposed by Dirac for systems with constraints \cite{Dirac}. In particular, in GR the Hamiltonian is a linear combination of constraints which, via Poisson brackets, generate diffeomorphism transformations, that are the fundamental symmetries of the theory. Dirac's proposal consists in requiring that those constraints are satisfied at the quantum level on the physical states of the system. In more detail, the geometric degrees of freedom in vacuo are described in LQG by pairs of canonical variables that consist of the components of a densitized triad and a gauge connection \cite{ashlewlqg, lqgThiemann}. Their respective fluxes through surfaces and holonomies form an algebra under Poisson brackets, which is the algebra that one wants to represent quantum mechanically over a Hilbert space, where the constraints of the theory should finally be imposed. 

A major obstacle that the different candidates for a theory of quantum gravity have to face, regardless of their nature, is the extreme difficulty that is found to confront them with experimental data. Most of the possible effects of a quantum spacetime are expected to occur in regimes of very high curvatures or energies. In this sense, the Universe that we observe appears to be very classical, and GR explains it almost perfectly. However, there are observational windows to regimes of the Universe in which traces of a phenomenology that exceeded Einstein's theory might be found. A relevant example is the so-called Cosmic Microwave Background (CMB). This approximately black-body radiation reaches us from regions so far away that provides information about how the Universe was like at the early times when it became transparent. Under certain circumstances, this information might as well contain some details about very previous stages of the Universe when the spacetime geometry could have experienced quantum effects \cite{Turok}, especially if the observable Universe had in those epoques a size of the order of the Planck scale. A second obstacle for most of the quantum gravity proposals is the complication to extract concrete predictions about those regimes where quantum effects may have been important. Therefore, from a physical point of view, it is greatly convenient to consider the specialization of those formalisms to more restricted scenarios that, even without contemplating all the phenomena that may be accounted for in the full quantum theory, are able to describe regions of the Universe of particular interest. With this motivation, approximately at the beginning of this century, it was suggested to apply LQG methods to cosmological systems that possess a finite number of degrees of freedom, owing to the presence of certain symmetries. This effort crystallized in the appearance of Loop Quantum Cosmology (LQC) \cite{bojo,abl,ashparam}, a branch of LQG aimed to deal with the quantum analysis of cosmological systems. 

The first cosmologies that were studied in LQC were homogeneous and isotropic universes of the Friedmann-Lema\^{\i}tre-Robertson-Walker (FLRW) type, that classically provide a good approximation to the behavior of the observed Universe in large scales. The quantization of this type of cosmological systems was consistently completed, providing satisfactory results both from a formal and from a physical point of view.  Probably the most remarkable of these results is the resolution of the Big Bang cosmological singularity, that is replaced in this formalism with a quantum bounce, usually called the Big Bounce. In addition to these investigations on homogeneous and isotropic spacetimes, other homogeneous cosmologies with a lower degree of symmetry have also been considered in LQC to discuss the role of anisotropies. In particular, a special attention has been devoted to the quantization of  Bianchi I models \cite{chiou,chiou2,mmp1,mmp2,awe1}.

Although homogeneity and isotropy are very successful hypotheses to describe our universe at large scales, it is necessary to give an explanation to the existence and evolution of the observed inhomogeneities. In fact, the temperature of the CMB itself presents anisotropies that contain information about the small inhomogeneities in the geometry and matter content of the primeval Universe. Such inhomogeneities should be ultimately responsible of having given rise to the structures that we observe today \cite{structures}. As we have pointed out, it has been recently proposed that the power spectrum of the CMB anisotropies might even encode information about quantum effects that were relevant in the very early stages of the Universe, if the scale of the region that we observe nowadays was of the Planck order at that time \cite{Ivan,Turok}. Other information about those epochs of the Universe with extremely high curvature might be present in non-gaussianities of the CMB, or in the spectrum of tensor cosmological perturbations. Even if the information that we could extract from just one of this kind of observations might be insufficient to falsify the predictions of a candidate theory of quantum cosmology, such as LQC, the combined set of a number of different types of observations might increase the statistical significance of a possible agreement with the predictions \cite{AshNe}. With this motivation in mind to investigate quantum effects of gravity in realistic cosmological spacetimes, a hybrid strategy was proposed a decade ago in LQC for the quantum description of scenarios that contemplate the presence of inhomogeneities, both geometric and in the matter content. On the one hand, this canonical strategy employs methods inspired by LQC for the representation of the homogeneous sector of the geometry. On the other, it uses Fock representations, typical of Quantum Field Theory (QFT), to describe the rest of degrees of freedom of the system. The combination of both techniques must result in a consistent quantization of the complete system. This formalism for the quantization of inhomogeneous spacetimes implicitly assumes that there is a regime in which the most important quantum gravitational effects are felt by the homogeneous sector of the system, an assumption that seems plausible in the early stages of our universe. In the light of this hybrid approach, the advantages of reaching an evolution of the inhomogeneities that is unitary in the regime of QFT in a curved spacetime, applicable when the behavior of the homogeneous sector can be considered approximately classical,  exceed the purely theoretical aspects and appear essential to allow robust physical results, that are not affected by the severe ambiguity that would imply the consideration of Fock representations that are not even unitarily equivalent among them. 

The hybrid quantization approach, using an LQC representation for the homogeneous geometry, was first implemented in one of the Gowdy cosmological models \cite{hybrid1,hybrid2,hybrid3,hybrid4}. These models describe spacetimes that, even if subject to certain symmetry conditions (the presence of two spatial Killing vectors), still include gravitational inhomogeneities \cite{gowdy1,gowdy2}. The hybrid quantization was completed for the model with three-torus spatial topology and linearly polarized gravitational waves \cite{hybrid1,hybrid-matter}. Although the physical Hilbert space was formally characterized, it is perhaps impossible to find analytically any of these states. Therefore, approximation techniques began to be developed for the operators that appear in the resulting constraints, valid for certain quantum states \cite{hybrid-matterapp}. On these states, the Hamiltonian constraint operator adopted a particularly simple approximate expression, formally corresponding to a homogenous and isotropic cosmology with different types of effective matter content, and possibly with higher-order curvature corrections, once the average volume of the Gowdy universe was identified with the isotropic volume \cite{hybrid-matter1,hybrid-matter2}. 

More interesting from a physical point of view is the application of the hybrid quantization approach to perturbed FLRW  spacetimes coupled to a scalar field \cite{hyb-pert1,hyb-pert2,hyb-pert3,hyb-pert4,hyb-pert5,hybr-pred,hybr-ten}. Within the framework of GR, a model of this kind can be employed to describe quite successfully the primordial Universe, including small inhomogeneities that, after undergoing an inflationary stage, are capable of explaining the experimental observations about the anisotropies of the CMB \cite{mukhanov1}. This application of hybrid LQC starts with a classical formulation in which the physical degrees of freedom of the cosmological perturbations are gauge invariants, i.e. quantities that do not vary under a perturbative diffeomorphism \cite{hyb-pert4,bardeen,MukhanovSasaki,sasaki,sasakikodama}. In fact, one can construct a canonical description of the perturbations that includes such gauge invariants as a subset of the canonical variables \cite{langlo,pintoneto1,pintoneto2,hyb-pert4}. However, the passage to a quantum treatment of the whole cosmological system requires that the homogeneous degrees of freedom, rather than being considered as a fixed background, are also included in this canonical description (see Refs. \cite{HalliwellHawking,shiraiWada} for considerations in WdW). This is actually possible at least at the lowest non-trivial order of perturbative truncation of the action \cite{hyb-pert4}. In this way, the system is well prepared for its canonical quantization following the hybrid quantization approach. 

Despite the attention paid recently to cosmological perturbations in LQC with scalar fields, it is also convenient to introduce other types of matter content that are known to exist in the Universe. This is the case of fermionic fields. The interest of contemplating the presence of these fields in early cosmological epochs goes beyond a formal analysis, because it is necessary to confirm that their inclusion does not affect significantly the quantum evolution of the cosmological scalar and tensor perturbations (which are bosonic in nature). 

In summary, the purpose of this work is to review the foundations of hybrid LQC and its application to inhomogeneous cosmological systems, with an emphasis put on the analysis of the possible quantum geometry effects on primordial perturbations. The final goal of the approach would be to extract predictions about modifications with respect to Einstein's theory with the hope that, in this era of precision cosmology, those modifications might be confronted with observations in order to falsify the formalism. We would like to remark that the focus of this review will be exclusively put on the hybrid approach. The literature already contains detailed reviews of homogenous LQC and of several other approaches dealing with inhomogeneities in LQC \cite{bojo,ashparam,BCMB,RoVi,AshBarr,Grain,GFT,AlesCian,Edward,agusingh,bojow}. This is the first extensive review specifically devoted to hybrid LQC. We will concentrate our discussion on the results achieved in the hybrid quantization, and we will mention only marginally other approaches in the conclusions, to comment on some distinctive properties of the hybrid proposal. For other strategies to cope with infinite dimensional systems in LQC, the existing reviews provide a fairly complete amount of information that the reader can directly consult.

The paper is organized as follows. We first review the foundations of LQC in Sec. 2. In the first subsection we explain the choice of Ashtekar-Barbero variables and some questions about the construction of the theory of LQG with them. In the remaining subsections of Sec. 2 we apply those variables to the study of homogeneous and isotropic universes, discussing their quantization and commenting in special detail the quantum Hamiltonian that one obtains for those models. We then pass to discuss the hybrid approach in LQC, studying in Sec. 3 the cosmological system that was first analyzed in this quantization scheme. In the rest of sections, we focus our attention on the more interesting case (from a physical point of view) of a perturbed homogeneous and isotropic spacetime, in order to explore how quantum gravity effects may have affected the cosmological perturbations in the primeval universe. With this aim, we first review in Sec. 4 the procedure to construct a canonical formulation for this cosmological system in terms of gauge invariants and gauge constraints for the perturbations, together with their momenta, and of suitable zero modes for the background. In Sec. 5 we consider the possible introduction of a Dirac field in the formalism. We then explain in Sec. 6 the implementation of the hybrid LQC approach in this canonical system. Next, in Sec. 7 we discuss how we can derive modified propagation equations for the perturbations starting with the quantum Hamiltonian constraint and introducing a convenient ansatz for the quantum states, as well as some plausible approximations. These modified equations for the gauge invariants can be studied to deduce predictions that ultimately might be confronted with observations. In doing this, a key piece of information are the initial conditions that one must choose, both for the background cosmology and for the perturbations, in order to fix their vacuum state. These issues are discussed in Sec. 8. Finally, in Sec. 9 we explore the possible determination, or at least restriction, of the viable choices of a vacuum for the gauge invariant perturbations that result from demanding a good behavior in the quantum Hamiltonian operators and the evolution of those perturbations, putting an emphasis on a procedure of asymptotic diagonalization of their Hamiltonians. Section 10 contains the conclusions. In the rest of this work, we adopt units such that $c=\hbar=1$, where $c$ is the speed of light in vacuo and $\hbar$ is Planck reduced constant.  Nonetheless, we maintain Newton constant explicitly in all our formulas. Owing to these conventions and to some convenient redefinitions of quantities with respect to the notation employed in previous works, special care must be taken when comparing numerical factors in our equations with those appearing in the published literature.

\section{Loop Quantum Cosmology}

Let us first introduce the formalism of LQC, applied in this work to spacetime systems that in Einstein's theory correspond to homogeneous cosmologies. We will focus our attention on a flat FLRW model, minimally coupled to a homogeneous scalar field. In this section, we will review in detail the case of a massless field, because then the quantum constraints can be solved exactly. Later in our discussion, when we consider inflationary cosmologies, we will introduce a potential in the action of the scalar field, that can be viewed as the inflaton field of the system. An FLRW spacetime is the model typically used to describe the evolution of an expanding homogeneous and isotropic universe in GR. Here, we will study only the case of flat spatial curvature. In the following section we will also generalize our analysis to globally hyperbolic spacetimes (that admit a foliation on compact Cauchy hypersurfaces)  of Bianchi I type.

LQC starts from a Hamiltonian formulation of the system under consideration, selecting as canonical variables for the geometry those used in LQG. In the Arnowitt-Deser-Misner (ADM) formulation of GR, given an arbitrary Cauchy hypersurface $\Sigma$, the dynamical degrees of freedom of the spacetime metric can be captured by the spatial metric  induced on $\Sigma$, $ h_{ab}$ (we use lower case letters from the beginning of the alphabet to denote spatial indices), and its variation along the normal surface vector. This variation is called the extrinsic curvature and is given by the tensor $ K_{ab} = \mathcal{L}_n {h}_{ab} / 2 $, where $\mathcal{L}_{ n}$ is the Lie derivative along the normal vector $n$. Taking the spatial metric $ h_{ab}$ as the configuration variable, a linear function of the extrinsic curvature determines its canonically conjugate momentum. Starting from this canonical pair for the geometry and canonical pairs corresponding to the matter content, if a Legendre transformation is carried out in the Hilbert-Einstein Lagrangian (with suitable boundary terms), one obtains a Hamiltonian that is a linear combination of first-class constraints. Their coefficients are non-dynamical Lagrange multipliers, provided by the lapse function $ N $ and the three components of the shift vector $ N^{a} $. Each of these constraints vanishes on the solutions of GR. They are the generators of the fundamental symmetries of GR: the spacetime difeomorphisms. More specifically, the constraint that is multiplied by the lapse function is called Hamiltonian or scalar constraint, and generates time reparametrizations, modulo a spatial difeomorphism. The three constraints that come multiplied by the components of the shift vector are called momentum constraints, and generate spatial difeomorphisms.

\subsection{Ashtekar-Barbero variables: Holonomies and fluxes}

The Hamiltonian description of GR can be reformulated in terms of geometric canonical variables that, involving a gauge connection, simplify the functional form of the constraints \cite{ashlqg}. Under the quantization scheme proposed by Dirac,  these variables may seem more convenient for developing the quantum theory. In addition, the introduction of a gauge connection allows the controlled use of structures that are well known in group theory, and that can facilitate the construction of a well-defined Hilbert space. These variables can be introduced as follows.

First, in the spatial sections we can make use of triads, which are defined as a local basis of vectors $e_{i}^{a}$  of the considered Cauchy hypersurface, and in terms of which the spatial metric can be expressed locally as
\begin{equation}\label{triads}
h_{ab}=e^{i}_{a}e^{j}_{b}\delta_{ij},
\end{equation}
where the co-triads $e^{i}_{a}$ are the inverse of $e^{a}_{i}$. Since the Kronecker delta is the Euclidean metric in three dimensions, the relationship \eqref{triads} is invariant under the transformation of the triadic basis under three-dimensional rotations, at each point of the Cauchy sections. Therefore, the use of co-triads for the description of the spatial metric automatically introduces additional local symmetry into the theory, provided by the group $SO(3)$. Any Cauchy hypersurface is thus supplied with a principal fiber structure of three-dimensional reference systems, with $SO(3)$ as the gauge group \cite{isham}. We employ lower case Latin letters from the middle of the alphabet for indices corresponding to components in a local triadic basis of orthonormal frames, or equivalently in a basis of the three-dimensional Lie algebra $\mathfrak{so}(3)$. A section of the bundle is a locally smooth assignment of an element of the group to each point of the manifold. Different choices of triads, related to each other point to point by gauge transformations, can then be understood as different sections of the bundle. 

In order to define a notion of horizontality between the different fibers, as well as the associated parallel transport, one introduces a gauge connection, characterized by a one-form on the spatial hypersurfaces with components that take values in the three-dimensional Lie algebra $\mathfrak{so}(3)$. We will call this connection $\Gamma_{a}^i$. Of all the possible connections, there is one that is uniquely determined by the densitized triad through the metricity condition
\begin{equation}\label{spinAB}
{}^{(3)}\nabla_{b}E^{a}_{i}+\epsilon_{ij}{ }^{k}\Gamma^{j}_{b}E^{a}_{k}=0,
\end{equation}
where $\epsilon_{ijk}$ is the totally antisymmetric Levi-Civit\`a symbol (its indices are raised and lowered using the Kronecker delta), ${}^{(3)}\nabla_{b}$ denotes the covariant derivative associated with the Levi-Civit\`a connection compatible with $h_{ab}$, and $E^{a}_{i}=\sqrt{h}e^{a}_{i}$ is the densitized triad, with $h$ equal to the determinant of the spatial metric. 

In adddition, taking into account that the extrinsic curvature in triadic form
\begin{equation}
K^{i}_{a}=K_{ab}e^{b}_{j}\delta^{ij}
\end{equation}
can be understood as a vector of $\mathfrak{so}(3)$ with respect to the gauge transformations, as well as a one-form in spacetime, it is possible to consider, instead of $\Gamma^{i}_{a}$, the so-called Ashtekar-Barbero connection \cite{barbero}:
\begin{equation}\label{connAB}
A^{i}_{a}=\Gamma^{i}_{a}+\gamma K^{i}_{a}.
\end{equation}
In this definition, $\gamma$ is a real non-vanishing number, of arbitrary value in principle, which is known as the Immirzi parameter \cite{immi}. 

The pair formed by this connection and the densitized triad, the so-called Ashtekar-Barbero variables, turns out to be canonical for GR:
\begin{equation}\label{ABcan}
\lbrace A^{i}_{a}(\vec{x}),E^{b}_{j}(\vec{y})\rbrace=8\pi G\gamma\delta^{b}_{a}\delta^{i}_{j}\delta^{3}(\vec{x}-\vec{y}),
\end{equation}
where $\delta^{3}(\vec{x}-\vec{y})$ is the three-dimensional Dirac delta. LQG starts with these variables in the attempt to construct a quantum theory of gravity.

Actually, in order to allow the coupling to the gravitational field of matter with a half-integer spin, $A^{i}_{a}$ is considered as a connection that takes values in the three-dimensional Lie algebra $\mathfrak{su}(2)$. That is, in practice the gauge group $SO(3)$ of the principal bundle is replaced by its double cover, $SU(2)$. 

In terms of the Ashtekar-Barbero variables, the Hamiltonian constraint $\mathcal{H}$ and the momentum constraints $\mathcal{H}_{a}$ of GR (in vacuo) take the form \cite{lqgThiemann}:
\begin{eqnarray}\label{hconstr}
\mathcal{H}&=&\frac{1}{16\pi G\sqrt{ h}}\left[\epsilon^{ij}{}_{k}F^{k}_{ab}-(1+\gamma^{2})(K_{a}^{i}K_{b}^{j}-K_{b}^{i}K_{a}^{j})\right]E^{a}_{i}E^{b}_{j}, \\ \label{momconstr}
\mathcal{H}_{a}&=&\frac{1}{8\pi G\gamma}F^{i}_{ab}E^{b}_{i},
\end{eqnarray} 
where $F^{i}_{ab}$ is the curvature of the Ashtekar-Barbero connection:
\begin{equation}\label{curvA}
F^{i}_{ab}=\partial_{a}A^{i}_{b}-\partial_{b}A^{i}_{a}+\epsilon^{i}{}_{jk}A^{j}_{a}A^{k}_{b}.
\end{equation}
Finally, the introduction of an additional gauge symmetry in the theory translates into the appearance of three new constraints, that generate spin rotations in $SU (2)$ (once this is considered as the cover of the group of three-dimensional rotations):
\begin{equation}
\mathcal{G}_{i}=\frac{1}{8\pi G\gamma}\left[\partial_{a}E^{a}_{i}+\epsilon_{ij}{}^{k}A^{j}_{a}E^{a}_{k}\right].
\end{equation}
Owing to their form as a divergence, given by the covariant derivative of the triadic field with respect to the connection $A^{i}_{a}$ \cite{isham} and contracted in spacetime indices, these three constraints resemble the Gauss law of electromagnetism, and, accordingly, they are usually called the Gauss constraints.  For the type of spacetimes with homogeneous spatial surfaces that we want to consider, and with a suitable choice of reference system, the Gauss and the momentum constraints are automatically satisfied, therefore involving no restriction on the system.

From a systematic point of view, the first step in the construction of a quantum theory of gravity, based on the introduced Hamiltonian formalism with a gauge connection, would be to find a representation, as operators on a Hilbert space, of an algebra that captures all the relevant information about the canonical pair \eqref{ABcan}. Now, the formulation of the quantum theory must reasonably be such that physical results turn out to be described by quantities that do not depend on the choice of $SU(2)$ gauge. With this purpose, it is convenient that the elements of the algebra of classical variables to be quantized do not vary, or vary as little as possible, under the $SU (2)$ transformations that change the sections of the bundle. A well-known construction in Yang-Mills theories that captures the gauge invariant information about the connection is the holonomy. Given a curve $\tilde{\gamma}$ on a spatial hypersurface $\Sigma$, the holonomy along it of the connection $A^{i}_{a}$ is defined as follows:
\begin{equation}\label{holonomy}
h_{\tilde\gamma}=\mathcal{P}\exp\int_{\tilde\gamma}A^{i}_{a}\tau_{i}dx^{a},
\end{equation}
where $\mathcal{P}$ denotes path ordering and $ \tau_ {i}$ provides a basis of the Lie algebra $\mathfrak{su}(2)$. Holonomies determine the parallel transport defined by the Ashtekar-Barbero connection between the $SU(2)$ fibers that are assigned to each point of the manifold. Given any section of the principal bundle and the curve $\tilde{\gamma}$, the holonomy dictates how this curve should be lifted to the fiber so that its tangent vector is parallelly transported \cite {isham}. Under a change of section, the holonomy is only affected by the gauge transformation evaluated at the end points of the curve. On the other hand, it is clear that its construction does not depend at all on any fixed spacetime structure, nor on the choice of coordinate system. All these properties of the holonomies make them good candidates to be the variables represented quantum mechanically in order to capture the relevant information about the configuration space of Ashtekar-Barbero connections. In LQG, one considers holonomies along edges $e$, typically piecewise analytic, defined as an embedding of the interval $[0,1]$ in our Cauchy hypersurface \cite{ashlewlqg}. The variables that represent the rest of the phase space must contain the densitized triad $E_ {i}^{a}$. Since this triad is a vector density in $\Sigma$, its Hodge dual can be directly integrated over two-dimensional surfaces $S$. The result is a flux through them, which again does not depend on any additional spacetime structure nor on  the choice of coordinates, 
\begin{equation}\label{flux}
E(S,f)=\int_{S}f^{i}E_{i}^{a}\epsilon_{abc}dx^{b}dx^{c},
\end{equation}
where $f^{i}$ is a function that takes values on the algebra $\mathfrak{su}(2)$ and can be treated as a vector with respect to gauge transformations. The space of holonomies $h_{e}$  and fluxes $E(S,f)$ forms an algebra under Poisson brackets that no longer possesses the distributional divergences of the canonical relations \eqref{ABcan}. This is the algebra chosen in LQG to represent quantum mechanically the canonical commutation relations of GR.

\subsection{Homogeneous cosmologies: Polymer quantization}

We will now summarize the methodology used in LQC for the quantization of flat FLRW cosmologies, following a strategy inspired by LQG. First, let us recall that the momentum and Gauss constraints are trivial in this hogeneous system, setting at convenience the reference system for the description of the spatial metric $h_{ab}$, as well as the internal gauge of the triads that determine it [see Eq. \eqref{triads}]. For simplicity, we will choose spatial coordinates adapted to the homogeneity of the spatial sections and homogeneous diagonal triads, proportional to the Kronecker delta $\delta_{i}^{a}$. For these triads, the connection $\Gamma ^ {i} _ {a} $ vanishes. We will also assume that the spatial hypersurfaces of the chosen foliation are compact, with a three-torus topology ($T^3$). This compactness ensures that spatial integrations do not give rise to infrared divergences (in non-compact cases, this problem can be handled by introducing fiducial structures \cite{abl,awe1}). In addition, restricting the study to compact hypersurfaces is very convenient if these cosmologies provide the homogeneous sector of other more general scenarios, because the application of the hybrid strategy for their quantization would use QFT techniques that are known to be well-posed and robust in the case of compact Cauchy sections.

Taking all these considerations into account, and choosing the compactification period in $T^3$ of each of the orthogonal directions adapted to homogeneity equal to $2\pi$, the geometric sector of flat FLRW cosmologies can be described using Ashtekar-Barbero variables that adopt for them the specific form
\begin{equation}\label{abflrw}
A^{i}_{a}=\frac{c}{2\pi}\,\delta^{i}_{a},\quad\quad E_{i}^{a}=\frac{p}{4\pi^2}\,\delta_{i}^{a}, \quad\quad \{c,p\}=\frac{8\pi G\gamma}{3}.
\end{equation}
For any global function of time $t$, the canonical variables $p (t)$ and $c (t)$ can be classically related with the scale factor $a (t)$, typically used in cosmology, and with its temporal derivative $\dot{a} (t) $ by the equations
\begin{equation}\label{cpscalefactor}
|c|= 2\pi  \gamma \left|\frac{\dot{a}}{N}\right|,\quad\quad |p|=4\pi^{2}a^{2}.
\end{equation}
Note that the geometric sector of the phase space has a finite dimension (equal to two), a fact that will greatly facilitate its quantum description. Inspired by the methodology of LQG, we construct holonomies that describe the degree of freedom $c$ characterizing the connection. Thanks to the symmetries of the spatial sections, it is sufficient to consider straight edges $e_{a}$ of length $2\pi\mu$, with $\mu \in \mathbb{R} $, in the three orthogonal directions adapted to the spatial homogeneity \cite{APS1}. The holonomies of the connection $A^{i}_{a}$ along these edges have the simple expression
\begin{equation}\label{holonomflrw}
h_{e_{a}}(\mu)=\cos\left(\frac{c\mu}{2}\right)I+2\sin\left(\frac{c\mu}{2}\right)\delta^{i}_{a}\tau_{i},
\end{equation}
where $I$ is the identity in SU(2). Similarly, spatial symmetries allow us to restrict all our considerations just to fluxes of the densitized triad through squares, formed by edges along two of the reference orthogonal directions adapted to homogeneity. These fluxes are then completely determined by the variable $p$, that hence describes the geometric sector of the momentum space. Holonomies, or equivalently their matrix elements, describe the rest of the geometric sector of the phase space. More specifically, the geometric configuration space consists of the algebra formed by functions that depend on the connection through finite linear combinations of the complex exponentials $\mathcal{N}_{\mu}(c)=\exp(i\mu c/2)$, with $\mu\in\mathbb{R}$. On the other hand, we recall that the mater content of our FLRW cosmology is given by a homogeneous (massless) scalar field $\phi$. This scalar field is minimally coupled to the geometry. We will call  $\pi_{\phi}$ its canonically conjugate momentum. The canonical algebra that we want to represent has then the following non-trivial Poisson brackets:
\begin{equation}\label{algflrw}
\{\mathcal{N}_{\mu}(c),p\}=i\frac{4\pi G\gamma}{3}\mu\mathcal{N}_{\mu}(c), \quad\quad \{\phi,\pi_{\phi}\}=1.
\end{equation}

In LQC, the quantum representation of this algebra parallels the strategy adopted in LQG.  In that theory, the geometric configuration space is described by means of the so-called cylindrical functions. These are functions that depend on the Ashtekar-Barbero connection through holonomies along graphs that are formed by a finite number of edges. The algebra of cylindrical functions is completed with respect to the supreme norm, obtaining a commutative $C^{*}$-algebra with identity element. Gel'fand's theory guarantees that this algebra is isomorphic to an algebra of continuous functions over a certain compact space, called the Gel'fand spectrum, that contains the smooth connections as a dense subspace. The Hilbert space for the representation of the algebra of holonomies and fluxes in LQG is then that of square integrable functions on the Gel'fand spectrum, with respect to a certain measure \cite{ashlewlqg,lqcconfig}.

In the homogeneous and isotropic scenarios that we are considering, on the other hand, the geometric sector of the configuration space, when completed with respect to the supreme norm, is the $C^{*}$-algebra of quasi-periodic functions over $\mathbb{R}$. The complex exponentials that describe the holonomies, $\mathcal{N}_{\mu}:\mathbb{R}\rightarrow S^{1}$, where $S^{1}$ is the circumference of unit radius, form a basis in it \cite{lqcconfig}. Its Gel'fand spectrum is the Bohr compactification of the real line, $\mathbb{R}_{B}$, that contains $\mathbb{R}$ as a dense subspace \cite{rudin}. The space $\mathbb{R}_{B}$ can be characterized as the set of all homomorphisms between the additive group of real numbers and the multiplicative group of complex unit module numbers. That is, every $x\in\mathbb{R}_{B}$ is a map $x:\mathbb{R}\rightarrow S^{1}$ such that
\begin{equation}
x(0)=1,\quad\quad x(\mu+\mu')=x(\mu)x(\mu'),\quad\quad \forall\mu,\mu'\in\mathbb{R}.
\end{equation}

This space  $\mathbb{R}_{B}$ admits a compact topological group structure \cite{lqcconfig}. All functions $F_{\mu}:\mathbb{R}_{B}\rightarrow S^{1}$ such that $F_{\mu}(x)=x(\mu)$, for any $\mu \in \mathbb{R}$, are continuous with respect to that topology. In addition, since it is a compact group, it admits a unique Haar measure $M_ {H}$, which is invariant under the group action. The Hilbert space for the representation of the algebra of holonomies and fluxes in homogeneous and isotropic LQC is then $L^{2}(\mathbb{R}_{B},M_{H})$. It follows that the set of functions $\{F_{\mu},\mu\in\mathbb{R}\}$ are an orthonormal basis of this Hilbert space \cite{lqcconfig}, that is therefore not separable. Using Dirac's notation, we will denote this basis as $\{|\mu\rangle,\mu\in\mathbb{R}\}$, where $\langle\mu|\mu'\rangle=\delta_{\mu\mu'}$ is the inner product on $L^{2}(\mathbb{R}_{B},M_{H})$. The quantum representation of the algebra \eqref{algflrw} that describes the gravitational sector of the phase space is \cite{abl,lqcconfig}
\begin{equation}\label{bohrrepflrw}
\hat{\mathcal{N}}_{\mu'}|\mu\rangle=|\mu+\mu'\rangle,\quad\quad\hat{p}|\mu\rangle=\frac{4\pi G\gamma}{3}\mu |\mu\rangle.
\end{equation}

This representation is often called polymer representation, and its Hilbert space is isomorphic to that of functions over $\mathbb{R}$ that are square summable with respect to the discrete measure. Making use of this isomorphism, it is clear that the states of the polymer Hilbert space must have support only on a countable number of points, and, when this number is finite, they are the direct analogue of the cylindrical functions of LQG. Besides, note that the representation of the basic operators that describe the holonomies is not continuous. As a consequence, the operator that would represent the Ashtekar-Barbero connection is not well defined, a fact that also occurs in LQG.

At this point of our discussion, it may be worth noticing that the construction of $L^{2}(\mathbb{R}_{B},M_{H})$ as the Hilbert space of the representation strongly depends on the choice of the Haar measure. In fact, it is possible to find another measure in $\mathbb{R}_{B}$ that results in a standard Schr$\ddot{\rm o}$dinger representation for the connections and triads \cite{lqcconfig}. This alternate representation, unlike the polymer one, is continuous and is employed in the more familiar WdW quantization of this cosmological system \cite{WdW}. However, owing to the discrete character of $M_{H}$, the two measures, and hence their corresponding quantum theories, are not equivalent. It is therefore not surprising that LQC can provide different predictions than traditional geometrodynamics about the quantum regimes of this cosmological model.

Finally, a standard continuous Schr$\ddot{\rm o}$dinger representation is chosen for the matter sector of the phase space, that can be described by the scalar field and its conjugate momentum. The corresponding Hilbert space is $L^{2}(\mathbb{R},d\phi)$, where the scalar field acts by multiplication and its momentum as the derivative $\hat{\pi}_{\phi}=-i\partial_{\phi}$.

\subsection{LQC: Hamiltonian constraint}

The Hilbert space obtained by the tensor product of the polymer space and $L^{2}(\mathbb{R},d\phi)$ does not necessarily contain the physical states of the quantum theory. They should still satisfy the Hamiltonian constraint, that is the only non-trivial constraint that exists on the system, and which should be imposed \`a la Dirac quantum mechanically \cite{Dirac}. For this reason, the elements of the considered Hilbert space are often called {\sl kinematic states}. The next step in our quantization is then the representation of the Hamiltonian constraint as an operator on the kinematic Hilbert space.  The gravitational part of this constraint is given by
\begin{align}\label{hchom}
-\frac{\pi^{2}}{2 G\gamma^{2}\sqrt{ h}}E^{a}_{i}E^{b}_{j}\epsilon^{ij}{ }_{k}F^{k}_{ab}.
\end{align}
Taking into account that the lapse function is homogeneous, we have already considered the integrated version of the constraint over the three spatial directions. In terms of the Ashtekar-Barbero variables introduced before for the geometric sector, the constraint ${\mathcal H}_{S}$ that we obtain for homogeneous and isotropic cosmologies, including the contribution of the homogeneous massless matter field $\phi$, has the following form
\begin{equation}\label{Cflrw}
{\mathcal H}_{S}=|p|^{-3/2} \left(\frac{\pi_{\phi}^{2}}{2}-\frac{3}{8\pi G\gamma^{2}}c^{2}p^{2} \right).
\end{equation}

The first evident obstruction for a polymer quantization of this constraint is the absence of an operator to represent the connection. However, this difficulty can be surpassed if the following classical identity is taken into account:
\begin{equation}\label{curvid}
F^{i}_{ab}=-2\lim\limits_{A_{\square}\rightarrow 0}\text{tr}\left(\frac{h_{{\square}_{ab}}}{A_{\square}}\tau^{i}\right), \qquad a \neq b,
\end{equation}
where the symbol $\text{tr}(\cdot)$ stands for the trace and $h_{{\square}_{ab}}$ is the holonomy along a certain circuit that encloses a coordinate area $A_{\square}$. 
For spacetimes such as flat FLRW and Bianchi I cosmologies, one can consider a rectangular circuit in the plane formed by the directions $a$ and $b$. Thus, in our specific flat FLRW case, this holonomy can be written as
\begin{equation}
h_{{\square}_{ab}}=h_{e_{a}}(\mu)h_{e_{b}}(\mu)h_{e_{a}}^{-1}(\mu)h_{e_{b}}^{-1}(\mu)
\end{equation}
and the enclosed coordinate area is $A^{FLRW}_{\square}=4 \pi^{2}\mu^{2}$. 

If these holonomies are represented polymerically, the limit contained in  expression \eqref{curvid} is not well defined, because neither is the connection operator. Therefore, in LQC, the enclosed coordinate area is not made to tend to zero, but instead one appeals to the existence of a minimum value, characterized by the minimum coordinate length $2\pi\bar{\mu}$ of the edges that enclose it: $A_{\square_{min}}^{FLRW}=4\pi^{2}\bar\mu^{2}$. It seems clear then that this prescription for the quantum representation of the curvature introduces a new scale. The arbitrariness in its choice can be fixed by recurring to full LQG, where the geometric area operator has a minimum non-zero eigenvalue $\Delta$. Drawing inspiration from this fact, in LQC one postulates that this value coincides with the geometric physical area corresponding to $A_{\square_{min}}^{FLRW}$. 

Recalling the (second) classical relation in  \eqref{cpscalefactor}, one concludes then that the minimum coordinate length should satisfy \cite{APS2}
\begin{equation}\label{impdyn1}
\bar\mu=\sqrt{\frac{\Delta}{|p|}}.
\end{equation}

Once we have determined the scale $\bar\mu$ (now turned into a dynamical variable) that sets the minimum coordinate area in the FLRW cosmological model, we have to represent the classical expression  \eqref{curvid} on the polymer Hilbert space, taking in it the limit $A^{FLRW}_{\square}\rightarrow A^{FLRW}_{\square_{min}}$. In practice, this prescription amounts to the replacement of $c$ with the function $\sin(\bar{\mu} c) / \bar\mu$ in the classical expression of the Hamiltonian constraint ${\mathcal{H}}_{S}$ before one proceeds to its quantum representation. The dependence of the constraint on the connection is thus captured by the complex exponentials $\mathcal{N}_{\pm 2 \bar\mu} (c) $, that in particular depend on $p$ through $\bar\mu$. We must then specify their representation on the polymer Hilbert space, since  the classical dependence of $\mathcal{N}_{\pm 2\bar\mu}(c)$ on $c$ and $p$ makes their construction ambiguous in terms of the operators that we have taken so far as basic for the FLRW geometry, namely $\hat{\mathcal{N}}_{\mu}$ and $\hat{p}$. With this aim, it is useful to introduce first the following operator, constructed by means of the spectral theorem \cite{reedsimon}:  
\begin{equation}\label{vflrw}
\hat{v}=\frac{\widehat{\text{sign}(p)}\widehat{|p|^{3/2}}}{2\pi G\gamma\sqrt{\Delta}},\qquad \hat{v}|\mu \rangle= \frac{\text{sign}[p(\mu)]|p(\mu)|^{3/2}}{2\pi G\gamma\sqrt{\Delta}}|\mu \rangle,
\end{equation} 
where $p(\mu)$ is the eigenvalue of $\hat{p}$ corresponding to the eigenstate $|\mu\rangle$, given in Eq. \eqref{bohrrepflrw}. The direct classical counterpart of this operator is proportional to the physical volume of the FLRW flat and compact universe. In addition, it has a Poisson bracket with $b = \bar{\mu}c$  equal to minus two. If we relabel the orthonormal basis $\{|\mu\rangle,\mu\in\mathbb{R}\}$ using the eigenvalues $v$ of $\hat{v}$, the operators $\hat{\mathcal{N}}_{\pm \bar{\mu}}$ are then defined so that their action is simply a constant translation, namely they simply shift the new label by a constant \cite{APS2}:
\begin{equation}\label{impflrw}
\hat{\mathcal{N}}_{\pm \bar{\mu}}|v\rangle=|v\pm 1\rangle.
\end{equation}
The square of these operators defines $\hat{\mathcal{N}}_{\pm 2\bar{\mu}}$.

The prescription that we have explained in order to represent the elements of holonomies that appear in the Hamiltonian of LQC is commonly called the {\sl improved dynamics} scheme\footnote{An alternate way to define the Hamiltonian constraint operator using a regularized expression for the extrinsic curvature has been explored recently \cite{Ma,DaporLieg,DaLiTomasz,AlejMena}.}. However, this scheme alone is not enough to complete the representation of the constraint ${\mathcal H}_{S}$. The presence of the zero eigenvalue in the spectrum of the operator $\hat{p}$ creates problems added to those already mentioned. Indeed, the Hamiltonian constraint depends on the inverse of the geometric variable $p$ via ratios that involve the determinant of the spatial metric. The quantum representation of this inverse cannot be defined in the kinematic Hilbert space using the spectral theorem, because zero is part of the discrete spectrum of $\hat{p}$. This difficulty can be circumvented again by appealing to the following classical identity, employed as well in LQG adapted to a more general context \cite{thiemann}:
\begin{equation}\label{invp}
\frac{\text{sign}(p)}{|p|^{1/2}}=\frac{|p|^{1/2}}{2\pi G\gamma\sqrt{\Delta}}\text{tr}\left(\tau_{i}\sum_{a}\delta^{i}_{a}h_{e_{a}}(\bar{\mu})\{h^{-1}_{e_{a}}(\bar{\mu}),|p|^{1/2}\} \right).
\end{equation}
The quantum operators that correspond to inverse powers of the triadic variable $p$ are then defined by representing, on the improved dynamics scheme, the variables appearing on the right-hand side of this equality. In particular, Poisson brackets are represented by $-i$ times the commutator of the corresponding operators.
If one follows this quantization procedure, the operator representing the homogeneous Hamiltonian constraint ${\mathcal H}_{S}$ in LQC is  \cite{mmo}
\begin{equation}\label{hconstrflrw}
\hat{{\mathcal H}}_{S}=\widehat{\left[\frac{1}{\sqrt{|p|}}\right]}^{3/2}{\hat{H}}_{S}\widehat{\left[\frac{1}{\sqrt{|p|}}\right]}^{3/2}, 
\end{equation}
where we have defined the densitized operator
\begin{equation}\label{hdensconstr}
{\hat{H}}_{S}=\frac{\hat{\pi}_\phi^2}{2}-\frac{\hat{\Omega}_{0}^2}{2}.
\end{equation}
Here
\begin{eqnarray}\label{Omegaop}
\hat{\Omega}_{0}&=&\frac{3\pi G}{2}\sqrt{|\hat v|}\left[\widehat{{\rm sign}(v)}{\widehat{\sin}}(b)+{\widehat{\sin}}(b)\widehat{{\rm sign}(v)}\right]
\sqrt{|\hat v|},\\
{\widehat{\sin}}(b)&=& \frac{1}{2i} \left(\hat{\mathcal{N}}_{2\bar\mu}-\hat{\mathcal{N}}_{-2 \bar\mu}\right).
\end{eqnarray}
In these definitions, we have used a prescription for the factor ordering of the involved operators proposed by Mart\'{\i}n-Benito, Mena Marug\'an,  and Olmedo \cite{mmo}, known with the initials of these authors as the MMO prescription. Its most characteristic feature is the symmetrization of the sign of the orientation of the triad with the holonomy elements in Eq. \eqref{Omegaop}. The operator $\hat{{\mathcal H}}_S$ defined in this way presents certain interesting properties thanks to its symmetric factor ordering, compared with the quantum constraint obtained with another, frequently adopted prescription, proposed by Ashtekar, Pawlowski, and Singh (APS) \cite{APS1,APS2,APS}. In particular, its action decouples the state of the polymer basis with $v = 0$  from its orthogonal complement. This allows for a neat densitization of the Hamiltonian constraint \cite{mmo}. Besides,  the action of the operator does not mix states with positive and with negative values of $v$ \cite{mmo} (namely, it does not change the orientation of the triad). We can then restrict the quantum analysis of the FLRW cosmologies to the linear subspace generated by states $|v\rangle$ with $ v \in \mathbb{R}^{+}$, for example. Actually, the action of the constraint leaves invariant smaller and separable subspaces that are called superselection sectors, and that are simpler with the MMO prescription than in the APS case owing to the separation between sectors of different triad orientation. In more detail, the action of $\hat{{\mathcal H}}_S$ (or, equivalently, of ${\hat{H}}_{S}$) turns out to preserve every one of the linear subspaces generated by states $|v\rangle$ with $v$ belonging to the semilattice $\mathcal{L}_{\bar{\varepsilon}}^+=\{\bar{\varepsilon}+4k, k\in\mathbb{N}\}$, entirely characterized by its smallest point $\bar{\varepsilon}\in(0,4]$. Notice that the value of $\bar{\varepsilon}$ is always strictly positive, and therefore the same is automatically true for the variable $v$ in each of the considered superselection sectors. Finally, a comment is due about the imposition of the (densitized) Hamiltonian constraint ${\hat{H}}_{S}$. Although the kernel of this operator is not a proper subspace of the kinematic Hilbert space, the quantum constraint can be rigorously imposed in our representation by its adjoint action, allowing in this way generalized solutions that should provide the physical Hilbert space once they are supplied with a suitable inner product, different from the kinematic one.

In fact,  it has been possible to characterize the resulting physical Hilbert space, together with complete sets of Dirac observables.  The resolution of the constraint is straightforward once one completes the spectral analysis of the operator ${\hat{\Omega}}_0^{2}$. It has been proven that this operator has a non-degenerate absolutely continuous spectrum equal to the positive real line \cite{mmo,mmp1}. The eigenvalue equation of the operator ${\hat{\Omega}}_0^{2}$ can be regarded as a second-order difference equation.  With the MMO prescription adopted in its definition, the generalized eigenfunctions turn out to be entirely determined by their value at $\bar{\varepsilon}$, point from which they can be constructed by solving the eigenvalue equation. Besides, up to a global phase, these eigenfunctions $e^{\bar{\varepsilon}}_{\delta}(v)$ are real, because the second-order difference equation associated with the action of the operator is a real equation. With the eigenfunctions at hand, one easily obtains the solutions to the Hamiltonian constraint, which take the form 
\begin{equation}
\xi(v,\phi)=\int_0^{\infty} d\delta\, e^{\bar{\varepsilon}}_{\delta}(v) \left[\xi_+(\delta) e^{i\sqrt{\delta}\phi}+ \xi_-(\delta) e^{-i\sqrt{\delta}\phi}\right].
\end{equation}
Therefore, physical states can be identified, for instance, with the positive frequency solutions $\xi_+(\delta)\exp{(i\sqrt{\delta}\phi)}$ that are square integrable over the spectral parameter $\delta\in\mathbb{R}^+$ \cite{mmo} (negative frequency solutions provide essentially the same Hilbert space). A complete set of Dirac observables is given by $\hat{\pi}_{\phi}$ and $|\hat{v}|_{\phi_0}$, where this latter operator is defined as the action of the operator $\hat{v}$ when the scalar field equals $\phi_0$. On the Hilbert space of physical states specified above, these observables are self-adjoint, property that in fact characterizes the inner product on the space of solutions described by the functions $\xi_+(\delta)$.

On the other hand, a numerical analysis of the dynamics, with respect to the homogeneous scalar field, of certain families of states with a semiclassical behavior at large volumes shows that they remain sharply peaked during the quantum evolution \cite{APS2}. Actually, the trajectories of their peaks coincide, in the regions of low matter density, with those dictated by Einstein's equations in the considered FLRW cosmology. However, when the matter density reaches values that are comparable to the Planck density $\rho_{\text{Planck}}$, the trajectory of the peak separates from the classical solution and turns to describe a transition from a universe in contraction to an expanding one (or vice versa) \cite{APS2}. In particular, the matter density reaches a critical value when it is equal to 0.41 $\rho_{\text{Planck}}$ (for the value of the Immirzi parameter that leads in LQG to the Bekenstein-Hawking law for the entropy of black holes \cite{bek-hawk-im,lewdoma,Meisn}). As we explained in the Introduction, this phenomenon of quantum nature that eludes the cosmological singularity of the Big Bang is known by the name of Big Bounce. There is also evidence that it occurs in LQC beyond the context of homogeneous and isotropic cosmologies, with indications that it is present as well in anisotropic cosmologies such as Bianchi I models \cite{bianchibounce}, or in inhomogeneous cosmologies such as the linearly polarized Gowdy model with toroidal spatial sections \cite{tarrio}.

\section{Hybrid LQC: The Gowdy model}

Historically, hybrid LQC was first developed for the Gowdy linearly polarized cosmological model, with spatial sections with the topology of a three-torus, $T^3$, and after carrying out a partial gauge fixing that removes all constraints on the system except for the zero mode of the Hamiltonian constraint and of one of the momentum constraints (these zero modes are the average of those constraints on the spatial sections, modulo a constant numerical factor). Gowdy models are inhomogeneous cosmological spacetimes with compact spatial sections and two axial Killing vector fields  \cite{gowdy1,gowdy2}. The case with three-torus spatial topology is the simplest one. The linear polarization restriction on the gravitational wave content of the model amounts to require that the two Killing vectors are hypersurface orthogonal. The Killing symmetries then imply that the physical degree of freedom still present in those gravitational waves can be thought of as varying in only one spatial direction. After the mentioned partial gauge fixing, the phase space of this Gowdy model can be identified with that of a Bianchi I spacetime containing a linearly polarized wave. Furthermore, given the spatial behavior of this wave, we can describe it as a scalar field $\chi$ defined on the circle, $S^1$, that corresponds to the cyclic spatial direction in which the wave propagates. In addition, we couple a massless scalar field $\Phi$ with the same symmetries as those of the geometry \cite{hybrid-matter}. The hybrid quantization of this model will therefore be based on the quantization of Bianchi I cosmologies according to the LQC formalism (and the MMO prescription, see Refs. \cite{hybrid1,hybrid2,hybrid3,hybrid4}), as well as on a suitable Fock representation of the matter field $\Phi$ and of the scalar field $\chi$ assigned to the linearly polarized gravitational wave. For convenience, we extract the zero mode $\phi$ of the matter field, that behaves as a homogeneous scalar field giving a non-trivial matter content to the Bianchi I cosmology, and assume that $\chi$ has vanishing zero mode (this assumption is only meant to simplify the discussion and involves no relevant conceptual consequences). The two sectors of the model, namely the homogeneous Bianchi I sector and the inhomogeneous scalar field sector, get mixed in a non-trivial way by the zero mode of the Hamiltonian constraint, that must be imposed on the considered system.

Let us describe the model in more detail. It is most convenient to choose coordinates $\{t,\theta,\sigma,\delta\}$ adapted to the Killing symmetries, such that $\partial_\sigma$ and $\partial_\delta$ are the Killing vectors. Then, the degrees of freedom of the model only depend on the time $t$ and on the cyclic spatial coordinate $\theta \in S^1$. Starting with the canonical formulation of GR, we can then peform a symmetry reduction to take into account the Killing symmetries, as well as a partial gauge fixing that removes the momentum and Hamiltonian constraints except for the zero modes of the latter and of the momentum constraint in the $\theta$-direction \cite{hybrid2}. In this way, we get a reduced phase space that is formed by four canonical pairs of degrees of freedom (corresponding to the zero modes of the model), a gravitational field (that describes the linearly polarized gravitational wave of the model, and that we consider devoid of zero mode), and the inhomogeneous part of the massless scalar field. The homogeneous sector, composed of the four pairs of zero modes,  coincides with the phase space of a Bianchi I cosmology coupled to a homogeneous massless scalar field $\phi$. Besides, using a Fourier transform, we decompose the gravitational and matter scalar fields of the model, $\chi$ and $\Phi$ respectively, passing to describe them in terms of their Fourier (non-zero) modes. These modes and the corresponding modes of the canonical momenta of the two fields form the inhomogeneous sector of the system. The obtained reduced phase space is subject to two constraints, that were not eliminated in the process of partial gauge fixing. One of them is the zero mode of the momentum constraint in the $\theta$-direction, $H_{\theta}$. This constraint generates rigid rotations in the circle coordinatized by $\theta$, and imposes a restriction that affects exclusively the inhomogeneous sector.  The other constraint that must be imposed is the zero mode $H_{S}$ of the Hamiltonian constraint of the system. This constraint $H_{S}$ is the sum of a homogeneous term $H_{\text{hom}}$, that is the Hamiltonian constraint of the Bianchi I model, and an additional term $H_{\text{inh}}$, that couples the homogeneous and inhomogeneous sectors and vanishes when the inhomogeneous sector is not present. 

The next step in our analysis consists in describing the Bianchi I cosmologies with three-torus spatial topology in terms of Ashtekar-Barbero variables, in order to quantize them by using LQC techniques. We can adopt a suitable internal gauge and adopt a diagonal form for the triads and connections. In this way, each of the Ashtekar-Barbero variables can be totally characterized by three homogeneous functions, that determine the diagonal components. We will call these functions $p_{a}$ and $c_{a}$, corresponding to the densitized triad and $\mathfrak{su}(2)$-connection, respectively, and with $a=\theta, \sigma,\delta$. The only non-trivial Poisson brackets for them are $\{c_{a}, p_{b}\}=8\pi \gamma G \delta_{ab}$, so they form canonical pairs. We call  ${\mathcal H}_{\text{kin}}^{\text{BI}} \otimes L^{2}(\mathbb{R},d\phi)$  the kinematic Hilbert space for the Bianchi I model in LQC, where ${\mathcal H}_{\text{kin}}^{\text{BI}}$ denotes the polymer representation space of the Bianchi I geometries in LQC and $L^{2}(\mathbb{R},d\phi)$ is the space of square integrable functions for the homogeneous scalar field, defined on the real line  \cite{awe1}. With this choice of Hilbert space, we adopt again a standard Schr$\ddot{\rm o}$dinger representation for the zero mode of the matter field, $\phi$, so that its canonical conjugate momentum acts as a derivative, $\hat{\pi}_{\phi}=-i\partial_{\phi}$. The construction of $\mathcal{H}_{\text{kin}}^{\text{BI}}$, on the other hand, is similar to that explained for the FLRW geometry in LQC, except for the fact that we now have three pairs of canonically conjugated Ashtekar-Barbero degrees of freedom instead of only one. The inner product on this Hilbert space is discrete, so that the triad operators $\hat{p}_{a}$ have a point spectrum equal to the real line. Defining the tensor product $\otimes_{a} |p_{a} \rangle $ (with $a=\theta, \sigma,\delta$) of the eigenvectors of each of the triad operators we obtain eigenstates $|p_\theta,p_\sigma,p_\delta\rangle$ that  form an orthonormal basis of the Hilbert space ${\mathcal H}_{\text{kin}}^{\text{BI}}$. 

On the other hand, we can extend the improved dynamics proposal from the FLRW geometries to the Bianchi I model as proposed in Ref. \cite{awe1}, introducing in this way minimum coordinate length scales $\bar{\mu}_a$ for each of the spatial directions. Explicitly, these length scales are
\begin{equation}
\bar{\mu}_{a}=\sqrt{\frac{\Delta|p_{a}|}{|p_{b}p_{c}|}},
\end{equation}
with $a\neq b \neq c$ and $a,b,c \in \{\theta,\sigma,\delta\}$. We then introduce the operators $\hat{\mathcal{N}}_{\pm\bar{\mu}_{a}}$ to represent the holonomy elements $\mathcal{N}_{\pm{\bar{\mu}}_{a}}=\exp(\pm i{\bar{\mu}}_{a}c_{a}/2)$ along an edge in the $a$-direction of coordinate length $2\pi{\bar{\mu}}_a$. These operators appear in the regularization of the curvature operator in the Hamiltonian constraint and are defined in a similar way as we did for the isotropic case in FLRW. 

The action of these holonomy operators on the states $|p_\theta,p_\sigma,p_\delta\rangle$ is rather complicated, since each of the length scales $\bar{\mu}_a$ depends not only on $p_a$, but also on the triad variables in the other directions. To simplify the expressions, it is convenient to relabel these states in the form $|v,\lambda_{\sigma},\lambda_{\delta}\rangle$, where 
\begin{eqnarray}
\lambda_{a}&=& \text{sign}(p_{a}) \frac{\sqrt{|p_{a}|}}{(4\pi G \gamma \sqrt{\Delta})^{1/3}} ,\\
v&=&2\lambda_{\theta}\lambda_{\sigma}\lambda_{\delta}.
\end{eqnarray} 
Apart from an orientation sign, $v$ is equal to $1/(2\pi G \gamma \sqrt{\Delta})$ multiplied by the physical volume of the Bianchi I universe, volume that we will often call also the homogeneous volume. The action of the holonomy operators $\hat{\mathcal{N}}_{\pm\bar{\mu}_{\theta}}$ just scale $\lambda_{\theta}$ in such a way that the label $v$ is shifted by the unit \cite{awe1}. In full detail, we have
\begin{equation}\label{repB}
\hat{\mathcal N}_{\pm\bar\mu_\theta}|v,\lambda_\sigma,
\lambda_\delta\rangle =|v\pm\text{sign}
(\lambda_\sigma\lambda_\delta),\lambda_\sigma,
\lambda_\delta\rangle.
\end{equation}
On the other hand, the holonomy operators $\hat{\mathcal{N}}_{\pm\bar{\mu}_{\sigma}}$ and $\hat{\mathcal{N}}_{\pm\bar{\mu}_{\delta}}$ additionally produce state-dependent scalings of $\lambda_{\sigma}$ and $\lambda_{\delta}$, respectively. For example, we have
\begin{equation}
\hat{\mathcal N}_{\pm\bar\mu_\sigma}|v,\lambda_\sigma,\lambda_\delta\rangle =|v\pm\text{sign}(\lambda_\sigma v), v^{-1}[v\pm \text{sign}(\lambda_\sigma v)]\lambda_\sigma,\lambda_\delta\rangle.
\end{equation}

To complete the ingredients that are needed for the hybrid quantization of the Gowdy model, we have to select a Fock quantization of the inhomogeneous sector.  Actually, it has been proven that it is possible to single out a unique Fock quantization (given by a Fock representation and a Heisenberg dynamics for the background independent part of the fields), up to unitary equivalence, by imposing certain natural conditions, that require that the symmetry generated by $H_{\theta}$ and the quantum evolution of the creation and annihilation operators be unitarily implementable \cite{men2,men3}. In particular, this result removes the freedom of choice among the infinite number of inequivalent Fock representations, that may lead to different physics. Besides, the unitarity of the Heisenberg dynamics also imposes a concrete parametrization for the non-zero modes of both the gravitational field $\chi$ and the matter field $\Phi$ in terms of the background variables (namely the zero modes). Following these criteria, we represent the inhomogeneous sector of our hybrid model in Fock spaces $\mathcal{F}^{\alpha}$ (with $\alpha=\chi,\Phi$) chosen in Refs. \cite{men2,men3}. An orthonormal basis of each of these Fock spaces is provided by the $n$--particle states $|\mathfrak{n}^{\alpha}\rangle=|\cdots, n_{-2}^{\alpha}, n_{-1}^{\alpha},n_1^{\alpha},n_2^{\alpha},\cdots\rangle $, where $n_l^{\alpha}$ denotes the occupation number of the field $\alpha$ in the mode $l\in\mathbb{Z}-\{0\}$. Let $\hat{a}_l^{(\alpha)}$ and $\hat{a}_{l}^{(\alpha)\dagger}$ denote the annihilation and creation operators of that mode, respectively.  We can then reach a kinematic Hilbert space for the hybrid quantization of the Gowdy model by taking the tensor product ${\mathcal H}_{\text{kin}}={\mathcal H}_{\text{kin}}^{\text{BI}}\otimes L^{2}(\mathbb{R},d\phi)\otimes \mathcal{F}^{\chi}\otimes \mathcal{F}^{\Phi}$. Excluding the zero mode of the matter field temporally from our considerations, an orthonormal basis for the Hilbert space of the rest of the system is formed by the states $|v,\lambda_{\sigma},\lambda_{\delta},\mathfrak{n}^\chi,\mathfrak{n}^{\Phi} \rangle$ for all real values of the three first labels and all sets $\mathfrak{n}^\chi$ and $\mathfrak{n}^{\Phi}$ of integers with a finite number of non-vanishing elements. 

Finally, we are in an adequate position to represent the constraints of the system as densely defined operators on this Hilbert space. Choosing normal ordering, the generator of the translations in the circle reads \cite{hybrid-matter}
\begin{equation}
\hat{H}_{\theta}=\sum_{l=1}^{\infty}  \sum_{\alpha=\chi,\Phi}  l \left(\hat{a}^{(\alpha)\dagger}_{l}\hat{a}^{(\alpha)}_{l}-\hat{a}^{(\alpha)\dagger}_{-l}\hat{a}^{(\alpha)}_{-l}\right).
\end{equation}
This constraint leads to the condition  
\begin{equation}\label{mom}
\sum_{l=1}^{\infty}  \sum_{\alpha=\chi,\Phi}  l(n^{\alpha}_{l}-n^{\alpha}_{-l})=0
\end{equation}
on $n$--particle states of the inhomogeneities. Those states that satisfy the condition span a proper Fock subspace $\mathcal{F_{\text{p}}}$ of $\mathcal{F}^{\chi}\otimes\mathcal{F}^{\Phi}$. Let us now consider the quantum Hamiltonian constraint. We choose again normal ordering for the creation and annihilation operators of the inhomogeneous sector, while for the part of the constraint that acts on the homogeneous sector we choose a convenient symmetrization inspired by the MMO prescription \cite{hybrid3}. Rational powers of the norm of the triad variables are represented adopting an algebraic symmetrization, which  decouples the states of zero homogeneous volume $v$ from their orthogonal complement. Like in the FLRW cosmology, this fact allows us to remove the states with vanishing homogeneous volume from our kinematic space, therefore eliminating the quantum kinematic analogues of  the classical singularities with $v=0$. Moreover, again like in the FLRW model, one finds that the action of the Hamiltonian constraint does not mix states with different orientations of any of the components of the triad or, equivalently, with different signs  of the variables $v$, $\lambda_{\sigma}$, and $\lambda_{\delta}$. Hence, as far as the constraints of the system are concerned, one can restrict all considerations, e.g., to the sector of strictly positive labels for the homogeneous geometry. Taking this into account, we redefine $\Lambda_{a}=\ln(\lambda_{a})$ for $a=\sigma,\delta$ so that the anisotropy labels continue to take values over the real line. 

The Hamiltonian constraint ${\hat H}_{S}={\hat H}_\text{hom}+{\hat H}_\text{inh}$ that one obtains with this hybrid quantization, after performing a densitization similar to that in the FLRW case, has the following form \cite{hybrid3,hybrid4}:
\begin{equation}\label{CBI}
{\hat H}_\text{hom}= \frac{\hat{\pi}_{\phi}^{2}}{2} -\frac{\pi G } {16} \sum_{a\neq b} \sum_{b} \widehat{\Theta}_{a}\widehat{\Theta}_{b} , \end{equation}
\begin{equation}\label{inh} 
{\hat H}_\text{inh}= 2\pi (4 \pi G \gamma \sqrt{\Delta} )^{2/3} \widehat{e^{2\Lambda_\theta}} {\hat H}_{F}+ \frac{\pi G^{4/3}   }{16 (4 \pi \gamma \sqrt{\Delta})^{2/3}} \widehat{e^{-2\Lambda_\theta}}\widehat D(\widehat{\Theta}_{\delta}+\widehat{\Theta}_{\sigma})^{2} \widehat D {\hat H}_{I}.
\end{equation}
Here $a,b \in \{\theta,\sigma,\delta\}$. As we have already commented, ${\hat H}_\text{hom}$ is the constraint operator for the Bianchi I model with a homogeneous massless scalar field in LQC, according to the MMO prescription \cite{mmp1}. On the other hand,  the operator $\pi G  \gamma \widehat{\Theta}_{a}$ is the representation of ${c_{a}p_{a}}$, which is a constant of motion in the classical theory. We have defined
\begin{equation}
\widehat{\Theta}_{a}=\frac{1}{2i}\widehat{\sqrt{|v|}}\left[\left(\hat{\mathcal{N}}_{2\bar{\mu}_{a}}-\hat{\mathcal{N}}_{-2\bar{\mu}_{a}}\right)\widehat{\text{sign}(p_{a})}+\widehat{{\text{sign}}(p_{a})}\left(\hat{\mathcal{N}}_{2\bar{\mu}_{a}}-\hat{\mathcal{N}}_{-2\bar{\mu}_{a}}\right)\right]\widehat{\sqrt{|v|}},
\end{equation}
similar to the operator ${\hat{\Omega}}_0$ introduced in the isotropic case [see Eq. \eqref{Omegaop}]. In addition, the operator $\widehat{D}$ represents the product of the volume by its inverse [which is regularized in the standard way within LQC; see Eq. \eqref{invp}]. Its action on the basis of volume eigenstates is 
\begin{equation}
\widehat{D}|v\rangle=v\left(\sqrt{|v+1|}-\sqrt{|v-1|}\right)^2|v\rangle.
\end{equation} 

The contribution of the inhomogeneities is captured by ${\hat H}_{\text{F}}$ and ${\hat H}_{I}$.  The operator ${\hat H}_{\text{F}}$ can be understood as a free-field Hamiltonian, that leaves invariant the $n$--particle states. It is given by
\begin{equation}
{\hat H}_{F}=\sum_{l=1}^{\infty} \sum_{\alpha=\chi, \Phi} l \left( \hat{a}^{(\alpha)\dagger}_{l} \hat{a}^{(\alpha)}_{l} + \hat{a}^{(\alpha)\dagger}_{-l} \hat{a}^{(\alpha)}_{-l} \right).
\end{equation}
The operator ${\hat H}_{I}$ may be interpreted as an interaction Hamiltonian that creates and annihilates an infinite collection of pairs of particles, while preserving the momentum constraint ${\hat H}_{\theta}$. Explicitly,
\begin{equation}
{\hat H}_{I}=\sum_{l=1}^{\infty} \sum_{\alpha=\chi, \Phi}   \frac{1}{l} \left( \hat{a}^{(\alpha)\dagger}_{l} \hat{a}^{(\alpha)}_{l} + \hat{a}^{(\alpha)\dagger}_{-l} \hat{a}^{(\alpha)}_{-l} +\hat{a}^{(\alpha)\dagger}_{l} \hat{a}^{(\alpha)\dagger}_{-l} + \hat{a}^{(\alpha)}_{l} \hat{a}^{(\alpha)}_{-l}\right).
\end{equation}
It is worth remarking that the inhomogeneities of both fields contribute to the constraints in exactly the same way. 

The action of the Hamiltonian constraint operator ${\hat H}_{S}$ does not relate all of the states with different values of $v\in \mathbb{R}^+$ and $\Lambda_{a}\in \mathbb{R}$, with $a= \sigma,\delta$. There are invariant subspaces in the Hilbert space spanned by those states. Each of these subspaces provides a superselection sector for the quantum theory. The superselection sectors in the homogeneous volume $v$ are semilattices of step four, $\mathcal{L}_{\bar{\varepsilon}}^+=\{\bar{\varepsilon}+4n, n\in\mathbb{N}\}$, determined by the initial point $\bar{\varepsilon}\in(0,4]$, exactly as in the FLRW model. Note that, again, the homogeneous volume is bounded from below by a strictly positive quantity in each of these sectors. The superselection sectors in $\Lambda_{a}$ are more complicated. If we fix some initial data $\Lambda_{a}^{\ast}$ and $\bar{\varepsilon}$, the values of $\Lambda_{a}$ in the corresponding sector (constructed by the repeated action of the Hamiltonian constraint) take the form $\Lambda_{a}=\Lambda_{a}^{\ast}+\Lambda_{\bar{\varepsilon}}$, where $\Lambda_{\bar{\varepsilon}}$ is any of the elements of a certain set $\mathcal{W}_{\bar{\varepsilon}}$ that is countable and dense in the real line \cite{hybrid3}:
\begin{equation}
\mathcal{W}_{\bar{\varepsilon}}=\left\{z\ln\left(\frac{\bar{\varepsilon}-2}{\bar{\varepsilon}}\right)+ \sum_{n,m\in\mathbb{N}}
k_m^n\ln\left(\frac {\bar{\varepsilon}+2n } {\bar{\varepsilon}+2m} \right) \right\}.
\end{equation}
Here, $k_m^n\in\mathbb{N}$ and $z\in\mathbb{Z}$ if $\bar{\varepsilon}>2$, while $z=0$ when $\bar{\varepsilon}\leq2$.

Given that the action of $\widehat{\Theta}_{a}$ is considerably complicated, it has not been possible to elucidate yet whether this operator is self-adjoint. In spite of this, it is common to assume that ${\hat H}_\text{hom}$ is essentially self-adjoint \cite{awe1} and that the same applies to ${\hat H}_{S}$ \cite{hybrid3,hybrid4}. Regardless of this, one can try to formally solve the constraints of the Gowdy model.  The solutions turn out to be completely determined by the data on the section of the $v$-space defined by $v=\bar{\varepsilon}$. Thanks to this fact, one can characterize the physical Hilbert space as  the Hilbert space of such initial data, with an inner product that can be determined by imposing reality conditions on a complete set of observables \cite{reality,reality2}. In this way, one arrives to the space $\mathcal{H}_{\text{phys}}=\mathcal{H}_{\text{phys}}^{\text{BI}}\otimes L^{2}(\mathbb{R},d\phi) \otimes \mathcal{F}_{\text{p}}$, where $\mathcal{H}_{\text{phys}}^{\text{BI}}$ is the physical Hilbert space for Bianchi I cosmologies derived in Ref. \cite{hybrid4}.

\section{Hybrid LQC: Cosmological Perturbations}

After testing the viability of the hybrid quantization strategy in the Gowdy model, the approach was also applied to the discussion of a much more relevant scenario in cosmology, namely the study of primordial cosmological perturbations in the very early stages of the Universe.  Using that the inflationary Universe is usually described as an FLRW cosmology that plays the role of a background where the perturbations develop and propagate, the idea was to quantize this background in the framework of LQC and treat the perturbations with the techniques of QFT in a curved spacetime. The hybrid approach then transforms the curved,  FLRW classical background into a quantum spacetime with which the quantum field excitations corresponding to the perturbations coexist and interact by means of the gravitational constraints. For simplicity and for a better control of the mathematical techniques of QFT, we will again assume that the spatial sections are compact, with a three-torus topology. On the other hand, in order to isolate the perturbative degrees of freedom that do not depend on  a possible perturbative diffeomorphism of the FLRW spacetime, that would result in a new identification of the background geometry, we will adopt a description in terms of perturbative gauge invariants. For cosmological scalar perturbations, one can employ the invariants introduced by Mukhanov and Sasaki (MS) \cite{MukhanovSasaki,sasaki,sasakikodama} (considered as a pair of canonical fields). Gauge invariants are also the tensor perturbations, as well as the degrees of freedom of a Dirac field if it is present \cite{fermilala} (treating this field entirely as a perturbation).  The description of the phase space of the perturbations can be completed with suitable redefinitions of the generators of the perturbative diffeomorphisms and canonical momenta of them. For the hybrid quantization, a piece of information that is most relevant as far as the inhomogeneities are concerned is the choice of a Fock representation for the gauge invariant fields. To restrict this choice and adopt a representation with especially appealing physical properties, we will still adhere to the criterion that the Fock quantization must allow a unitary implementation of the spatial symmetries of the model and of the Heisenberg dynamics associated with the creation and annihilation operators. With these ingredients, we will proceed to construct a hybrid quantum theory for the perturbed system. On this system, we will see that the only non-trivial constraint turns out to be the zero mode of the Hamiltonian constraint. We will then discuss its quantum imposition. Moreover, we will show how to extract from it (with a convenient ansatz and plausible approximations) Schr$\ddot{\rm o}$dinger equations for the perturbations, as well as effective equations to describe the propagation of the perturbations on the FLRW geometry subject to quantum effects. These equations can be used to study modifications to the power spectra of the cosmological perturbations, originated from quantum gravitational effects. The program that we have outlined will be implemented in this and the following five sections. 

We start by constructing a convenient canonical description of the system formed by the FLRW cosmology and its perturbations that contains a complete set of gauge invariants. As in the case studied in our introduction to LQC, the FLRW spacetimes that we will consider possess compact sections with the topology of a three-torus. Their geometry can be described by a scale factor $a$ and its canonical momentum $\pi_a$ (or equivalently by the pair of variables $c$ and $p$ that determine the Ashtekar-Barbero variables in LQC). With the same choice of reference system for this cosmological background that we employed in the previous expositions about LQC, the coordinate volume of the spatial sections equals $8\pi^3$.  As before, these spacetimes will contain a homogeneous scalar field, $\phi$, responsible of the expansion and that consequently will play the role of an inflaton. This inflaton can be interpreted as the zero mode of a generally inhomogeneous scalar field $\Phi$, interpretation that will be especially useful at the moment of introducing perturbations in the system. On the other hand, the main difference with respect to our previous studies is that we will now allow the possible existence of a potential ${\mathcal V}(\phi)$ for this inflaton.  

The FLRW system is subject only to a non-trivial homogeneous Hamiltonian constraint, as we have discussed above. It can be written as $H_0=0$ where\footnote{We reserve the notation $H_S$ for the constraint of the whole perturbed model.}
\begin{equation}\label{H0}
H_{0}= \frac{1}{16\pi^3 a^3} \left(\pi_{\phi}^2- \frac{4\pi G}{3}a^2 \pi_{a}^2+128 \pi^6 a^{6} {\mathcal V}(\phi) \right).
\end{equation}

Let us next introduce perturbations in this system, both for the geometry and for the matter scalar field. It is also possible to introduce a Dirac field to describe fermions, regarded as perturbations of the FLRW cosmology \cite{fermilala}. We postpone the consideration of these fermions to the next section. We can separate the metric and inflaton perturbations into scalar, vector, and tensor, depending on their behavior under the symmetries of the spatial sections (notice that these symmetries provide the Euclidean group in the limit in which the sections become non-compact). In addition, using that the spatial Laplacian (and the Dirac operator) corresponding to the auxiliary Euclidean metric ${}^0h_{ab}$ (with unit determinant) defined on our toroidal sections respect these symmetries, we can expand the different perturbations in eigenmodes of this differential operator. Moreover, since the spatial sections are compact, these modes are discrete. In this way, we can deal with the spatial dependence of the perturbations by considering infinite sequences of modes. For instance, choosing again (orthogonal) spatial coordinates of period equal to $2\pi$, we expand the scalar perturbations of the metric and the matter field in a Fourier basis of sines and cosines,
\begin{equation}
Q^{\vec{k},+} (\vec\theta)= \sqrt 2\cos\left(\vec{k}\cdot\vec{\theta}\right),\quad\quad Q^{\vec{k},-} (\vec{\theta})= \sqrt 2\sin\left(\vec{k}\cdot\vec{\theta}\right).
\end{equation}
Here, the vector notation $\vec{\theta}$ stands for the spatial coordinates $(\theta,\sigma,\delta)$, and the Euclidean scalar product has been denoted with a dot symbol. Each mode is characterized by a wavevector $\vec{k} \in \mathbb{Z}^3-\{0\}$, with strictly positive first non-vanishing component. Note that, in this way, we are not including the zero mode, that is part of the degrees of freedom considered in the FLRW background. The eigenvalue of the spatial Laplacian corresponding to $\vec{k}$ is $-\omega_k^2=- \vec{k} \cdot \vec{k}$. Scalar perturbations are described then by the corresponding Fourier coefficients of the scalar field $\Phi$ (without the zero mode, namely the inflaton), the trace and traceless scalar parts of the spatial metric $h_{ab}$ (without the FLRW contribution), the lapse $N$ (without its homogeneous part), and the scalar part of the shift $N^{a}$. 

Similarly, tensor perturbations are described by the Fourier-like coefficients of the tensor part of the spatial metric, with two possible polarizations. These coefficients arise from the expansion in terms of the real tensor harmonics $G^{\vec{k},\varepsilon,\pm}_{ab}$, eigentensors of the spatial Laplacian \cite{Mikel}.  As above, the tuple $\vec{k}$ can take here any value in $\mathbb{Z}^3 - \{0\}$, with positive first non-vanishing component, while $\varepsilon$ is the dichotomic label that specifies the polarization, and the superscripts $\pm$ indicate whether the harmonic is even or odd under a periodic translation of $\vec{\theta}$, as in the scalar case. Vector perturbations, on the other hand, are described by the remaining parts of the shift and the spatial metric, that can be expanded in Fourier-like coefficients in terms of eigenvectors $S^{\vec{k}}_a$ of the Laplace operator and of tensors obtained from those by spatial covariant derivatives \cite{HalliwellHawking}. Here, $\vec{k}$ is again any non-vanishing tuple of integers. It is convenient to parametrize all of these mode coefficients as follows:
\begin{eqnarray}\label{3metric}
h_{ab}(t,\vec\theta) &=& a^2(t) \; {}^0h_{ab}(\vec\theta )\left[1+2\sum_{\vec{k},\pm}a_{\vec{k},\pm} (t) Q^{\vec{k},\pm}(\vec\theta)\right] \nonumber\\
&+&6 a^2(t)\sum_{\vec{k},\pm}b_{\vec{k},\pm}(t)\left[\frac1{\omega_{k} ^2} Q^{\vec{k},\pm}_{|ab}(\vec\theta)+\frac13{}^0h_{ab}(\vec\theta)Q^{\vec{k},\pm}(\vec\theta)\right]\\
&+&  a^2(t) \sum_{\vec{k}} c_{\vec{k}}(t)  S^{\vec{k}}_{(a|b)}(\vec\theta)+2\sqrt{6} a^2(t) \sum_{\vec{k},\varepsilon,\pm}d_{\vec{k},\varepsilon,\pm}(t) G^{\vec{k},\varepsilon,\pm}_{ab}(\vec\theta) , \\
\label{lapse}
N(t,\vec\theta) &=& N_0(t)+ \frac{6\pi^2}{G}a^3(t) \sum_{\vec{k},\pm}g_{\vec{k},\pm}(t)Q^{\vec{k},\pm}(\vec\theta), \\ 
\label{shift}
N_a(t,\vec\theta) &=& a^{2}(t)\sum_{\vec{k},\pm}\frac1{\omega_k^2}l_{\vec{k},\pm}(t)Q^{\vec{k},\pm}_{|a}(\vec\theta) + a(t) \sum_{\vec{k}} v_{\vec{k}}(t) S^{\vec{k}}_{a}(\vec\theta),\\
\label{field}
\Phi(t,\vec\theta) &=& \phi(t)+\sqrt{\frac{3}{4\pi G}}\sum_{\vec{k},\pm}f_{\vec{k},\pm}(t) Q^{\vec{k},\pm} (\vec\theta).
\end{eqnarray}
A vertical bar stands for the spatial covariant derivative with respect to the Euclidean metric ${}^0h_{ab}$, and a parenthesis enclosing two spatial indices indicates symmetrization. Thus,  the scalar perturbations are determined by $a_{\vec{k},\pm}$, $b_{\vec{k},\pm}$, $g_{\vec{k},\pm}$,  $l_{\vec{k},\pm}$, and $ f_{\vec{k},\pm}  $, whereas the tensor perturbations are described by the coefficients $d_{\vec{k},\varepsilon,\pm}$, and the vector perturbations by $c_{\vec{k}}$ and $v_{\vec{k}}$. We have normalized some of these coefficients in a convenient way to absorb several factors in the formulas that we will use in our discussion. 

Inserting these expressions in the Hilbert-Einstein action minimally coupled to the scalar field $\Phi$ (with suitable boundary terms) and truncating the result at quadratic order in the coefficents of the perturbations, it is possible to reach a Hamiltonian formulation for our system \cite{HalliwellHawking,Mikel}. In this formulation, the above coefficients for the perturbations either play the role of Lagrange multipliers of some of the constraints, or form a canonical set together with the FLRW scale factor, the inflaton, and suitable momenta for all of them. In other words, at the order of our truncation in the action, the system formed by the FLRW cosmology and its perturbations is a totally constrained system that admits a canonical symplectic structure \cite{hyb-pert4}. It is worth emphasizing that, at the considered truncation order, we are treating exactly the zero modes that determine the FLRW background, so that the perturbations that we have expressed explicitly do not contain zero modes. On the other hand, with the kind of matter content considered in our discussion, it is possible to show that the vector perturbations do not include any physical degree of freedom, but are pure gauge. To simplify our exposition, we will therefore eliminate them from our analysis in the following.

The perturbed system that we have constructed is subject to two types of constraints. On the one hand, the perturbations of the momentum and Hamiltonian constraints  lead to a collection of constraints that are linear in the perturbations, and that appear accompanied by Lagrange multipliers that are also linear perturbative factors. Explicitly, the mode ${H}^{\vec{k},\pm}_{\uparrow 1}$ of the linear perturbative momentum constraint has Lagrange multiplier given by  the coefficient $l_{ \vec{k},\pm}$ of the perturbations of the shift vector, while the mode $H^{\vec{k},\pm}_{1}$ of the linear perturbative scalar constraint adopts as Lagrange multiplier the coefficient  $g_{\vec{k},\pm}$ of the perturbation of the lapse. These constraints depend exclusively on the scalar perturbations of the metric, once the vector perturbations have been gauged away. A different type of constraint is the zero mode of the Hamiltonian constraint, which can be considered a global restriction on the system formed by the FLRW cosmology and the perturbations. This constraint has a Lagrange multiplier given by the homogeneous lapse function $N_0$, and is the sum of two contributions: a term that reproduces what would have been the constraint $H_{ 0 }$ of the FLRW cosmology in the absence of perturbations, and an additional term $H_2$ that contains the perturbative contribution and that is quadratic in the perturbations. This latter term is composed in turn of two parts, ${}^{(s)}H_{2}$ and ${}^{(T)}H_{2}$, respectively formed by the contributions of the scalar and the tensor perturbations. In this way, the total Hamiltonian takes the expression
\begin{equation}
H =  N_0 \left[  H_{ 0 }+   {}^{(s)}H_{2}+ {}^{(T)}H_{2} \right] +  \sum g_{\vec{k},\pm} H^{\vec{k},\pm}_{1} +  \sum l_{ \vec{k},\pm}  H^{\vec{k},\pm}_{\uparrow 1} .
\end{equation}
Moreover, the quadratic perturbative contributions to the zero mode of the Hamiltonian constraint can be decomposed as the sum of the contributions of each of the modes of the perturbations as follows:
\begin{equation}
{}^{(s)}H_2= \sum_{\vec{k},\pm} {}^{(s)}H^{ \vec{k},\pm}_{2}, \quad\quad
{}^{(T)}H_2=\sum_{\vec{k},\varepsilon,\pm} {}^{(T)}H_2^{\vec{k},\varepsilon,\pm}.
\end{equation}

The variables that we have chosen to describe the perturbative degrees of freedom have the drawback that they do not commute with the linear perturbative constraints, even when the FLRW cosmology is taken as a fixed entity with vanishing Poissson brackets. As a consequence, those variables would change if one performs a perturbative diffeomorphism, that  would alter the form of the FLRW background without affecting the physics. To avoid this problem with the physical identification of the background, it is most convenient to use a set of variables that indeed commutes with the linear perturbative constraints when the zero modes are frozen in the computation of Poisson brackets. This leads us to consider gauge invariants for the perturbations. In the case of flat spatial sections, the gauge invariant degrees of freedom of the scalar perturbations are usually described in cosmology employing MS invariants, because they are straightforwardly related to the co-moving curvature perturbations. The variables that we have introduced for the tensor perturbations, on the other hand, are directly gauge invariant, and we will only redefine them linearly to re-express their dynamical contribution to the Hamiltonian in a convenient way.

With this motivation, we are going to introduce a change of variables for the perturbations, from the canonical set that we have been using to a new set formed by the following variables \cite{langlo,hyb-pert4}: 
\begin{itemize}
\item The mode coefficients of the MS gauge invariant field, $\nu_{\vec{k},\pm}$. Explicitly, they are given by the formula \cite{hyb-pert2,hyb-pert3}
\begin{equation}\label{Mmode}
\nu_{\vec{k},\pm} = \sqrt{\frac{6\pi^2}{G}}a \left[f_{\vec{k},\pm}+\sqrt{\frac{3}{4\pi G}}\frac{\pi_{\phi}}{ a\pi_a} (a_{\vec{k},\pm}+b_{\vec{k},\pm})\right].
\end{equation}
We notice that these coefficients mix the scalar perturbations of the metric and the perturbations of the matter scalar field. 
\item The mode coefficients of the tensor perturbations conveniently rescaled \cite{hybr-ten}:
\begin{equation}
\tilde{d}_{\vec{k},\varepsilon,\pm}=\sqrt{\frac{6\pi^2}{G}}a\,d_{\vec{k},\varepsilon,\pm}.
\end{equation}
This rescaling simplifies the dependence of the Hamiltonian on the tensor perturbations.
\item The mode coefficients $\pi_{\nu_{\vec{k},\pm}}$ and $\pi_{\tilde{d}_{\vec{k},\varepsilon}^{\pm}}$ of the canonical momenta of the above fields, defined also as gauge invariants. There exists a certain ambiguity in the specification of these momenta, since once can always add a contribution that is linear in the configuration fields, multiplied by any function of the FLRW background. A convenient criterion to fix this contribution is to require that the time derivative of each of these momenta, as dictated by Hamilton's equations, is proportional to the corresponding configuration variable. This condition amounts to demand that the Hamiltonian that generates the dynamics of the scalar and tensor perturbations should not contain cross terms between the configuration fields and their momenta, and turns out to determine the latter of these variables completely.
\item An Abelianization of the linear perturbative constraints. Actually, at the order of our perturbative truncation in the action, it is possible to modify these constraints on the scalar perturbations with terms that are linear in those perturbations and such that the new constraints that one obtains commute under Poisson brackets among them, as well as with the MS field and its momentum, after freezing the zero modes. To achieve this Abelianization, it suffices to introduce the replacement 
\begin{equation} 
{H}^{\vec{k},\pm}_{1} \rightarrow {\breve{H}}^{\vec{k},\pm}_{1} = {H}^{\vec{k},\pm}_{1}- \frac{18\pi^2}{G} a^3 H_{0} a_{\vec{k},\pm}.
\end{equation}
This new linear perturbative scalar constraint is used together with ${H}^{\vec{k},\pm}_{\uparrow 1} $ as additional variables in our canonical set.
\item Suitable momenta of the Abelianized linear perturbative constraints. As far as those constraints generate gauge transformations consisting of perturbative diffeomorphisms, their momenta can be interpreted as variables that parametrize possible gauge fixations for the perturbations.  A especially simple choice is
\begin{equation}\label{pigauge}
\pi_{{\breve{H}}^{\vec{k},\pm}_{1}}=\frac{1}{a\pi_{a}}(a_{\vec{k},\pm}+ b_{\vec{k},\pm}), \quad \quad \pi_{{H}^{\vec{k},\pm}_{\uparrow 1}}=-3 b_{\vec{k},\pm} .
\end{equation}
\end{itemize}

Remarkably, the introduced change of variables for the scalar and tensor perturbations can be completed into a canonical transformation for the entire system (that is, without freezing the background), at the considered truncation order, by modifying the zero modes with terms that are quadratic in the perturbations \cite{hyb-pert4,pintoneto1,pintoneto2}. For this, we can proceed as follows. We substitute the old perturbative variables in the Legendre term of the action (or, equivalently, in the symplectic potential) as functions of the new ones and, after several integrations by parts and convenient identifications of factors, we find new zero modes that keep the canonical form of the Legendre term up to perturbative contributions that are negligible in our truncation scheme. The new configuration variables obtained in this way adopt the generic expression
\begin{equation}\label{iota}
{\tilde w}^\iota_q = w^\iota_q + \frac{1}{2} \sum_{m,\vec{k},\pm} \left[ X^{\vec{k},\pm}_{q_m} \frac{{\partial X}^{\vec{k},\pm}_{p_m}} {\partial { w^\iota_p} } - \frac{{\partial  X}^{\vec{k},\pm}_{q_m}}{\partial { w^\iota_p}} X^{\vec{k},\pm}_{p_m} \right],
\end{equation}
where we have called $ \left\{ w_q^\iota \right\} = \left\{ a , \phi \right\}$ the configuration variables of the zero mode sector, $ \left\{ w^p_\iota \right\}$ are their momenta ($\iota=1,2$), and $\{ X^{\vec{k},\pm}_{q_m}, X^{\vec{k},\pm}_{p_m}  \}$ are the old variables for the scalar and tensor perturbations, each of which is given by a different value of the label $m$. A tilde on top of any of these canonical quantities indicates its new counterpart, defined according to the above procedure. 

In the case of the momentum variables for the zero modes, the change is given by a formula of the same kind, but with a flip of sign in the term that provides the corrections quadratic in the perturbations,
\begin{equation}
{\tilde w}^\iota_p = w^\iota_p - \frac{1}{2} \sum_{m,\vec{k},\pm} \left[ X^{\vec{k},\pm}_{q_m} \frac{{\partial X}^{\vec{k},\pm}_{p_m}} {\partial { w^\iota_q} } - \frac{{\partial  X}^{\vec{k},\pm}_{q_m}}{\partial { w^\iota_q}} X^{\vec{k},\pm}_{p_m} \right].
\end{equation}

The availability of a canonical set for the entire perturbed system, formed by the FLRW cosmology and the perturbations, is of the greatest importance. In particular, it makes possible an easy implementation of the hybrid strategy following canonical quantization rules. But, in order to proceed to this quantization, we still have to determine the form of the zero mode of the Hamiltonian constraint (the only non-linear perturbative constraint of the system) in terms of the new canonical set, keeping the quadratic truncation order. In order to do this, we notice that, since the change of zero modes is quadratic in the perturbations, an expansion of the FLRW contribution $H_0$ around the new zero modes leads immediately to the desired constraint if we only include the first derivative contribution. Let us introduce the compact notation
\begin{eqnarray}
&& \{ w^\iota\} = \{ w^\iota_q,w^p_\iota \},\quad\quad
\{ {\tilde{w}}^\iota \} = \{ {\tilde{w}}^\iota_q,{\tilde{w}}^p_\iota \},\\
&& \left\{ {\tilde{X}}^{\vec{k},\pm}_{m} \right\} = \left\{{\tilde{X}}^{\vec{k},\pm}_{q_m}, {\tilde{X}}^{\vec{k},\pm}_{p_m}  \right\}. \end{eqnarray}
Then, according to our comments, the expression of the new global scalar constraint at our truncation order is \cite{hyb-pert4,hybr-ten}
\begin{equation}\label{qcorrect}
H_{0}+ \sum_\iota \left( w^\iota - {\tilde w}^\iota \right) \frac{\partial H_{0}}{\partial {{ w}^\iota}} + \sum_{\vec{k},\pm} {}^{(s)}H^{\vec{k},\pm}_{2} + \sum_{\vec{k},\varepsilon,\pm} {}^{(T)}H^{\vec{k},\varepsilon,\pm}_{2},
\end{equation}
with the phase space dependence of $H_{0}$, its derivatives, ${}^{(s)}H^{\vec{k},\pm}_{2} $, and ${}^{(T)}H^{\vec{k},\varepsilon,\pm}_{2}$ evaluated directly at $({\tilde w}^\iota, {\tilde{X}}^{\vec{k},\pm}_{m})$. Namely, in formula \eqref{qcorrect}, the evaluation of the Hamiltonian functions  must be made as if one identified the old and the new set of variables. As a consequence, the contribution of each of the modes of the perturbations to the new global scalar constraint is
\begin{equation}\label{qmodecorrect}
{}^{(s)}{\breve{H}}^{\vec{k},\pm}_{2}\sim {}^{(s)}{H}^{\vec {k},\pm}_{2} + \sum_\iota {}^{(s)}\Delta{\tilde w}^\iota_{\vec{k},\pm} \frac{\partial H_{0}} {\partial {{ w}^\iota}} , \end{equation} \begin{equation}\label{qmodecorrectt}
{}^{(T)}{\breve{H}}^{\vec{k},\varepsilon,\pm}_{2} \sim {}^{(T)}{H}^{\vec{k},\varepsilon,\pm}_{2} + \sum_\iota {}^{(T)}\Delta{\tilde w}^\iota_{\vec{k},\varepsilon,\pm} \frac{\partial H_{0}} {\partial {{ w}^\iota}},
\end{equation}
where the symbol $\sim$ indicates equality modulo the Abelianized linear constraints and up to the relevant perturbative order in our truncation. Besides, we have called
\begin{equation}
w^\iota- {\tilde w}^\iota = \sum_{\vec{k},\pm} {}^{(s)}\Delta {\tilde w}^\iota_{\vec{k},\pm}+\sum_{\vec{k},\varepsilon,\pm} {}^{(T)}\Delta {\tilde w}^\iota_{\vec{k},\varepsilon,\pm},
\end{equation}
where the superscripts $(s)$ and $(T)$ stand for the quadratic contributions of scalar and tensor nature, respectively. It is possible to prove that the sum of contributions in the left-hand side of Eq. \eqref{qmodecorrect} gives precisely the MS Hamiltonian \cite{hyb-pert4}, i.e. the Hamiltonian that generates the dynamical evolution (in proper time) of the MS field on the FLRW background. Likewise, the sum ${}^{(T)}{\breve{H}}^{\vec{k},\varepsilon,\pm}_{2}$ of the tensor contributions to the constraint provides a dynamical Hamiltonian of harmonic oscillator type for the tensor perturbations on the FLRW cosmological background.

It is worth pointing out that, in the definition of these perturbative Hamiltonians, we can replace the squared momentum of the inflaton with $\pi_{\phi}^2-16\pi^{3}a^3 H_0$. This is so because all the new terms proportional to $H_0$ that are produced in this way are quadratic in the perturbations and can thus be absorbed in a redefinition of the zero mode of the lapse function up to a modification of the total scalar constraint that is at least quartic in those perturbations. Hence, such a modification is negligible at our truncation order. The freedom available in using this replacement can be fixed by requiring that the perturbative contribution to the Hamiltonian constraint be at most linear in the inflaton momentum, because one can always use the commented replacement to decrease the polynomial order in $\pi_{\phi}$ by two units until one reaches either a linear contribution of the inflaton momentum or a term that is independent of it. The total Hamiltonian of the system then becomes \cite{hyb-pert4}
\begin{equation}\label{totalhst}
H = {\bar N}_0 \left[   H_{0}+  \sum_{\vec{k},\pm} {}^{(s)}{\breve H}^{\vec{k},\pm}_{2} + \sum_{\vec{k},\varepsilon} {}^{(T)}{\breve{H}}_2^{\vec{k},\varepsilon,\pm} \right]+ \sum_{\vec{k},\pm} {\breve g}_{\vec{k},\pm} {\breve H}^{\vec{k},\pm}_{1}+ \sum_{\vec{k},\pm} {\breve l}_{\vec{k},\pm} {H}^{\vec{k},\pm}_{\uparrow 1},
\end{equation}
where ${\bar N}_0$ is the suitably redefined homogeneous lapse function that differs from the original one, $N_0$, in perturbative terms that are quadratic. Similarly, ${\breve g}_{\vec{k},\pm}$ and   ${\breve l}_{\vec{k},\pm}$ are Lagrange multipliers that differ from the original ones, ${g}_{\vec{k},\pm}$ and   ${l}_{\vec{k},\pm}$ respectively, by linear perturbative contributions. Their explicit expressions can be found in Ref. \cite{hyb-pert4}, but they are not especially relevant for the rest of our discussion.

Notice that this total Hamiltonian, imposed as a collection of constraints on the system, would include backreaction at the considered perturbative order. As we have commented, the contribution of the scalar perturbations to the global Hamiltonian constraint is nothing but the MS Hamiltonian, which is a sum of terms that are quadratic in the MS configuration variables and of quadratic terms in their momenta, but without terms that mix these two types of variables. This is a consequence of our choice of MS momentum field, as we explained when we introduced the new perturbative variables. A similar behavior is found in the contribution of the tensor perturbations to the Hamiltonian constraint. In more detail,
\begin{eqnarray}\label{Hscal}
{}^{(s)}{\breve H}^{\vec{k},\pm}_{2} &=& \frac{1}{2 {\tilde a} } \left[\left( \omega_{k}^2 + s^{(s)} + r^{(s)} \pi_{\tilde{\phi}}\right) \nu_{\vec{k},\pm}^2 + \pi_{\nu_{\vec{k},\pm}}^2 \right], \\
{}^{(T)}{\breve{H}}_2^{\vec{k},\varepsilon,\pm} &=& \frac{1}{2 {\tilde a}}\left[\left( \omega_{k}^2 + s^{(T)}\right) {\tilde{d}}_{\vec{k},\varepsilon,\pm}^2 + \pi_{{\tilde{d}}_{\vec{k},\varepsilon,\pm}}^2 \right].\label{Htens}
\end{eqnarray}
Here, $s^{(s)}+r^{(s)} \pi_{\tilde{\phi}}$ and $s^{(T)}$ play the role of effective background dependent masses for the scalar and the tensor perturbative modes, respectively. The expressions of these background functions are
\begin{eqnarray}
{s}^{(s)} &=& \frac{H_0^{(2)}}{ 32\pi^6{\tilde a}^4} \left(\frac{38\pi G}{3}- 9 \frac{ H_0^{(2)}}{{\tilde a}^2 \pi_{\tilde a}^2} \right)+  {\tilde a}^2 \left({\mathcal V}^{'' }({\tilde \phi})-\frac{16\pi G}{3} {\mathcal V}({\tilde \phi}) \right), \\
{s}^{(T)} &=& \frac{G}{ 48\pi^{5}} \frac{H_0^{(2)}}{{\tilde a}^4}-\frac{16\pi G}{3}{\tilde a}^2 {\mathcal V}({\tilde \phi})\\
r^{(s)} &=& - 12  \frac{\tilde{a}}{\pi_{\tilde{a}}} {\mathcal V}^{'}({\tilde{\phi}}).
\end{eqnarray} 
The prime symbol denotes de derivative of the potential ${\mathcal V}$ with respect to the inflaton ${\tilde \phi}$, and  
\begin{equation}\label{Hhomnew}
H_0^{(2)} = \frac{4\pi G}{3}{\tilde a}^2 \pi_{\tilde a}^2 - 128 \pi^6 {\tilde a}^6 {\mathcal V}({\tilde \phi}).
\end{equation}
We notice that
\begin{equation}\label{sctenma}
{s}^{(s)}={s}^{(T)} + \frac{9H_0^{(2)}}{ 32\pi^6{\tilde a}^4} \left(\frac{4\pi G}{3}-  \frac{ H_0^{(2)}}{{\tilde a}^2 \pi_{\tilde a}^2} \right)+ {\tilde a}^2 {\mathcal V}^{'' }({\tilde \phi}).
\end{equation}
In particular, substituting Eq. \eqref{Hhomnew}, we see that ${s}^{(s)}={s}^{(T)}$ when the inflaton potential ${\mathcal V}$ vanishes.

In total, after a convenient change of densitization similar to that explained in homogeneous and isotropic LQC (via multiplication by the homogeneous physical volume $V=8\pi^3\tilde{a}^3$), we obtain a Hamiltonian constraint that can be written in the form
\begin{equation}\label{scalar}
{H}_S = \frac{1}{2} \left[ {\pi}_{\tilde{\phi}}^2 - {H}_0^{(2)}- { \Theta}_e- {\Theta}_o {\pi}_{\tilde{\phi}} \right],
\end{equation}
where we have introduced the notation   
\begin{eqnarray}\label{Thetas}
\Theta_e&=& \sum_{\vec{k}, \pm} {}^{(s)} \Theta_e^{\vec{k},\pm}+ \sum_{\vec{k}, \varepsilon,\pm} {}^{(T)} \Theta_e^{\vec{k},\varepsilon,\pm},\\ \Theta_o&=& \sum_{\vec{k}, \pm} {}^{(s)} \Theta_o^{\vec{k},\pm},\\
{}^{(s)}\Theta_e^{\vec{k},\pm} &=& - \left[ (\vartheta_e \omega_k^2 + {}^{(s)}\vartheta_e^q ) \nu_{\vec{k},\pm}^2 + \vartheta_e {\pi}_{\nu_{\vec{k},\pm}}^2 \right], \\
{}^{(T)}\Theta_e^{\vec{k},\varepsilon,\pm} &=& - \left[ (\vartheta_e \omega_k^2 + {}^{(T)}\vartheta_e^q ) {\tilde d}_{\vec{k},\varepsilon,\pm}^2 + \vartheta_e {\pi}_{{\tilde d}_{\vec{k},\varepsilon,\pm}} ^2 \right], \\
{}^{(s)}\Theta_o^{\vec{k},\pm} &=&  - {}^{(s)}\vartheta_o \nu_{\vec{k},\pm}^2 ,
\end{eqnarray}
that explicitly separates the linear term in the inflaton momentum. Clearly, we have the identities
\begin{equation}\label{varthetas}
{}^{(s)}\vartheta_e^{q} =  \vartheta_e {s}^{(s)} ,\quad\quad
{}^{(T)}\vartheta_e^{q} =  \vartheta_e {s}^{(T)},  \quad\quad
{}^{(s)}\vartheta_o =  \vartheta_e r^{(s)},
\end{equation}
with $\vartheta_e = 8\pi^3{\tilde a}^2$.

We note that there is no tensor contribution to $ \Theta_o$. It is also worth remarking that all the  $\vartheta$-functions are independent of the particular mode that one considers. Besides, the part of the Hamiltonian 
constraint that contains the perturbative contribution is the same for the scalar and for the  tensor perturbations except for the difference in their background dependent mass. This shows up in the appearance of the terms $ {}^{(s)}\Theta_o^{\vec{k},\pm}$ and in the difference between  ${}^{(s)}\vartheta_e^{q}$ and ${}^{(T)}\vartheta_e^{q}$.

\section{Hybrid LQC: Inclusion of fermions}

In the matter content of our cosmological system, we can also include  fermionic fields, e.g. a Dirac field. Their pressence does not modify much our treatment if we consider them as perturbations, including the possible fermionic zero modes, so that they do not alter the dynamics of the homogeneous background cosmology in the linearized theory. Since the Dirac action is quadratic in the fermionic field, as a perturbation it couples directly only to the background FLRW geometry, but not to the perturbations of the metric, nor to the matter scalar field \cite{fermilala}. Moreover, for the same reason, the fermionic field does not contribute to the linearized perturbative constraints, that arise from the perturbation of the Hamiltonian and momentum constraints. As a consequence, the fermionic field can be treated as a gauge invariant perturbation at the considered order of truncation. This simplifies the formulation considerably.

If we adopt a Weyl representation \cite{DEathHall},  the Dirac field can be described by a pair of two-component spinors of definite chirality. We will call $\varphi^A$ and $\chi_{A'}^{*}$ the respective left-handed and right-handed spinors associated with the field. Capital Latin letters from the beginning of the alphabet, both primed and unprimed, take values equal to 1 or 2, corresponding to the two components of the chiral spinors. These indices will be raised and lowered using the antisymmetric symbols $\epsilon^{AB}$ and $\epsilon_{AB}$ (with e.g. $\epsilon_{12}=1$) , as well as their counterparts for right-handed chirality. It is most convenient to adopt an internal gauge such that the spatial part of the tetrad has vanishing temporal Lorentz components, namely $e^a_0=0$. As a consequence of this gauge fixation on the spin structure in four dimensions, the two-component spinors of the Dirac field can be viewed as families of cross-sections of a spinor bundle defined on the compact spatial sections. On the other hand, the Hamiltonian formalism of the Dirac field is initially complicated by the existence of second-class constraints that relate the field with its momentum. Nevertheless, one can eliminate these constraints and capture the canonical anticommutation relations of the Dirac field in anticommutators of its two-component spinors.  To take into full account this anticommuting character, we will treat these components as Grassmann variables \cite{berezin}. 

In a similar way as we did with the metric and the scalar field perturbations, we can decompose the spinors of the Dirac field in modes. Since the spatial differential operator that appears naturally in the dynamical equation of our fermionic field is the Dirac operator constructed with the auxiliary Euclidean triad ${}^{0}e^a_i$ on the toroidal spatial sections of our model (with $ {}^{0}e_{a}^{i}  \;{}^{0}e_{b}^{j}\delta_{ij} ={}^{0}h_{ab}$ being the Euclidean metric introduced above),  it is logical to treat the spatial dependence of the field by an expansion in eigenmodes of this Dirac operator. The spectrum of this operator is discrete, owing to the compactness of the sections. The eigenvalues are $\pm\omega_k$, where $\omega_k^2 = \vec{k} \cdot \vec{k}$ and $\vec{k} \in \mathbb{Z}^3$ is any tuple of integers. We are assuming a trivial spin structure on the spatial sections. Otherwise, the definition of $\omega_k$ would include a constant displacement of $\vec{k}$ characteristic of the specific spin structure chosen for the fermions \cite{dtorus}. Using these modes, we can express the two-component spinors of the Dirac field in the form
\begin{eqnarray} \label{modeexpansion}
\varphi_A(x)&=&\frac{1}{(2\pi)^{3/2}{\tilde a}^{3/2}}\sum_{\vec{k},(\pm)}\left[m_{\vec{k}}w^{\vec{k},(+)}_{A}+r^{*}_{\vec{k}}w^{\vec{k},(-)}_{A}\right],\\
\chi^{*}_{A'}(x)&=&\frac{1}{(2\pi)^{3/2}{\tilde a}^{3/2}}\sum_{\vec{k},(\pm)}\left[s^{*}_{\vec{k}}\left(w^{\vec{k},(+)}\right)^{*}_{A'}+t_{\vec{k}}\left(w^{\vec{k},(-)}\right)^{*}_{A'}\right].
\end{eqnarray}
Here $w^{\vec{k},(\pm)}_{A}$ are the left-handed Dirac eigenspinors with respective eigenvalue equal to $\pm\omega_k$. With our choice of the auxiliary Euclidean triad, and recalling that we have assumed a trivial spin structure, these eigenspinors take the expression
\begin{equation}\label{eigens}
w^{\vec{k},(\pm)}_{A}=u^{\vec{k},(\pm)}_{A}e^{i  \vec{k}\cdot\vec{\theta}}, 
\end{equation}
where the spinors $u^{\vec{k},(\pm)}_{A}$ are constant and normalized (including a choice of phase) so that 
\begin{eqnarray}
&&\left(u^{\vec{k},(\pm)}\right)^{*}_{1'} u^{\vec{k},(\pm)}_1+\left(u^{\vec{k},(\pm)}\right)^{*}_{2'} u^{\vec{k},(\pm)}_2=1, \\
&& \int d^3 \theta\,w^{\vec{k}',(+)}_{A}\epsilon^{AB}w^{\vec{k},(-)}_{B}=0,\\ && \int d^3 \theta\,w^{\vec{k}',(\pm)}_{A}\epsilon^{AB}w^{\vec{k},(\pm)}_{B}=2\pi\delta_{\vec{k}',-\vec{k}} ,
\end{eqnarray}
with $d^3\theta$ denoting the volume element $d\theta d\sigma d\delta$. The two last equations are not valid for zero modes. In that case, one can directly define $u^{\vec{0},(\pm)}_A$ as the spinors with 
\begin{eqnarray}
u^{\vec{0},(+)}_1&=&1, \quad\quad  u^{\vec{0},(-)}_1=0, \\
u^{\vec{0},(+)}_2&=&0, \quad\quad u^{\vec{0},(-)}_2=1.
\end{eqnarray} 
On the other hand, the complex conjugate of Eq. \eqref{eigens} provides a basis of right-handed modes, with the chirality of $\chi^{*}_{A'}$.

Each of the coefficients $m_{\vec{k}}$, $s_{\vec{k}}$, $t_{\vec{k}}$, and $r_{\vec{k}}$ forms a Grassmann canonical pair with its respective complex conjugate \cite{DEathHall}. Furthermore, in this sense they provide a canonical set together with the variables introduced in the previous section for the metric and scalar field perturbations and for the FLRW cosmology, once we have adopted a description of the cosmological perturbations in terms of gauge invariants \cite{fermilala}. For convenience, in the following we will employ the notation $(x_{\vec{k}},y_{\vec{k}})$ to refer to any of the ordered pairs of coefficients $(m_{\vec{k}},s_{\vec{k}})$ or $(t_{\vec{k}},r_{\vec{k}})$. 
 
As we have already commented, there is no fermionic term in the linear perturbative constraints of our system, so that the only contribution of the Dirac field to the total Hamiltonian is included in the zero mode of the Hamiltonian constraint. This contribution is given by the Dirac Hamiltonian, evaluated at the variables for the cosmological zero modes defined in the previous section and at the fermionic perturbations determined by the variables $(x_{\vec{k}},y_{\vec{k}})$, as far as the respective dependence on the FLRW cosmology and the Dirac field is concerned. Motivated by previous works on this subject \cite{DEathHall}, the first approach to the treatment of fermions in hybrid LQC was to carry out a change of fermionic variables that produces a diagonalization of the Dirac Hamiltonian. To reach this diagonalization, the change of variables must depend on the FLRW geometry, a situation that is similar to that studied when we introduced gauge invariants to describe the relevant degrees of freedom of the scalar perturbations. This has two consequences, as we know. First, the zero modes have to be corrected to maintain the canonical structure in the set of variables that describe the whole of the cosmological system. Second, the fermionic contribution to the Hamiltonian constraint gets an additional term, up to quadratic order in the perturbations, owing to the background dependence of the change of variables. Even if this change was designed to diagonalize the Dirac Hamiltonian, it will generally not diagonalize the new fermionic contribution, and therefore the resulting fermionic Hamiltonian will still contain interacting terms. Actually, the new fermionic variables have Poisson brackets of the creation-annihilation type, so that we can view these fermionic interactions as the creation or annihilation of pairs of particles. Finally, we will treat the fermionic zero modes on their own, keeping their description in terms of the original  variables $(x_{\vec{0}},y_{\vec{0}})$ to avoid problems with the particularization of our formulas to a vanishing Dirac eigenvalue (i.e., for $\omega_k=0$).  

Explicitly, and leaving aside those zero modes, the new variables are given by 
\begin{eqnarray}
\label{avariable}
a_{\vec{k}}^{(x,y)}&=& \sqrt{ \frac{\xi_k -\omega_k}{2\xi_k} }  x_{\vec{k}}+\sqrt{ \frac{\xi_k +\omega_k}{2\xi_k}} {y}_{-\vec{k}}^{*}, \nonumber\\
\left(b_{\vec{k}}^{(x,y)}\right)^{*}&=& \sqrt{ \frac{\xi_k +\omega_k}{2\xi_k}} x_{\vec{k}} - \sqrt{ \frac{\xi_k -\omega_k}{2\xi_k} } y_{-\vec{k}}^{*},	
\end{eqnarray}
where 
\begin{equation}
\label{xik}
\xi_k=\sqrt{\omega_k^2+M^2 {\tilde{a}}^2},
\end{equation}
with $M$ denoting the mass of the Dirac field. Notice that the sum of the square modulus of the coefficients in each of the above linear combinations of the variables  $(x_{\vec{k}},y_{\vec{k}})$ equals the unit. This ensures that the transformation is canonical in the fermionic phase space \cite{fermilala}. In a Fock representation with a standard interpretation, the operators representing $a_{\vec{k}}^{(x,y)}$ and  $b_{\vec{k}}^{(x,y)}$ would annihilate particles and antiparticles, respectively, while their adjoints (representing the complex conjugate variables) would create them. 

The fact that our change of fermionic variables depends only on the scale factor implies that we only need to modify the momentum of that background variable in order to recover a canonical set for the entire cosmological system. The modification of the momentum $\pi_{\tilde a}$ consists in adding to it the following terms that are quadratic in the fermionic perturbations, obtained in a similar way as it was explained in the previous section for the scalar and tensor perturbations \cite{fermilala}:
\begin{equation}
\label{modifiedpiDEH}
- \frac{i M}{2 }   \sum_{\vec{k}\neq \vec{0},(x,y) }  \frac{ \omega_k  }{{\xi}_k^2}\left[ a_{\vec{k}}^{(x,y)} b_{\vec{k}}^{(x,y)} + \left(a_{\vec{k}}^{(x,y)}\right)^{*} \left(b_{\vec{k}}^{(x,y)}\right)^{*}\right]. 
\end{equation}
For simplicity in our notation, we will denote the new momentum of the scale factor with the same symbol as before. From the context, it must be clear in our discussion whether we are referring to the original momentum or to the momentum that has been changed with the addition of fermionic contributions. On the other hand, we also notice that the variables for the scalar and tensor perturbations need not be altered at this stage, because our change of fermionic variables is independent of them.

In terms of this new canonical set, the total Hamiltonian has the same expression \eqref{totalhst} as before except for two things. First, its dependence on $\pi_{\tilde{a}}$ must be evaluated at the new momentum of the FLRW scale factor, which includes the fermionic modification. And second, the zero mode of the Hamiltonian constraint includes one additional contribution ${}^{(F)}H_2$  which is due to fermions,
\begin{equation}
H_0+{}^{(s)}H_2+{}^{(T)}H_2+{}^{(F)}H_2=0.
\end{equation}
In consonance with our comments above, this fermionic contribution is given by the sum of the Dirac Hamiltonian ${}^{(F)}H_D$, once it is expressed in terms of the new fermionic variables, and an interaction term ${}^{(F)}H_I$, arising from the correction to $H_0$ caused by the change of momentum for the scale factor [like in Eq. \eqref{qcorrect}]. In detail, their expressions are
\begin{eqnarray}
{}^{(F)}H_2&=&{}^{(F)}H_D+{}^{(F)}H_I, \\
{}^{(F)}H_D &=&  {}^{(F)}H_{\vec{0}} +  \frac{1}{2 {\tilde a}}\sum_{\vec{k}\neq 0, (x,y)}  \xi_k  \left[   \left(a_{\vec{k}}^{(x,y)}\right)^{*} a_{\vec{k}}^{(x,y)} -  a_{\vec{k}}^{(x,y)} \left( a_{\vec{k}}^{(x,y)}\right)^{*} \right] \nonumber \\
&+&  \frac{1}{2 {\tilde a}}\sum_{\vec{k}\neq 0, (x,y)}  \xi_k \left[  \left(b_{\vec{k}}^{(x,y)}\right)^{*} b_{\vec{k}}^{(x,y)}  - b_{\vec{k}}^{(x,y)} \left(b_{\vec{k}}^{(x,y)}\right)^{*} \right]  ,\\
{}^{(F)}H_{\vec{0}}&=& M \left[ {s}_{\vec{0}} r_{\vec{0}}^{*} + r_{\vec{0}}  s_{\vec{0}}^{*} + m_{\vec{0}}  t_{\vec{0}}^{*} + t_{\vec{0}}  m_{\vec{0}}^{*} \right]\!,\\
{}^{(F)}H_I &=&- \frac{i\pi_{\tilde a} GM }{ 12\pi^2 {\tilde a}}     \sum_{\vec{k}\neq 0, (x,y)} \frac{\omega_k}{ \xi_k ^2 } \left[ a_{\vec{k}}^{(x,y)} b_{\vec{k}}^{(x,y)} + \left( a_{\vec{k}}^{(x,y)}\right)^{*}  \left( b_{\vec{k}}^{(x,y)}\right)^{*} \right] .
\end{eqnarray} 

 \section{Hybrid quantization of cosmological perturbations: Implementation}
 
Once we have at our disposal a canonical set of variables for the description of our perturbed cosmological model in which the variables that describe the perturbations are either gauge invariants, perturbative gauge generators, or associated gauge degrees of freedom, we are in an appropriate situation to face the quantization of the system.  We will carry out this quantization adopting the hybrid approach within the framework of LQC. As we have already commented, this hybrid strategy is based on the hypothesis that the most relevant effects of quantum geometry for cosmology are those that affect the FLRW substrate, namely the behavior of the scale factor, while the purely quantum geometric effects on the perturbations can be approximately ignored, and handle the quantum description of those anisotropies and inhomogeneities using techniques directly related with the formalism of QFT in a curved spacetime, generalized to the case in which such a spacetime is quantum mechanical as well. In practice, we quantize the FLRW cosmology using the methods of LQC and the perturbations (essentially) with Fock quantization methods, then combine both types of quantum descriptions by adopting a tensor product representation space for the system, and finally impose on it the diffeomorphism constraints that are present in the system. We will see that, among them, the only intrincate constraint is the zero mode of the Hamiltonian constraint, that relates in a complicated way the FLRW cosmology with the scalar and tensor gauge invariant perturbations (as well as with the fermionic ones if we also consider a Dirac field). Thus, the hybrid quantization is non-trivial precisely because of the imposition of this constraint.

The linear perturbative constraints obtained from the Abelianization of the perturbations of the diffeomorphism constraints can be imposed straightforwardly by representing them as derivative operators with respect to their canonically conjugate degrees of freedom (or, if one considers an integrated version of these constraints, as operators that displace the values of such canonically conjugate degrees of freedom, resulting in transformations that should be symmetries). With this representation, the states that satisfy such constraints \`a la Dirac are simply those that are independent of the gauge degrees of freedom $\pi_{{\breve{H}}^{\vec{k},\pm}_{1}}$ and $\pi_{{H}^{\vec{k},\pm}_{\uparrow 1}}$ [see Eq. \eqref{pigauge}]. In other words, physical states depend only on (a complete set of compatible) zero modes and gauge invariants. Note that this result is obtained without the need to introduce any perturbative gauge fixing. Physical states must still satisfy one constraint, that is the only one remaining at this stage, namely the zero mode of the Hamiltonian constraint, as we anticipated.

The desired quantum formulation is then reached by choosing the representation space of the improved dynamics scheme of homogeneous and isotropic LQC, ${\mathcal H}^{grav}_{LQC}$, for the perturbatively corrected volume and its momentum (namely, the zero modes of the FLRW geometry once they have been suitably modified with terms that are quadratic in the perturbations in  order to maintain the canonical symplectic structure of the entire cosmological system). For the perturbatively corrected inflaton and its momentum, we use a standard Schr$\ddot{\rm o}$dinger representation $L^2(\mathbb{R}, d{\tilde{\phi}}) $. On the other hand, for the MS and tensor gauge invariants, we adopt Fock representations $ {\mathcal F}_s$ and $ {\mathcal F}_T$, chosen within a unique privileged family of unitarily equivalent representations that are characterized by \cite{uniquenessflat,uniquenessscale}:
\begin{itemize}\label{unique}
\item The invariance of the vacuum under the symmetries of the spatial hypersurfaces.
\item A unitarily implementable Heisenberg evolution of the creation and annihilation operators,  in the context of QFT in a curved background. This Heisenberg evolution is determined by the dynamics of the gauge invariant modes that we have picked out for the description of the perturbations.
\end{itemize}
In addition, if the system contains a Dirac field, viewed as a fermionic perturbation, we employ for it a  Fock representation $ {\mathcal F}_D$ in the equivalence class of the one that is naturally associated with the previously introduced choice of creation and annihilationlike variables proposed by D'Eath and Halliwell \cite{DEathHall}. This again belongs to a uniquely distinguished class of unitarily equivalent representations characterized by the same two conditions that we have listed above, together with the requirement of recovering a standard notion of particles and antiparticles. It is worth emphasizing that the choice of Fock representation (or of a family of unitarily equivalent representations) does not determine a concrete choice of vacuum state. Any Fock state in our representation is valid for this purpose. Therefore, in order to fix a unique vacuum state, more restrictions are needed, either in the form of additional requirements about the physical properties of the subsequent quantum theory or in the form of conditions on a particular spatial section able to specify the state there.

Let us continue with our hybrid approach, thus adopting as representation space the tensor product ${\mathcal H}^{grav}_{LQC} \otimes  L^2(\mathbb{R}, d{\tilde{\phi}}) \otimes {\mathcal F}_s \otimes {\mathcal F}_T \otimes {\mathcal F}_D$. We construct our quantum representation so that the zero modes commute with the perturbations, as it happens under Poisson brackets in the classical theory, and so that all functions of the inflaton $\tilde{\phi}$ act by multiplication.  As we have commented, the zero mode of the Hamiltonian constraint results in a non-trivial coupling between the various factors of our tensor product. The quantization proposed for this constraint is based on the representation adopted in homogeneous LQC\footnote{The quantum constraint that corresponds to the alternate regularization proposed for homogeneous LQC in Ref. \cite{DaporLieg} and its associated dynamics have been studied in Refs. \cite{CGGQMM,GQMMSP}.}. For the FLRW contribution $H_0$ we adopt the same prescription as in LQC. In particular, we adhere to the improved dynamics proposal, so that quantities that depend on the momentum of the scale factor are represented in terms of holonomies defined employing squares with a fiducial length that guarantees that the physical area enclosed by them coincides with the area gap $\Delta$, determined by the area spectrum of LQG. Using the resulting basic operators of homogeneous LQC, as well as the homogeneous physical volume operator $\hat{V}= 2\pi G \gamma \sqrt{\Delta} | \hat{v} | $ and the regularized inverse volume operator obtained from them, we get
\begin{equation}
{\hat {H} }_0^{(2)}={\hat{ \Omega}}_0^2 - 2 {\hat{V}}^2  {\mathcal V}({\tilde \phi}) .
\end{equation}
We recall that ${\mathcal V}({\tilde \phi}) $ is the inflaton potential, and that ${\hat{ \Omega}}_0$ was defined in Eq. \eqref{Omegaop}.

As for the functions of zero modes of the FLRW cosmology that appear in the perturbative part of the constraint, we adopt a symmetric factor ordering that tries and respects, as far as possible, the assignation of representation from homogeneous and isotropic LQC. In more detail, we adopt the following rules for their quantum representation:
\begin{itemize}
\item We symmetrize \`a la Weyl the representation of the product  $\Theta_o \pi_{\phi}$, to deal with the presence of functions of $\phi$ in $\Theta_o$ that do not commute with the inflaton momentum. 
\item We adopt an algebraic symmetrization for factors of the form $V^r g(b)$, that are promoted to the operators $\hat V^{r/2}{\hat g}_{\text{LQC}}\hat V^{r/2}$, where $r$ is any real number and $g(b)$ a function of the variable $b$, with ${\hat g}_{\text{LQC}}$ its operator counterpart in the improved dynamics scheme of LQC. This algebraic symmetric factor ordering is adopted as well for powers of the inverse volume. 
\item Even powers of  $-{\tilde a} \pi_{\tilde a}\sqrt{4\pi G/3}$ are promoted to even powers of the operator ${\hat{\Omega}}_0$, which represents this quantity in LQC, whereas odd powers, let's say, of order $2z+1$, with $z$ any integer, are represented as  $|{\hat{\Omega}}_0|^z {\hat{\Lambda}}_0|{\hat{\Omega}}_0|^z$. Here $|\hat\Omega_0|$ is the square root of the positive operator ${\hat{\Omega}}_0^2$, and ${\hat{\Lambda}}_0$ is defined exactly as ${\hat{\Omega}}_0$, but with holonomies of double length. The result can be obtained by dividing the right-hand side of expression \eqref{Omegaop} by 2, and replacing $b$ in that expression with $2b$. The operator ${\hat{\Lambda}}_0$ defined in this way only shifts $v$ in multiples of four units, and hence preserves the superselection sectors of the homogeneous and isotropic geometry.
\end{itemize}

With these prescriptions, we arrive at the following operator representation of the functions \eqref{varthetas} that appear in the densitized Hamiltonian constraint:
\begin{eqnarray}\label{hatvarthetas}
{\hat{\vartheta}}_e &=& 2\pi{\hat{V}}^{2/3},\\ 
{}^{(s)}{\hat{\vartheta}}_e^q &=& \frac{2G}{3}{ \widehat{  \left[  \frac{1}{V} \right]}}^{1/3} {\hat{H}}_0^{(2)} \left( 19 - 18  {\hat{\Omega}}_0^{-2} {\hat{H}}_0^{(2)} \right) {\widehat{ \left[ \frac{1} {V} \right]}}^{1/3}
+ \frac{{ \hat{V}}^{4/3}}{ 2 \pi }  \left( {\mathcal V}^{''}(\tilde \phi)-  \frac{16\pi G}{3} {\mathcal V}({\tilde\phi}) \right),\\ 
{}^{(T)}{\hat{\vartheta}}_e^q &=& \frac{2G}{3}{ \widehat{  \left[  \frac{1}{V} \right]}}^{1/3} {\hat{H}}_0^{(2)} {\widehat{ \left[ \frac{1} {V} \right]}}^{1/3}-\frac{8G}{3}{ \hat{V}}^{4/3}{\mathcal V}({\tilde\phi})
\\
{}^{(s)}{\hat{\vartheta}}_o &=& 12 \sqrt{\frac{G}{3\pi}}  {\mathcal V}^{'}({\tilde \phi})  {\hat{V}}^{2/3} | {\hat{\Omega}}_0 |^{-1} {\hat{\Lambda}}_0  | {\hat{\Omega}}_0 |^{-1} {\hat{V}}^{2/3}.
\end{eqnarray}  
According to these formulas, the counterpart of relation \eqref{sctenma} between the scalar and tensor background dependent masses is
\begin{equation}
{}^{(s)}{\hat{\vartheta}}_e^q ={}^{(T)}{\hat{\vartheta}}_e^q+12G{ \widehat{  \left[  \frac{1}{V} \right]}}^{1/3} {\hat{H}}_0^{(2)} \left( 1 -   {\hat{\Omega}}_0^{-2} {\hat{H}}_0^{(2)} \right) {\widehat{ \left[ \frac{1} {V} \right]}}^{1/3}
+ \frac{{ \hat{V}}^{4/3}}{ 2 \pi }{\mathcal V}^{''}(\tilde \phi).
\end{equation}
It is worth remarking that ${}^{(s)}{\hat{\vartheta}}_o $ is proportional to the derivative of the inflaton potential, so that one expects its contribution to be negligible when the dependence of the potential on the inflaton is not important. The operators representing the phase space functions \eqref{Thetas} can be constructed with the above operators and the Fock representation adopted for the modes of the MS and the tensor gauge invariants. In a completely similar manner, one can construct an operator representation for the fermionic contribution $H_F$ to the densitized zero mode of the Hamiltonian constraint  (obtained from the original one by multiplication with the homogeneous volume), that depends only on the FLRW geometry and the Dirac field, but not on the inflaton nor on its momentum. For more details about this fermionic contribution, we refer the reader to Refs. \cite{hyb-pert4,fermilala}.  In this way, we finally get
\begin{equation}\label{scalarD}
{\hat{H}}_S = \frac{1}{2} \left[ {\hat{\pi}}_{\tilde{\phi}}^2 - {\hat{H}}_0^{(2)}- { \hat{\Theta}}_e- \frac{1}{2} \left( {\hat{\Theta}}_o {\hat{\pi}}_{\tilde{\phi}} +  {\hat{\pi}}_{\tilde{\phi}}{\hat{\Theta}}_o \right) + {\hat{H}}_F \right].
\end{equation}

\section{Hybrid LQC: Modified perturbation equations}

Although we have been able to handle all the constraints of our perturbed model except the zero mode of the Hamiltonian constraint, this constraint is still so intrincate that, in the presented form, it does not seem possible to obtain its general solution analytically. In order to investigate the properties of the physical states, we will now introduce an ansatz that contemplates a situation of special interest. We will consider states in which the dependence on the FLRW geometry and on each of the gauge invariant fields can be separated. In this separation, all parts are allowed to depend on the inflaton. In more detail, from now on we analyze states of the form
\begin{equation}
\xi(v,{\tilde{\phi}}) \psi_s(N_s,{\tilde{\phi}})  \psi_T(N_T,{\tilde{\phi}})  \psi_F(N_F,{\tilde{\phi}}),
\end{equation}
where we have adopted the abstract notation $N_s$, $N_T$, and $N_F$ to denote the dependence on the degrees of freedom of the  corresponding Fock space, via a set of occupation numbers in the respective basis of $n$-particle states. In addition, $\xi(v,{\tilde{\phi}})$ designates a state in the kinematic Hilbert space of homogeneous and isotropic LQC, such that it  is normalized and evolves unitarily with respect to ${\tilde{\phi}}$ as  
\begin{equation}
\xi(v,{\tilde{\phi}})= {\hat{U}}(v,{\tilde{\phi}} )\chi (v),
\end{equation} 
where $\hat{U}$ is an evolution operator with generator ${\hat {\widetilde{H}}}_0$ that is close to the unperturbed one, determined by ${\hat{H}}_0^{(2)}$. This last condition can be understood as the requirement that the action of ${\hat{H}}_0^{(2)}-({\hat{\widetilde H}}_0)^{2}-[{\hat \pi}_{\tilde{\phi}},{\hat{\widetilde H}}_0]$ on $\xi(v,{\tilde{\phi}})$ be at most of the order of the perturbative contributions when imposing the Hamiltonian constraint. Moreover, in the following, for simplicity, we will assume that this term is actually negligible in the action of the Hamiltonian constraint on the considered state, assumption that can always be checked for consistency once $\xi(v,{\tilde{\phi}})$ is specified.

On this family of states, we still must impose \`a la Dirac the Hamiltonian constraint operator ${\hat{H}}_S$, that couples the FLRW background cosmology with the gauge invariant perturbations. To get solutions in physically relevant regimes, we can employ certain approximations that facilitate the resolution of the constraint. First, we consider regimes in which the transitions in the FLRW geometry mediated by the Hamiltonian constraint can be ignored as negligible on $\xi(v,{\tilde{\phi}})$. In this situation, the relevant part of the Hamiltonian constraint is provided by its expectation value on $\xi(v,{\tilde{\phi}})$ over the FLRW geometry (with the integration measure of the inner product of LQC). This expectation value provides a constraint equation on the gauge invariant perturbations of the form
\begin{equation}\label{constraintpert}
{\hat \pi}_{\tilde{\phi}}^2 \psi + \left(2 \langle {\hat {\widetilde H}}_0 \rangle_{\xi}-\langle {\hat \Theta}_{o} \rangle_{\xi}\right) {\hat \pi}_{\tilde{\phi}} \psi = \left[ \langle {\hat \Theta}_{e} + \frac{1}{2} ({\hat \Theta}_{o} {\hat{\widetilde H}}_{0} +  {\hat{\widetilde H}}_{0} {\hat \Theta}_{o}) - {\hat{H}}_F \rangle_{\xi} + \frac{1}{2} \langle  [ {\hat \pi}_{\tilde{\phi}} -{\hat{\widetilde H}}_{0},{\hat \Theta}_{o} ]  \rangle_{\xi}  \right] \psi ,
\end{equation} 
where we have called $\psi=\psi_s(N_s,{\tilde{\phi}})  \psi_T(N_T,{\tilde{\phi}})\psi_F(N_F,{\tilde{\phi}})$. In what follows, we will neglect the perturbative operator $\langle {\hat \Theta}_{o} \rangle_{\xi}$ when compared to $\langle {\hat {\widetilde H}}_0 \rangle_{\xi}$ on the left-hand side of this equation, according to our perturbative scheme.

Suppose for the moment that in Eq. \eqref{constraintpert} we can also neglect the first term, equal to the second derivative of the wave function of the perturbations with respect to the inflaton. This can be regarded as a kind of Born-Oppenheimer approximation, in the sense that one neglects the variation of certain degrees of freedom of the considered quantum state in comparison with the variation of others. Explicitly, we ignore the variation of the perturbations with respect to the inflaton in favor of the variation of the FLRW state, that is given in average by the expectation value $\langle {\hat {\widetilde H}}_0 \rangle_{\xi} $. Additionally, it is worth noticing that the last term in the constraint equation \eqref{constraintpert}  affects only the scalar perturbations, because ${\hat \Theta}_{o}$ depends only on them. Let us assume that this term for the scalar perturbations is negligibly small. Taking into account that, in our representation, $ {\hat \pi}_{\tilde{\phi}}$ acts as the derivative with respect to the explicit dependence on the inflaton ${\tilde{\phi}}$ (multiplied by $-i$), and that ${\hat{\widetilde H}}_{0}$ has been chosen to be close to the generator of the homogeneous and isotropic quantum dynamics with respect to the inflaton, the term under consideration can be understood as the total derivative of the operator ${\hat \Theta}_{o}$ with respect to the inflaton, both in its explicit and in its implicit dependence. Thus, we expect that the analyzed contribution to the scalar perturbations can be ignored when the variation with respect to the inflaton is not significantly relevant. With these two approximations, the studied constraint amounts to the sum of a set of Schr$\ddot{\rm o}$dinger equations, one for each of the considered perturbations (scalar, tensor, and fermionic). Specifically, we get the following equations for the gauge invariant perturbations:
\begin{eqnarray}
{\hat \pi}_{\tilde{\phi}} \psi_s &=& \frac{ \langle 2 {}^{(s)}{\hat \Theta}_{e} + ( {\hat \Theta}_{o} {\hat {\widetilde H}}_{0} + {\hat {\widetilde H}}_{0}  {\hat \Theta}_{o}) \rangle_{\xi} }  {4 \langle {\hat {\widetilde H}}_0 \rangle_{\xi} } \psi_s ,\\
{\hat \pi}_{\tilde{\phi}} \psi_T &=& \frac{ \langle {}^{(T)}{\hat \Theta}_{e}  \rangle_{\xi} }  {2 \langle {\hat {\widetilde H}}_0 \rangle_{\xi} } \psi_T ,\\
{\hat \pi}_{\tilde{\phi}} \psi_F &=& -\frac{ \langle  {\hat{ H} }_F  \rangle_{\xi} }  {2 \langle {\hat {\widetilde H}}_0 \rangle_{\xi} } \psi_F.
\end{eqnarray}
Note that the separation of variables can actually be made mode by mode in each of the gauge invariant perturbations, since these modes are not coupled by the Hamiltonian constraint.

Had we not neglected the contribution of ${\hat{H}}_0^{(2)}- ({\hat{\widetilde H}}_0)^{2}-[{\hat \pi}_{\tilde{\phi}},{\hat{\widetilde H}}_0]$, but considered instead that its action on the wave function of the FLRW geometry is of the same order as that of the perturbative contributions, we would have obtained an equation similar to the constraint \eqref{constraintpert} although with an additional term, given by the expectation value on $\xi(v,{\tilde{\phi}})$ of the discussed difference of operators.  Then, we should have added to the right-hand side of each Schr$\ddot{\rm o}$dinger equation a backreaction term, which could only depend on the inflaton. The balance between these backreaction terms $C^{(\xi)}(\tilde{\phi})$ would require that
\begin{equation}
\frac{ \langle ({\hat{\widetilde H}}_0)^{2}-{\hat{H}}_0^{(2)}+[{\hat \pi}_{\tilde{\phi}},{\hat{\widetilde H}}_0]\rangle_{\xi}  }{ 2\langle {\hat {\widetilde H}}_0 \rangle_{\xi} }= C^{(\xi)}_s(\tilde{\phi}) + C^{(\xi)}_T(\tilde{\phi}) + C^{(\xi)}_F(\tilde{\phi}),
\end{equation}
where the subscript on the backreaction tells us whether the term corresponds to the scalar ($s$), tensor ($T$), or fermionic ($F$) contribution. From this balance, we see that the sum of all the backreaction terms gives us information, in mean value and within our approximations, about how much the state  $\xi(v,{\tilde{\phi}})$ departs from an exact solution of the unperturbed homogeneous and isotropic cosmology in LQC.

Moreover, let us return to Eq. \eqref{constraintpert} and let us assume now {\emph{only}} that the gauge invariant perturbations admit a direct (effective) counterpart of the Heisenberg dynamics that results for their operator analogs from this Hamiltonian constraint equation, something that seems reasonable because the considered Hamiltonian is quadratic in the perturbative variables. Then, it is immediate to realize that we get a set of modified propagation equations for the MS modes, the tensor perturbations, and the fermionic perturbations. For instance, the modified MS equations are
\begin{equation}
d^2_{\eta_{\xi}} \nu_{\vec{k},\pm} = - \nu_{\vec{k},\pm} \left[  \omega_k^2 +  \frac{ \langle 2  {}^{(s)}{\hat \vartheta}_{e}^q +  ({}^{(s)}{\hat \vartheta}_{o}{\hat {\widetilde H}}_0 + {\hat {\widetilde H}}_0 {}^{(s)}{\hat \vartheta}_{o})+[ {\hat \pi}_{\tilde {\phi}} -{\hat{\widetilde H}}_{0},{}^{(s)}{\hat \vartheta}_{o} ]  \rangle_{\xi} } {  2\langle {\hat \vartheta}_{e} \rangle_{\xi} }  \right],
\end{equation}
where the conformal time $\eta_{\xi}$ is defined by the equation 
\begin{equation}\label{confytime}
\langle {\hat {\widetilde H}}_0 \rangle_{\xi} d\eta_{\xi}=  \langle \hat {\vartheta}_{e} \rangle_{\xi} d {\tilde{\phi}}.
\end{equation}
Therefore, this time depends on the state $\xi(v,{\tilde{\phi}})$ of the FLRW geometry. Similarly, for the modes of the tensor perturbations we obtain
\begin{equation}
d^2_{\eta_{\xi}} {\tilde d}_{\vec{k},\varepsilon,\pm} = -  {\tilde d}_{\vec{k},\varepsilon,\pm} \left[ \omega_k^2 + \frac{ \langle   {}^{(T)}{\hat \vartheta}_{e}^q  \rangle_{\xi} } { \langle {\hat \vartheta}_{e} \rangle_{\xi} }\right].
\end{equation}
The conformal time is the same as for the scalar perturbations, thanks to the fact that the operator ${\hat \vartheta}_{e}$ that multiplies the squared momenta in the Hamiltonian constraint (and that represents the squared scale factor, up to a constant) coincides both for tensor and scalar gauge invariants and, furthermore, for all the modes of these perturbations. In a similar way, an equation with quantum geometry corrections and in the same conformal time can be obtained as well for the femionic perturbations (see Ref. \cite{fermilala}).

In the above propagation equations, the ratio of expectation values on the right-hand side gives the quantum corrected mass for the specific gauge invariant perturbation under consideration. We notice that this corrected mass is actually mode independent, because this is the case for the corresponding operators. Also, note that the equations contain no dissipative term. Much more important, the deduced effective equations are hyperbolic in the ultraviolet regime, regardless of the concrete behavior of the quantum state for the FLRW geometry, provided that our approximations are valid.  

In order to extract predictions from the above equations about quantum geometry effects on the primordial perturbations, one needs to compute the expectation values that give the corrected masses for the MS and tensor perturbations. There are several possible strategies to reach this goal. Let us list three of these strategies, in decreasing order of accuracy but increasingly easier to implement. First, one could compute the quantum expectation values numerically. For this, one may try and ignore the backreaction (checking the validity of this approximation afterwards) and integrate numerically the quantum evolution of the FLRW state with respect to the inflaton. With the FLRW state obtained in this way, one can calculate with numerical methods the desired expectation values at each given value of $\tilde \phi$. The more difficult part of this program is the integration of the FLRW dynamics in the presence of non-trivial inflaton potentials. Second, taking into account the commented complication that the potential introduces,  one can compute the evolution of the FLRW state not numerically, but in an interaction picture in which the potential (or part of it) is regarded as an interaction added to the homogeneous and isotropic Hamiltonian of LQC \cite{hyb-pert5}, and treated as a perturbation of that Hamiltonian using a Dyson series expansion \cite{Galindo}. And third, for suitable FLRW states, one can directly adhere to the effective dynamics description of LQC, integrating numerically only the trajectory of the peak of the state, rather that the quantum dynamics strictly speaking. Furthermore, this integration can be simplified by identifying regimes with universal behavior in the evolution from the bounce for the background solutions of interest in LQC \cite{Ivan,Universe,Wang1,Wang2}. For instance, the most interesting situations to get quantum geometry corrections on the primordial spectra that can be observed nowadays are found for background solutions that are kinetically dominated around the bounce, so that the potential there has little influence. This allows us to introduce further simplifications in the integration of the FLRW trajectories that, at the end of the day, facilitate the calculation of the quantites that determine the studied masses of the perturbations. 

Most of the work in the literature has indeed been done assuming an effective dynamics for the description of the FLRW cosmology in LQC. Even if, with this approximation, the problem of computing the evolution of the primordial perturbations is handleable, the results (and hence the predictions obtained from them) depend critically on the initial conditions that one chooses for the FLRW background in this effective dynamics, as well as on the initial conditions that determine the state of the perturbations subject to the propagation equations that we have derived. We will discuss these issues in the next section.

Let us point out that, adopting this effective dynamics for the description of the FLRW states, it has been proven \cite{mass} that the corrected mass that appears in the modified propagation equations for the scalar and tensor perturbations is positive around the Big Bounce, at least for the most interesting ranges of energy density contribution of the inflaton potential. Since the Big Bounce is precisely the region where the quantum effects on the geometry are more significant, one would expect that the largest departures from the classical situation described by GR cosmology happen there. This positivity of the quantum corrected mass is important to be able to define adiabatic vacua as initial states around the bounce for all the perturbative modes \cite{Universe,mass,MBO}. A negative mass involves a breakdown of the adiabatic approximation around the bounce at least for values of $\omega_k$ that are not sufficiently large, invalidating the construction of adiabatic states as natural candidates for a vacuum at frequencies that can be of physical interest, for instance because they cover part of the observed spectrum in the CMB. Moreover, the positivity of the mass at the bounce is not shared by other proposals for the quantization of cosmological perturbations within the framework of LQC, like the so-called dressed metric approach that has been put forward by Agullo, Ashtekar, and Nelson \cite{AAN3,AAN1,AAN2}.

Finally, in our discussion above, and owing to the compactness of the spatial sections, the modes that we have considered possessed a discrete spectrum of Laplace eigenvalues, $\omega_k$, that play the role of frequencies in the modified propagation equations for the perturbations even after the introduction of quantum geometry corrections. Nonetheless, it is possible to reach the continuum limit for this set of frequencies in the following form. One first extracts a length scale of reference from the scale factor. All observable quantities are defined with respect to this reference scale, that becomes physically irrelevant. One may choose as such scale the value of the scale factor today, or at the moment of the bounce, for instance. Then, the desired continuum limit is reached as the limit in which we make the reference scale tend to infinity. We refer the reader to Ref. \cite{continuum} for more details.

\section{Initial conditions}

As we have commented, even if we have succeeded in deriving propagation equations for the primordial perturbations that contain modifications caused by quantum geometry effects and even if we assume FLRW states that can be described within the effective dynamics approach to LQC, in order to extract predictions about the primordial cosmological perturbations we need to specify the particular FLRW effective solution that plays the role of a background and, in addition, the vacuum state that determines the conditions on the perturbations. Both pieces of information can be supplied by giving convenient initial data on a certain spatial section. An appealing possibility is to choose this section precisely at the Big Bounce. We will concentrate our discussion on this case. Other possibilities are equally valid, for instance a section in the asymptotic past, if the effective dynamical evolution previous to the bounce connects with a manageable asymptotic region \cite{Wang3}. 

Let us consider first the initial conditions for the FLRW background, solution to the effective dynamics of LQC. The FLRW cosmology is described by two pairs of canonical zero modes, namely four variables. But we have chosen to impose initial conditions at the bounce, where the time derivative of the scale factor vanishes, reducing the liberty in one degree of freedom\footnote{This is only a reflection of the fact that, as it happens in classical FLRW cosmology with homogeneous matter content, one of Hamilton's equations of motion contains redundant information in effective LQC.}. In addition, the Hamiltonian constraint associated with the effective dynamics reduces the degrees of freedom in one more variable.  Moreover, we have also commented that we can employ the value of the scale factor at the bounce as a reference scale, depriving it of physical relevance. In practice, this allows us to set that value equal to the unit, for instance. In total, we see that only one variable must be fixed at the bounce by the initial conditions there. We choose the value of the inflaton as this piece of initial data. On the other hand, we can consider also as free data the parameters that determine the inflaton potential. Focusing our attention on the most studied case of a quadratic potential, we find only one parameter, given by the inflaton mass. From this perspective, the FLRW effective background turns out to be completely fixed if we provide the value of the inflaton at the bounce and the value of the inflaton mass.

\begin{figure}
		\centering
		\includegraphics[width=13 cm]{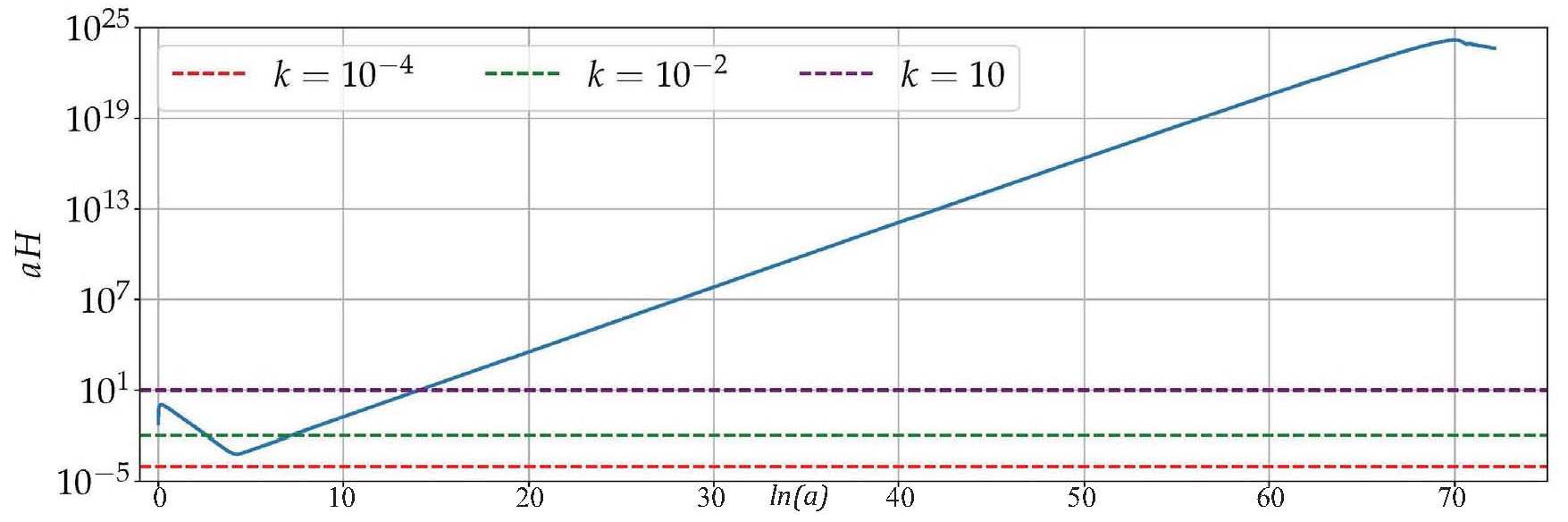}
		\caption{Solution of the rescaled Hubble parameter $aH$ in the effective dynamics of LQC for a matter content given by an inflaton with mass equal to $1.2\times 10^{-6}$ and a value at the bounce equal to $0.97$ (both quantities in Planck units). The plot shows several wavenumbers to illustrate the different numbers of intersections that are possible.}
		\label{fig1} 
\end{figure}

Actually, we are only interested in effective solutions that lead to power spectra for the perturbations that are compatible with the observations, but that still retain some quantum geometry corrections. One expects that, if these corrections have survived, they should be present in the region of large angular scales or its nearby region, because it is only in this region that the agreement between GR and observations may not be completely solid \cite{planck,planck-inf}. This requirement determines a relatively narrow interval of values for the initial condition on the inflaton $\phi_B$ and the inflaton mass $m$, around $\phi_B=0.97$ and $m=1.2\times 10^{-6}$ (in Planck units).  For this latter choice of specific values, we show in Fig. \ref{fig1} the evolution of the Hubble parameter $H$ multiplied by the scale factor. This rescaled Hubble parameter $aH$ vanishes at  the bounce and then increases in a very short superinflationary epoch in which $H$ grows to a value of the Planck order. This happens so fast that the scale factor remains almost constant in the process. Since we have taken the scale factor at the bounce equal to one, then $aH$ at its maximum should be of the order of one as well in Planck units (like $H$). This maximum sets a scale, that we denote $K_{LQC}$ in terms of wavenumbers, and that should be of Planck order according to our previous arguments. From that moment on, the rescaled Hubble parameter starts to decrease until it reaches a minimum. Besides, the quantum corrections in the effective dynamical equations of the background become negligibly small, and the effective trajectory gets totally adapted to a GR cosmological solution.

For solutions that allow for quantum geometry corrections in the spectra of the perturbations at large scales, the inflaton dynamics around the bounce is  dominated by its kinetic energy density, which is of Planck order, with an ignorable contribution of the potential. Since the potential is almost negligible, the effective solution behaves as if the scalar field were massless, situation in which the inflaton momentum is a constant of motion and the kinetic energy density decreases rapidly, as $a^{-6} $. The kinetic energy density continues to diminish until it becomes of the order of the potential. Given that $\dot{a}$ increases when the potential drives the evolution of the scale factor (both in GR and in the effective dynamics of LQC), the coincidence between the kinetic energy density and the potential of the inflaton occurs approximately when $aH$ reaches its minimum. On the other hand, during the evolution from the bounce to this minimum of $aH$, the inflaton typically increases only by a few orders of magnitude. As a result, the potential, quadratic in the inflaton, varies as well only in a few orders. Taking this into account, and since the potential at the bounce is $m^2\phi_B^2/2$, with values around $10^{-12}$ in Planck units, when the kinetic and potential energies coincide we expect a density in the range $[10^{-12},10^{-9}]$. This energy density determines yet another scale in the system, that we call $K_{K-P}$ expressed as a wavenumber.

In total, the influence of the effective background solution on the perturbations is characterized by two (wavenumber) scales, that we have already mentioned, $K_{LQC}$ and $K_{K-P}$.  In a first approximation to the problem, these scales determine the regions where the quantum geometry effects may cause departures from the standard model of inflation in GR. As we have argued, $K_{LQC}$ is of the order of the unit, because it is related to quantum gravity phenomena. To estimate $K_{K-P}$ in our solutions, note that during the epoch of kinetic dominance, the energy density decreases as $a^{-6}$ from a value of the Planck order to values in the interval $[10^{-12},10^{-9}]$ as we have pointed out. Recalling the Hamiltonian constraint of effective LQC (or of FLRW cosmology in GR, once one is away from the immediate vecinity of the bounce), one concludes that $H^2$ must be proportional to the discussed energy density in the considered region, and thus decrease during kinetic dominance also as $a^{-6}$. Consequently, $aH$ must evolve as $a^{-2}$, decreasing from the Planck order as the cubic root of the energy density, and hence reaching values in the range $[10^{-4},10^{-3}]$. In Fig. \ref{fig1} we see that wavenumbers larger than $K_{LQC}$ only intersect $aH$ once in the evolution. Note that this intersection is the moment when the associated physical length $a/k$ coincides with the Hubble scale $1/H$, and therefore can be taken as the moment of horizon crossing. For modes between $K_{LQC}$ and $K_{K-P}$, there exist three intersections. Essentially, the modes exit the horizon immediately after the bounce, reenter in the phase of kinetic dominance, and exit again during the inflationary expansion. We expect these modes to be severely affected by the quantum geometry effects around the bounce, and that their evolution differs considerably from that experienced in GR for a background solution with the same behavior in the classical region. Finally, modes with wavenumbers below $K_{K-P}$ do not exit the horizon during inflation but much before, a fact from which one may expect important departures from the predictions of standard inflation.
	
After the potential equals the kinetic energy density, the latter rapidly decreases while the potential becomes essentially constant and generates inflation. We are interested in effective background solutions such that the modes that experience quantum geometry effects (roughly, those with wavenumbers between $K_{LQC}$ and $K_{K-P}$)  are entering the horizon today. If they had entered much long ago, they would correspond to scales of the power spectra where there is no discrepancy with GR, raising the problem of explaining the absence of departures from the Einsteinian predictions or implying that quantum gravity effects are too tiny to be observable in those circumstances. On the other hand, if they had not entered the horizon again (nor were about to do it), they could not be observed in the power spectra. The interesting situation is when the modes are entering the horizon at present, as we have said. But the effective FLRW backgrounds for which this happens turn out to experience a short-lived inflation. As a consequence, during the first moments of the inflationary expansion, there is still some influence of the kinetic energy density, producing departures from a genuine slow-roll behavior. This will affect the power spectra if the modes that were exiting the horizon at those moments are observed today (see e.g. the discussion of Ref. \cite{linde} in the framework of GR). Therefore, the slow-roll approximation will not be good, at least, for modes that exited the horizon during those first stages of inflation \cite{linde}. Such modes are precisely those close to the scale $K_{K-P}$. Hence, those modes will experience two types of corrections from a standard inflationary scenario with slow-roll in GR: quantum geometry effects, accumulated  around the bounce, and short-lived inflation effects. One of the most important challenges for LQC nowadays is to be able to separate these two kinds of effects and prove that it is possible to identify and falsify the quantum modifications in cosmological observations.
	
In summary, the LQC modifications in the FLRW background with respect to the standard inflationary solutions of GR may have a relevant influence on modes between the typical scale of the quantum geometry effects and the scale $K_{K-P}$, close to the onset of inflation. If these include the modes that are now re-entering the horizon, so that the scale of the Universe that we observe today was at the very early stages in the range affected by the quantum effects, some traces of those quantum modifications may have survived in the CMB in spite of the later inflationary expansion, and they might be observable. The fact that the background which those modes feel effectively differs substantially from a de Sitter expanding phase should imply that the natural vacuum for them ought to differ from the standard Bunch-Davies vacuum \cite{BD}. As a result, the power spectra of the perturbations at those scales changes from the conventional predictions based on the choice of a Bunch-Davies state. Suppose that the new vacuum state is related to the standard one by a Bogoliubov transformation that does not mix modes with different values of $\omega_k$, something that is ensured if the invariance under the symmetries of the spatial sections is respected. Let us call $\alpha_k$ and $\beta_k$ the coefficients of this Bogoliubov transformation, with $|\alpha_k|^2-|\beta_k|^2=1$. Recall that the beta-coefficient determines the antilinear part of the Bogoliubov transformation, i.e. the part that mixes creation and annihilation operators. These coefficients can be determined, e.g., at the initial time chosen in our analysis, if we know there the initial data that specify the two bases of solutions of the gauge invariant field equations, $\{{\tilde \mu}_k\}$ and $\{\mu_k\}$, that characterize respectively the new and the old vacua. Then, if the primordial power spectrum of the standard vacuum is $\mathcal{P_{\mathcal{R}}}(k)$, the power spectrum of the new vacuum state does not need to be calculated from scratch: it is given by the formula
\begin{equation}\label{powspectr}
{\tilde{\mathcal{P}}_{\mathcal{R}} (k)}= \left[|\alpha_k|^2 +|\beta_k|^2 + 2|\alpha_k||\beta_k|\cos\left(\phi^\alpha_{k}-\phi^\beta_{k} + 2\phi^{\mu}_{k}\right)\right] \mathcal{P_{\mathcal{R}}}(k).
\end{equation}
Here, $\phi^\alpha_{k}$ and $\phi^\beta_{k}$ are the phases of the respective Bogoliubov coefficients, treated as complex numbers, and $\phi^\mu_{k}$ is the phase of the solution $\mu_k$ evaluated at the time of computation of the power spectrum (typically by the end of inflation).

The second problem related with the choice of initial data is, therefore, the selection of conditions that determine the vacuum state of the perturbations. Clearly, from the above formula, a change of vacuum state may result in a radical variation of the power spectrum. The predictive power of the formalism is lost unless we have a way to select a vacuum as the preferred state for the gauge invariant perturbations. While, in situations like de Sitter inflation, the high degree of symmetry of the background can help us in picking out a unique state, invariant under the symmetries and with a local Minkowskian behavior, this does not seem possible in more general situations, like those experienced by the modes affected by quantum geometry effects in the kind of effective backgrounds that appear in LQC. In these circumstances, several proposals have been suggested in order to single out a unique Fock state that could then be viewed as privileged in the system. 

Among these proposals, the attempt to use adiabatic states \cite{adiabatic1,adiabatic2} has received a considerable attention \cite{Ivan,AAN1,AAN2,MBO}. Nonetheless, their construction may find some obstructions, especially if the effective mass in the propagation equations of the perturbations becomes negative. We have seen that this does not occur in the hybrid approach in the region of important quantum geometry effects, at least for (effective) solutions with kinetic dominance in the energy balance of the inflaton \cite{mass}. Nonetheless, this is not the case for other approaches like the dressed metric quantization if one considers scales that are not sufficiently small \cite{mass}. Besides, the power spectra of adiabatic states, computed numerically, often present large oscillations, and even if these oscillations are averaged, they typically result in an increase of power that does not seem to fit properly with observations if the scales affected by the quantum geometry effects are inside the Hubble horizon today \cite{Ivan}.

Ashtekar and Gupt have put forward a different proposal for the vacuum state \cite{AGvacio1,AGvacio2}. In the region with relevant LQC effects, they have required that the quantum Weyl curvature satisfy a bound which is the lowest value compatible with the uncertainty principle and stable under evolution. This condition selects a ball of states. Among them, the vacuum of the perturbations is chosen by imposing another condition at  the end of inflation, ensuring that the dispersion in the field operators be minimized \cite{AGvacio2}. In the dressed metric approach, this proposal has been shown to lead to primordial power spectra that, though still highly oscillatory, seem in very good agreement with observations after being averaged \cite{AGvacio2,AshNe}. Nonetheless, the direct relation of this vacuum with adiabatic states is not known.
	
Another interesting proposal for a vacuum state is the so-called non-oscillating vacuum, suggested by Mart\'{\i}n-de Blas and Olmedo \cite{MBO}. The proposal is to select the state that minimizes the integral 
\begin{equation}\label{IO}
		\int_{\eta_{0}}^{\eta_{f}}d\bar{\eta}\left|\frac{d\big( |\mu_{k}|^{2}\big)}{d\bar{\eta}}\right|
\end{equation}
in a certain interval of conformal time, usually the interval from the time of the initial spatial section to a time well inside the inflationary regime. For instance, in our typical class of effective backgrounds, this can be a time when the kinetic energy density of the inflaton becomes so negligible that the inflationary expansion is completely driven by the potential. Since the primordial power spectrum for each mode is proportional to the square norm of the associated mode solution $\mu_k$, the proposal picks out a state that minimizes the power oscillations in a definite sense. In general, the determination of this vacuum state needs numerical methods, since the criterion of choice is posed as a variational problem that involves the calculation of an integral.  For simple cases, the proposal can be handled analitycally and has been proven to provide a conventional choice of vacuum state. Thus, it selects the Poincar\'e vacuum for flat spacetime in the presence of a scalar field, either massless or with a quadratic potential. In addition, for de Sitter spacetime, the proposal selects the Bunch-Davies vacuum \cite{MBO}. The primordial and angular power spectra for this vacuum state have been calculated numerically, both for scalar and tensor perturbations \cite{hybr-pred,hybr-ten}. The results are compatible with observations, and they even open the possibility of explaining some of the features of the spectra that perhaps may be in tension with GR \cite{hybr-pred,Universe}, at large angular scales or for multipoles around $l=20$ \cite{planck,planck-inf}.

\section{Choice of vacuum state for the perturbations: Splitting of phase space variables}

The problem of selecting a vacuum for the perturbations appears in our formalism because the requirements that we have imposed to determine the Fock representation of the gauge invariant perturbations in the hybrid approach at most select a family of representations that are unitarily equivalent, but not a privileged state. Any of those representations, or equivalently any Fock state in the considered family, can be chosen as the vacuum. This leaves a large freedom in the selection of a vacuum for the perturbations, and so in the initial conditions that characterize it. The proposals that we have commented at the end of the previous section are some of the attempts to fix this freedom, but there is yet no general consensus about how to settle the question. Besides, although some of those proposals lead to power spectra that are compatible with observations, they happen to rely on numerical and/or minimization techniques.

In fact, we can consider families of representations related among them by unitary transformations which depend on the background. The Heisenberg dynamics of the creation and annihilation operators associated with each of these representations, even if unitarily implementable as provided by a composition of unitary transformations, would differ between them, given that part of the evolution is removed by assigning it to the background sector of phase space. What is more, by means of this type of unitary transformations with dependence on the background, we can change the splitting of the phase space degrees of freedom between the zero modes that describe the background and the modes of the gauge invariant perturbations. Actually, there exist many ways of separating the phase space into a homogeneous sector and an inhomogeneous one using canonical transformations that mix them. The specific splitting that one adopts determines the properties of the resulting quantization. In particular, the representation of the Hamiltonian constraint and its ultraviolet features strongly depend on this choice.

This fact can be employed to improve the behavior of the field operators that represent the perturbative terms, ameliorating the need for the introduction of regularization procedures typical of QFT. Indeed, as they stand, the actions of the MS, tensor, and fermionic Hamiltonians that appear in the constraint \eqref{scalarD} are ill defined with a standard choice of Fock representation for the corresponding perturbations. Moreover, the backreaction is in general divergent. For instance, by constructing the unitary operator that implements the Heisenberg dynamics of the fermionic variables (in the context of QFT in a quantum mechanically corrected background), one can compute the backreaction of the fermions, $C_F^{(\xi)}(\tilde{\phi})$, and show that it is not absolutely convergent \cite{fermilala}.

These problems can be solved, or at least alleviated, by introducing new gauge invariants prior to quantization, defined by using canonical transformations that depend on the zero modes. Considering the system as a whole, we have freedom in
\begin{itemize}
\item Changing the dynamical separation between the FLRW geometry and the gauge invariant perturbations  via canonical transformations. 
\item Chosing the Fock vacuum for the perturbations, within the hybrid scheme, regarding this vacuum as the state from which one defines the Fock representation as a cyclic one.
\end{itemize}
All this ambiguity can be encoded in choices of the form
\begin{equation}\label{annrelat}
a_{\vec{k},\pm} = f_{\vec{k},\pm}(\tilde{a},\pi_{{\tilde{a}}}, {\tilde{\phi}}, \pi_{\tilde{\phi}} ) \nu_{\vec{k},\pm}+ g _{\vec{k},\pm}(\tilde{a},\pi_{{\tilde{a}}}, {\tilde{\phi}}, \pi_{\tilde{\phi}} ) \pi_{\nu_{\vec{k},\pm}}
\end{equation}
for the MS annihilationlike variables. Here,   
\begin{equation}\label{ccrs}
f_{\vec{k},\pm} {g}_{\vec{k},\pm}^{*}- g_{\vec{k},\pm} {f}_{\vec{k},\pm}^{*}=-i,
\end{equation}
so that the introduced variables satisfy, when one freezes the background, canonical commutation relations with the corresponding  MS creationlike variables, defined by the complex conjugate of relation \eqref{annrelat}. For the tensor variables, on the other hand, one is led to consider analogous families of creation and annihilationlike variables, characterized by two functions ${f}_{\vec{k},\varepsilon,\pm}$ and ${g}_{\vec{k},\varepsilon,\pm}$ that satisfy a condition similar to Eq. \eqref{ccrs}. Generally, one is interested exclusively in canonical transformations of the gauge invariant variables that depend (apart that on the cosmological zero modes)  only on the frequency $\omega_k$ of the mode, but not on other details about the wavevector $\vec{k}$, nor on the sine or cosine character of the Fourier mode or the polarization of the tensor mode. For those cases, we would adopt the simpler notation $f_k$ and $g_k$ for the functions that define the creation and annihilationlike variables.

In the case of fermions, the ambiguity is captured in the freedom to define annihilationlike variables for particles and creationlike variables for antiparticles as follows:
\begin{eqnarray}
a_{\vec{k}}^{(x,y)} &=& f_1^{\vec{k}} (\tilde{a},\pi_{{\tilde{a}}},\tilde{\phi}, \pi_{\tilde{\phi}} ) x_{\vec{k}}+ f_2^{\vec{k}}(\tilde{a},\pi_{{\tilde{a}}},\tilde{\phi}, \pi_{\tilde{\phi}} ) y^{*}_{-\vec{k}},\\
\left(b_{\vec{k}}^{(x,y)}\right)^{*}&=& g_1^{\vec{k}}(\tilde{a},\pi_{{\tilde{a}}},\tilde{\phi}, \pi_{\tilde{\phi}} ) x_{\vec{k}}+ g_2^{\vec{k}}(\tilde{a},\pi_{{\tilde{a}}},\tilde{\phi}, \pi_{\tilde{\phi}} ) y^{*}_{-\vec{k}},
\end{eqnarray}
with 
\begin{equation}
f_2^{\vec{k}}= e^{i F_2^{\vec{k}}} \sqrt{ 1- \left|f_1^{\vec{k}}\right|^2}, \quad\quad g_1^{\vec{k}}= e^{i J_{\vec{k}}}\left(f_2^{\vec{k}}\right)^{*},\quad\quad  g_2^{\vec{k}}= - e^{i J_{\vec{k}}} \left(f_1^{\vec{k}}\right)^{*}.
\end{equation}
In the same spirit that we have commented above, one is usually interested only in cases in which the functions $f_1^{\vec{k}}$, $g_1^{\vec{k}}$, $g_1^{\vec{k}}$, and $g_2^{\vec{k}}$ depend on $\vec{k}$ only via $\omega_k$. Notice that the creation and annihilationlike variables \eqref{avariable} that we used for the fermions in Sec. 5 were of this kind. We will restrict to this type of cases in the following.

As we already know, a change from the gauge invariant variables that we have adopted for our system to any of the above sets of creation and annihilationlike variables for the perturbations can  be completed into a canonical set for the full cosmological model. It suffices to correct again the zero modes with contributions that are quadratic in perturbations in the way that we discussed in Sec. 4. In addition, in terms of the new canonical set, the resulting MS, tensor, and fermionic Hamiltonians are the old ones plus some known corrections. These new contributions contain, in general, both diagonal products of annihilation and creationlike variables, and terms that are responsible for the creation and destruction of pairs. The asymptotic behavior of these latter interaction terms when $\omega_k \rightarrow \infty $ is what tells us if the quantization of the Hamiltonians is well defined on the vacuum, assuming normal ordering. In all cases $f_k$, $g_k$, $f_1^{k}$, $f_2^{k}$, $g_1^{k}$, and $g_2^{k}$ can be chosen so that the dominant powers of $\omega_k$ in the interaction terms that prevent a nice behavior of the Hamiltonian operators on Fock space are eliminated.

Moreover, it is possible to remove, order by order in inverse powers of  $\omega_k$, all the asymptotic contribution to the interaction terms in the Hamiltonians.
For example, let us consider the scalar perturbations. The MS Hamiltonian gets asymptotically diagonalized with \cite{uniquenessThiemann}
\begin{equation}
\omega_k g_k = i f_k \left[1-\frac{1}{2 \omega_k^2}\sum_{n=0}^{\infty}\left(\frac{-i}{2\omega_k}\right)^{n}\gamma_{n}\right].
\end{equation}
The functions $\gamma_n$ are determined by the recursion relation
\begin{equation}
\gamma_{n+1}= \tilde{a}\, \{H_{0},\gamma_n\}+4s^{(s)}\left[\gamma_{n-1}+\sum_{l=0}^{n-3}\gamma_l \gamma_{n-(l+3)}\right]-\sum_{l=0}^{n-1}\gamma_l \gamma_{n-(l+1)},\qquad \forall n\geq0,
\end{equation}
where $\gamma_0=s^{(s)}+r^{(s)}\pi_{\tilde\phi}$ is just the background dependent mass for the MS field.  Creation and annihilationlike variables are then asymptotically fixed, up to a phase, since from the canonical commutation relations it generally follows that
\begin{equation}
2 |f_k|^2= - \frac{|h_k|^2}{ \text{Im}(h_k) }, \quad\quad {\rm where} \quad \quad h_k= \frac{f_k}{g_k}.
\end{equation}

Similar asymptotic characterizations to diagonalize the field Hamiltonians can be obtained for the tensor perturbations and for the Dirac field \cite{fermidiagonal}. Actually, in all of these cases, the first few terms in the asymptotic expansion are enough to construct variables with well-defined Hamiltonians (and finite backreaction contributions to the quantum constraint). 

On the other hand, the phases that still remain free in the creation and annihilationlike variables can be determined univocally by means of further physical considerations. Specifically, it seems natural to demand that the background dependence extracted from the dynamics of the original perturbations by our choice of those phases is the minimum allowed, and that the resulting asymptotically diagonal Hamiltonians are positive, as functions of the background. 

Given that our analysis has been carried out asymptotically for modes with large wavenumbers, the question arises of what happens for other kinds of modes and, in particular, if the asymptotic expansions provided by the Hamiltonian diagonalization for large $\omega_k$ can uniquely specify a choice of creation and annihilationlike variables for all the modes. Let us consider, e.g., the scalar perturbations. In fact, the interaction terms in the Hamiltonian for each possible value of $\omega_k$ are completely eliminated if and only if
\begin{equation}\label{diagallk}
\omega_k^2+ s^{(s)}+r^{(s)}\pi_{\tilde\phi}+ h_k^2 - \tilde{a} \{h_k, H_0\}=0. 
\end{equation}
This is a semilinear partial differential equation for which the complex solutions satisfy
\begin{equation}
\text{Im}(h_k)^2=\omega_k^2+s^{(s)}-\frac{\text{Im}(h_k)^{\prime\prime}}{2\text{Im}(h_k)}+\frac{3}{4}\left[\frac{\text{Im}(h_k)^{\prime}}{\text{Im}(h_k)}\right]^2,
\end{equation}
where the prime stands for the operation of taking the Poisson bracket $\tilde{a}\{.,H_{0}\}$. It is worth commenting that, in the linearized context of QFT in curved spacetimes, our asymptotic characterization above can be shown to lead in a unique way to the Minkowski vacuum in the case of constant mass, and to the Bunch-Davies vacuum when the homogeneous background is taken as the de Sitter solution \cite{uniquenessThiemann}. Thus, in these linearized contexts, the procedure of asymptotic diagonalization is able to uniquely fix a solution to Eq. \eqref{diagallk} for all wavenumbers, and this solution reproduces the natural choice of vacuum state in the considered scenarios. Furthermore, the corresponding asymptotic expansions for the fermionic creation and annihilationlike variables that diagonalize the Hamiltonian have been proven to determine as well a unique choice for all scales in the linearized de Sitter context, even if it is known that those expansions have zero radius of convergence in this case \cite{fermidesitter}.

In summary, the asymptotic diagonalization of the Hamiltonian of the perturbations may provide a procedure to determine a vacuum state, and therefore to fix initial conditions for the primordial perturbations in such a way that they are optimally adapted to the dynamics dictated by the Hamiltonian constraint of the total system. Moreover, recent investigations \cite{enmmp} support a close analytical relation between the vacuum state that would be selected in this manner and the NO vauum proposed by Mart\'{\i}n-de Blas and Olmedo \cite{MBO}, at least in the context of hybrid LQC.

\section{Conclusions}

In this work, we have reviewed the hybrid approach to LQC. This approach to the quantum description of gravitational systems with local degrees of freedom within the framework of the loop quantization program tries to provide, in a controlled way, a formalism for the study of inhomogeneous cosmological scenarios that, yet, display some symmetries that simplify the physics, or in which the inhomogeneities can be described in a perturbative way over a highly symmetric background. In this way we have been able to analyze linearly polarized gravitational waves in Gowdy cosmologies with toroidal compact sections, and scalar, tensor, and fermionic perturbations at quadratic order in the action around an FLRW spacetime in the presence of an inflaton. In particular, for these cosmological perturbations and at the considered truncation order, we have found a canonical set for the full system composed of gauge invariant perturbations (including the MS field), linear perturbative constraints and gauge variables conjugate to them, and zero modes that contain the relevant information about the background FLRW cosmology. In a hybrid quantization of this canonical system, physical states depend only on the quantum FLRW background and on gauge invariant perturbations. Starting from the zero mode of the Hamiltonian constraint, that couples these perturbations with the FLRW background, and adopting a suitable ansatz for the quantum states of interest, we have been able to derive propagation equations for the perturbations in the primeval stages of the Universe. These equations differ slightly from those of GR by the inclusion of quantum corrections, corrections that we have succeeded to explicitly derive with our hybrid strategy taking fully into account the quantum behavior of the FLRW substrate, and therefore beyond the level of an effective description of this background within homogeneous and isotropic LQC. 

In order to quantize differently the system within the framework of LQC, but still adhering to the idea of developing a QFT for the perturbations on a quantum spacetime, one can follow the so-called dressed metric approach, put forward in Refs. \cite{AshLewaDress,AAN1,AAN2}, instead of the hybrid approach. Indeed, as in the hybrid proposal, the dressed metric approach adopts also the philosophy of combining a loop representation for the homogeneous sector of the (truncated) phase space and a Fock representation for the tensor and MS perturbations (and possible fermionic perturbations, if they are present). Again, in the dressed metric approach one also introduces an ansatz for the quantum states of cosmological interest in which the dependence on the homogeneous geometry and on the perturbations factorizes. In this ansatz, all partial wavefunctions are allowed to depend on the inflaton field $\phi$. However, in the dressed  metric case there is no Hamiltonian constraint that affects the perturbations, since the whole of the truncated cosmology is not treated as a constrained symplectic system. Instead, one has the Hamiltonian constraint of the homogeneous FLRW model, and the Hamiltonian functions \eqref{Hscal} and \eqref{Htens} that, classically, generate the dynamics of the perturbations. Consequently, the approach requires that the homogeneous part of the states be an exact solution of the FLRW model in LQC, and then uses this solution to define the quantum dynamics on the phase space of the gauge invariant perturbations \cite{AAN1,AAN2}. In this way, the perturbations behave as test fields that see a dressed metric determined by certain expectation values of operators of the homogeneous geometry, which incorporate the most relevant quantum effects. 

In spite of the similarities between the hybrid and the dressed metric approaches, the effective equations that they provide for the propagation of the gauge invariant perturbations are somewhat different even if backreaction is neglected. The discrepancy appears only in the term of the time dependent mass in the propagation equations \cite{mass}. At the end of the day, this can be traced back to the differences in the treatment of the phase space of the perturbed FLRW cosmologies in the hybrid and the dressed metric proposals. As we have emphasized, in the hybrid case the whole phase space is treated as a symplectic manifold, and accordingly it is described in terms of canonical variables. This applies, in particular, to the expression deduced for the time dependent mass. On the contrary, in the dressed metric formalism, one evaluates the time dependent mass directly on the FLRW metric dressed with quantum corrections. For states such that this metric satisfies the effective dynamics of LQC, the time derivatives involved in the corresponding expression of the time dependent mass are then computed along an effective trajectory of homogeneous and isotropic LQC. The difference then arises because of the departure of the standard classical relation between the time derivatives of the scale factor and its canonical momentum (inherent to the hybrid approach) with respect to the alternative effective relation in LQC (employed in the dressed metric case) \cite{mass}. Remarkably, this difference is specially important around the bounce, precisely the region where the quantum corrections on the propagation of the perturbations are expected to be relevant.

Several other approaches have also been suggested for the investigation of cosmological perturbations within LQC. For a comprehensive summary of such approaches, we refer the reader to the reviews listed in Refs. \cite{BCMB,RoVi,AshBarr,Grain,GFT,AlesCian,Edward,agusingh}. They include the deformed constraint algebra approach \cite{Bojo0,Bojo1,CLB,Bojo2}, the group field theory models \cite{Edward2,OG,gielen2}, and the quantum reduced loop gravity scheme \cite{alesci1,alesci2}. Additionally, different ways of addressing backreaction effects of the perturbations on the background within canonical quantum cosmology have been recently explored using techniques from space adiabatic perturbation theory \cite{susanvegan}. Our attention here has been exclusively put on the hybrid approach in order to fill a gap in the literature, as this is the first extensive review of this proposal that includes a detailed description of the application to primordial perturbations.

A remarkable fact of the hybrid quantization is that, while inhomogeneities and background degrees of freedom are treated as parts of a single constrained system, the imposition of the quantum constraints is consistent and does not give rise to anomalies. This statement holds both in the Gowdy model and for cosmological perturbations. The precise relation of these constraints with the full set of four-dimensional spacetime diffeomorphisms is a different issue that calls for more detailed investigations. As presented in Sec. III, the Gowdy model is not only a symmetry reduction of Einstein gravity, but it is also a partially gauge-fixed system in which only two global constraints remain, namely the zero mode of the Hamiltonian constraint and the zero mode of the momentum constraint in the angular direction on which the metric fields depend\footnote{For other alternative quantizations of the Gowdy model developed recently within the framework of LQC, see e.g. Refs. \cite{bojogow,gowlmedo}.}.  It is worth emphasizing that these are only two constraints, and not two constraints per point (neither of the spatial section nor in the considered angular direction). The aforementioned momentum constraint generates rigid translations in the corresponding angle, while the Hamiltonian one generates global time reparameterizations. These two constraints of the model actually display vanishing Poisson brackets between them and, with the adopted quantization, their corresponding operators commute. For cosmological perturbations, the constraint algebra has to be consistent just up to the order of the perturbative truncation used in our treatment. We have shown that the linear perturbative diffeomorphisms and Hamiltonian constraints admit an Abelianization at this truncation order, and we have represented them directly as part as our canonical elementary variables. The only remaining constraint in the system is a global one, given by the zero mode of the Hamiltonian constraint, that includes contributions from the background and from the perturbations. Notably, its only dependence on the perturbations is via gauge invariants, and therefore commutes with the linear perturbative constraints both classically and in the quantum theory. Indeed, we recall that in the hybrid quantization the Mukhanov-Sasaki and tensor perturbations are represented as operators that commute with the linear perturbative constraints. In other words, at the level of our perturbative truncation and with our hybrid strategy, the algebra of the quantum constraints of our perturbed system does not present anomalies.

The physical relation of these constraints with the four-dimensional diffeomorphisms algebra and the extent to which recent claims about problems with general covariance in LQC  \cite{bojow,bojocr,bojocr2} affect the system at the considered perturbative order deserve further study. These claims have been inspired in part by the deformed constraint algebra approach, which in particular predicts processes of effective signature change in high curvature regimes \cite{BM,ebounce,ebounce2}. In this respect, let us point out, for instance, that some of the perturbative canonical variables used in the hybrid approach are defined with fields that are non-local functions of the spatial metric, inasmuch as they can only be obtained by taking inverse derivatives. This is a common situation even in standard cosmological perturbation theory \cite{mukhanov1,bardeen,langlo,MukhanovSasaki,sasaki,sasakikodama}. Moreover, in terms of the background variables employed in our formulation, the metric functions include corrections that are quadratic in the perturbations already at the studied truncation order. A representation of these metric components as quantum operators has yet to be constructed, but it is clear that now issues such as the non-degenerate Lorentzian character of the metric become intrincate questions from a quantum perspective. Even the square scale factor, that in absence of pertubations is \emph{strictly} positive in each superselection sector of homogeneous LQC with the quantization prescription adopted here
\footnote{In fact, this strict positivity is in tension with the assumption inherent in certain representation independent analyses in LQC that the canonical variable describing the densitized triad is supported over the whole real line.},
might in principle turn negative by the effect of the perturbations. Nonetheless, none of these unexplored questions on the quantum geometric structure changes the hyperbolic ultraviolet behavior that we have found for the propagation equations of the perturbative modes.

Even if we have succeeded in deriving such mode equations, that rule the evolution of the primordial perturbations in the hybrid approach, we have seen that this is not yet enough to extract predictions that can be confronted with observations. For this purpose, we also need two types of initial data, namely initial values to fix the FLRW background and conditions to choose a unique vacuum state for the perturbations. With respect to the FLRW cosmology, we have seen that it suffices to provide, e.g., the value of the inflaton at the bounce, apart from the parameters that determine the inflaton potential. In the case of a quadratic potential, we have found values for the inflaton on the bounce section and for the inflaton mass such that the modes affected by quantum geometry effects are those that are re-entering the Hubble horizon nowadays, situation that is the most interesting possibility in terms of observational plausibility in the CMB. Concerning the vacuum state of the perturbations, we have commented on various proposals to select it that lead to power spectra that seem compatible with the observational data.

To go beyond those proposals and find a criterion to select the vacuum that is rooted on the hybrid strategy, that combines loop and Fock representations, we have put an additional emphasis on the choice of splitting between the homogeneous and isotropic sector of phase space and the gauge invariant perturbations. This freedom can be employed to reach a Hamiltonian constraint with nice properties, at least as far as its action on the perturbations is concerned. Requiring such good physical and mathematical properties turns out to restrict the possible quantum dynamics of the perturbative gauge invariants, as well as the Fock representation chosen for them. In turn, this can be regarded as a limitation in the admissible choices of vacuum state. In particular, we have shown that a criterion such as the diagonalization of the Hamiltonian of the gauge invariant perturbations, based on its asymptotic structure, might be able to provide a unique vacuum state, to which one may partcularize in the future the discussion of the effects of quantum geometry in cosmology to extract concrete and distinctive predictions.

\section*{Conflict of Interest Statement}

The authors declare that the research was conducted in the absence of any commercial or financial relationships that could be construed as a potential conflict of interest.

\section*{Author Contributions}

The two authors have contributed equally to this review, revisiting the original material, writing the manuscript, and deciding its final redaction.

\section*{Funding}

This work was supported by Grant No. MINECO FIS2017-86497-C2-2-P from Spain. 

\section*{Acknowledgments}

The authors want to thank all the researchers that have contributed to the development of Hybrid Loop Quantum Cosmology, a community that includes Laura Castell\'o Gomar, Jer\'onimo Cortez, Mikel Fern\'andez-M\'endez, Alejandro Garc\'{\i}a-Quismondo, Mercedes Mart\'{\i}n Benito, Daniel Mart\'{\i}n-de Blas, Javier Olmedo, Santiago Prado, Paula Tarr\'{\i}o, and Jos\'e M. Velhinho. They also want to thank Abhay Ashtekar, Alejandro Corichi, Kristina Giesel, Jerzy Lewandowski, Hanno Sahlmann, Thomas Thiemann, and Anzhong Wang for inspiration and discussions.


\begin{thebibliography}{299}

\bibitem{ashlqg} Ashtekar A. New variables for classical and quantum gravity. Phys. Rev. Lett. (1986) 57:2244.

\bibitem{ashlewlqg}  Ashtekar A, Lewandowski  J. Background independent quantum gravity: A status report. Classical Quantum Gravity (2004) 21:R53.

\bibitem{lqgThiemann} Thiemann T.  Modern Canonical Quantum General Relativity. Cambridge, UK: Cambridge University Press (2007). 

\bibitem{WdW} DeWitt BS. Quantum theory of gravity I. The canonical theor. Phys. Rev. (1967) 160:1113.

\bibitem{Hall} Halliwell JJ. ``Introductory lectures on quantum cosmology''. In: Coleman S, Hartle JB, Piran T, Weinberg S, editors. Proceedings of the 1990 Jerusalem Winter School on Quantum Cosmology and Baby Universes. Singapore: World Scientific (1991) p. 159-243.

\bibitem{yangmills} Yang CN, Mills RL. Conservation of isotopic spin and isotopic gauge invariance. Phys. Rev. D (1954) 96:191.

\bibitem{Dirac}  Dirac PAM. Lectures on Quantum Mechanics. New York: Belfer Graduate School Monograph Series (1964). 

\bibitem{Turok} Di Tucci A, Feldbrugge J, Lehners J-L, Turok N. Quantum incompleteness of inflation. Phys. Rev. D (2019) 100:063517. 

\bibitem{bojo} Bojowald M. Loop Quantum Cosmology. Living Rev. Rel. (2008) 11:4. 

\bibitem{abl} Ashtekar A, Bojowald M, Lewandowski J. Mathematical structure of loop quantum cosmology. Adv. Theor. Math. Phys. (2003) 7:233. 

\bibitem{ashparam} Ashtekar A, Singh P. Loop quantum cosmology: A status report. Classical Quantum Gravity  (2011) 28:213001. 

\bibitem{chiou} Chiou DW. Loop quantum cosmology in Bianchi type I models: Analytical investigation. Phys. Rev. D (2007) 75:024029. 

\bibitem{chiou2} Chiou DW. Effective dynamics, big bounces, and scaling symmetry in Bianchi type I loop quantum cosmology. Phys. Rev. D (2007) 76:124037.

\bibitem{mmp1} Mart\'{\i}n-Benito M, Mena Marug\'an GA, Paw{\l}owski T. Loop quantization of vacuum Bianchi I cosmology. Phys. Rev. D (2008) 78:064008.

\bibitem{mmp2} Mart\'{\i}n-Benito M, Mena Marug\'an GA, Paw{\l}owski T. Physical evolution in loop quantum cosmology: The example of vacuum Bianchi I. Phys. Rev. D (2009) 80:084038.

\bibitem{awe1} Ashtekar A, Wilson-Ewing E. Loop quantum cosmology of Bianchi type I models. Phys. Rev. D (2009) 79:083535. 

\bibitem{structures} Liddle AR, Lyth DH. Cosmological Inflation and Large-Scale Structure. Cambridge, UK: Cambridge University Press (2000). 

\bibitem{Ivan} Agullo I, Morris NA. Detailed analysis of the predictions of loop quantum cosmology for the primordial power spectra. Phys. Rev. D (2015) 92:124040. 

\bibitem{AshNe} Ashtekar A, Gupt B, Jeong D, Sreenath V. Alleviating the tension in CMB using Planck-scale physics.  Phys. Rev. Lett. (2020) 125:051302.

\bibitem{hybrid1} Mart\'{\i}n-Benito M, Garay LJ, Mena Marug\'an GA. Hybrid quantum Gowdy cosmology: Combining loop and Fock quantizations. Phys. Rev. D (2008) 78:083516.

\bibitem{hybrid2} Mena Marug\'an GA, Mart\'{\i}n-Benito M. Hybrid quantum cosmology: Combining loop and Fock quantizations. Int. J. Mod. Phys. A  (2009) 24:2820. 

\bibitem{hybrid3} Garay LJ, Mart\'{\i}n-Benito M, Mena Marug\'an GA. Inhomogeneous loop quantum cosmology: Hybrid quantization of the Gowdy model. Phys. Rev. D  (2010) 82:044048.

\bibitem{hybrid4} Mart\'{\i}n-Benito M, Mena Marug\'an GA, Wilson-Ewing E. Hybrid quantization: From Bianchi I to the Gowdy model. Phys. Rev. D  (2010) 82:084012.

\bibitem{gowdy1} Gowdy RH. Gravitational waves in closed universes. Phys. Rev. Lett. (1971) 27:826.

\bibitem{gowdy2} Gowdy RH. Vacuum spacetimes with two-parameter spacelike isometry groups and compact invariant hypersurfaces: Topologies and boundary conditions. Ann. Phys.  (1974) 83:203.

\bibitem{hybrid-matter} Mart\'{\i}n-Benito M, Mart\'{\i}n-de Blas D, Mena Marug\'{a}n GA. Matter in inhomogeneous loop quantum cosmology: The Gowdy T3 model. Phys. Rev. D  (211) 83:084050.

\bibitem{hybrid-matterapp}  Mart\'{\i}n-Benito M, Mart\'{\i}n-de Blas D, Mena Marug\'{a}n GA. Approximation methods in loop quantum cosmology: From Gowdy cosmologies to inhomogeneous models in Friedmann-Robertson-Walker geometries. Classical Quantum Gravity  (2014) 32:075022.

\bibitem{hybrid-matter1} Elizaga Navascu\'es B, Mart\'{\i}n-Benito M, Mena Marug\'{a}n GA. Modeling effective FRW cosmologies with perfect fluids from states of the hybrid quantum Gowdy model. Phys. Rev. D (2015) 91:024028.

\bibitem{hybrid-matter2} Elizaga Navascu\'es B, Mart\'{\i}n-Benito M, Mena Marug\'{a}n GA. Modified FRW cosmologies arising from states of the hybrid quantum Gowdy model. Phys. Rev. D  (2015) 92:024007.

\bibitem{hyb-pert1} Fern\'andez-M\'endez M, Mena Marug\'an GA, Olmedo J. Hybrid quantization of an inflationary universe. Phys. Rev. D (2012) 86:024003.

\bibitem{hyb-pert2}  Fern\'andez-M\'endez M, Mena Marug\'an GA, Olmedo J. Hybrid quantization of an inflationary model: The flat case. Phys. Rev. D (2013) 88:044013.

\bibitem{hyb-pert-eff} Fern\'andez-M\'endez M, Mena Marug\'an GA, Olmedo J. Effective dynamics of scalar perturbations in a flat Friedmann-Robertson-Walker spacetime in loop quantum cosmology. Phys. Rev. D (2014) 89:044041.

\bibitem{hyb-pert3} Castell\'o Gomar L, Fern\'andez-M\'endez M, Mena Marug\'an GA, Olmedo J. Cosmological perturbations in hybrid loop quantum cosmology: Mukhanov--Sasaki variables. Phys. Rev. D (2014) 90:064015.

\bibitem{hyb-pert4} Castell\'o Gomar L, Mart\'{\i}n-Benito M, Mena Marug\'an GA. Gauge-invariant perturbations in hybrid quantum cosmology, JCAP (2015) 06:045.

\bibitem{hyb-pert5} Castell\'o Gomar L, Mart\'{\i}n-Benito M, Mena Marug\'an GA. Quantum corrections to the Mukhanov-Sasaki equations. Phys. Rev. D (2016) 93:104025.

\bibitem{hybr-pred} Castell\'o Gomar L, Mena Marug\'an GA, Mart\'{\i}n de Blas D, Olmedo J. Hybrid loop quantum cosmology and predictions for the cosmic microwave background. Phys. Rev. D (2017) 96:103528.

\bibitem{hybr-ten} Ben\'{\i}tez Mart\'{\i}nez F, Olmedo J. Primordial tensor modes of the early universe. Phys. Rev. D (2016) 93:124008.

\bibitem{mukhanov1} Mukhanov V. Physical Foundations of Cosmology. Cambridge, UK: Cambridge University Press (2005).

\bibitem{bardeen} Bardeen JM. Gauge-invariant cosmological perturbations. Phys. Rev. D (1980) 22:1882.

\bibitem{MukhanovSasaki} Mukhanov V. Quantum theory of gauge-invariant cosmological perturbations. Zh. Eksp. Teor. Fiz. (1988) 94:1 [Sov. Phys. JETP (1988) 67:1297].

\bibitem{sasaki} Sasaki M. Gauge invariant scalar perturbations in the new inflationary universe. Prog. Theor. Phys. (1983) 70:394.

\bibitem{sasakikodama} Kodama H, Sasaki M. Cosmological perturbation theory. Prog. Theor. Phys. Suppl.  (1984) 78:1.

\bibitem{langlo} Langlois D. Hamiltonian formalism and gauge invariance for linear perturbations in inflation. Classical Quantum Gravity  (1994) 11:389.

\bibitem{pintoneto1} Pinho EJC, Pinto-Neto N. Scalar and vector perturbations in quantum cosmological backgrounds. Phys. Rev. D  (2007) 76:023506.

\bibitem{pintoneto2} Falciano FT, Pinto-Neto N. Scalar perturbations in scalar field quantum cosmology. Phys. Rev. D (2009) 79:023507.

\bibitem{HalliwellHawking} Halliwell JJ, Hawking SW. Origin of structure in the Universe. Phys. Rev. D  (1985) 31:1777.

\bibitem{shiraiWada} Shirai I, Wada S. Cosmological perturbations and quantum fields in curved space. Nucl. Phys. B  (1988) 303:728.

\bibitem{BCMB} Banerjee K, Calcagni G, Mart\'\i n-Benito M. Introduction to loop quantum cosmology. SIGMA (2012) 8:016.

\bibitem{RoVi} Rovelli  C, Vidotto F. Covariant Loop Quantum Gravity: An Elementary Introduction to Quantum Gravity and Spinfoam Theory. Cambridge, UK: Cambridge University Press (2015).

\bibitem{AshBarr} Ashtekar A, Barrau A. Loop quantum cosmology:  From pre-inflationary dynamics to observations. Classical Quantum Gravity (2015) 32:234001.

\bibitem{Grain} Grain J.  The perturbed universe in the deformed algebra approach of loop quantum cosmology. Int. J. Mod. Phys. D (2016) 25:1642003.

\bibitem{GFT} Gielen S, Sindoni L. Quantum cosmology from group field theory condensates: A review. SIGMA (2016) 12:082. 

\bibitem{AlesCian} Alesci E, Cianfrani F. Quantum reduced loop gravity and the foundation of loop quantum cosmology. Int. J. Mod. Phys. D (2016) 25:1642005.

\bibitem{Edward} Wilson-Ewing E. Testing loop quantum cosmology. Comptes Rendus Physique (2017) 18:207.

\bibitem{agusingh} Agullo I, Singh P. Loop quantum cosmology: A brief review. In:  Ashtekar A, Pullin J, editors. 100 Years of General Relativity, Vol. 4. Loop Quantum Gravity: The First 30 Years. Singapore: World Scientific (2017) p. 183-240.

\bibitem{bojow} Bojowald M. Critical evaluation of common claims in loop quantum cosmology. Universe (2020) 6:36.

\bibitem{isham} Isham CJ.  Modern Differential Geometry for Physicists, 2nd edition. Singapore: World Scientific (1999).

\bibitem{barbero}  Barbero JF. Real polynomial formulation of general relativity in terms of connections. Phys. Rev. D (1995) 51:5507.

\bibitem{immi} Immirzi G. Quantum gravity and Regge calculus. Nucl. Phys. Proc. Suppl. (1997) 57:65.

\bibitem{APS1}  Ashtekar A, Paw\l{}owski T, Singh P. Quantum nature of the big bang: An analytical and numerical investigation. Phys. Rev. D (2006) 73:124038.

\bibitem{lqcconfig} Velhinho JM. The quantum configuration space of loop quantum cosmology. Classical Quantum Gravity (2007) 24:3745.

\bibitem{rudin} Rudin W. Fourier Analysis on Groups. New York: Interscience Publishers (1962).

\bibitem{APS2} Ashtekar A, Paw{\l}owski T, Singh P. Quantum nature of the big bang: Improved dynamics. Phys. Rev. D (2006) 74:084003.

\bibitem{reedsimon} Reed M, Simon B. Methods of Modern Mathematical Physics I: Functional Analysis. San Diego: Academic Press (1980).

\bibitem{Ma} Yang J, Ding Y, Ma Y. Alternative quantization of the Hamiltonian in loop quantum cosmology II: Including the Lorentz term. Phys. Lett.  (2009) B682:1.

\bibitem{DaporLieg} Dapor A, Liegener K. Cosmological effective Hamiltonian from full loop quantum gravity dynamics. Phys. Lett. (2018) B785:506.

\bibitem{DaLiTomasz} Assanioussi M, Dapor A, Liegener K, Paw\l{}owski T. Emergent de Sitter epoch of the loop quantum cosmos: A detailed analysis. Phys. Rev. D (2019) 100:084003.

\bibitem{AlejMena} Garc\'{\i}a-Quismondo A, Mena Marug\'an GA. The Mart\'{\i}n-Benito-Mena Marug\'an-Olmedo prescription for the Dapor-Liegener model of loop quantum cosmology. Phys. Rev. D (2019) 99:083505.

\bibitem{thiemann} Thiemann T. Anomaly-free formulation of non-perturbative four-dimensional Lorentzian quantum gravity. Phys. Lett.  (1996) B380:257.

\bibitem{mmo} Mart\'{\i}n-Benito M, Mena Marug\'an GA, Olmedo J. Further improvements in the understanding of isotropic loop quantum cosmology. Phys. Rev. D (2009) 80:104015.

\bibitem{APS} Ashtekar A, Paw{\l}owski T, Singh P. Quantum nature of the big bang. Phys. Rev. Lett. (2006) 96:141301.

\bibitem{bek-hawk-im} Ashtekar A, Baez JC, Krasnov K. Quantum geometry of isolated horizons and black hole entropy. Adv. Theor. Math. Phys. (2001) 4:1.

\bibitem{lewdoma} Domagala M, Lewandowski J. Black hole entropy from quantum geometry. Classical Quantum Gravity (2004) 21:5233.

\bibitem{Meisn} Meissner KA. Black hole entropy in loop quantum gravity. Classical Quantum Gravity (2004) 21:5245.

\bibitem{bianchibounce} Gupt B, Singh P. Quantum gravitational Kasner transitions in Bianchi-I spacetime. Phys. Rev. D  (2012) 86:024034.

\bibitem{tarrio} Tarr\'{\i}o P, Fern\'andez-M\'endez M, Mena Marug\'an GA. Singularity avoidance in the hybrid quantization of the Gowdy model. Phys. Rev. D (2013) 88:084050.

\bibitem{men2} Corichi A, Cortez J, Mena Marug\'{a}n GA, Velhinho JM. Quantum Gowdy T3 Model: A Uniqueness Result. Classical Quantum Gravity  (2006) 23:6301.

\bibitem{men3}  Cortez J, Mena Marug\'{a}n GA, Velhinho JM. Uniqueness of the Fock quantization of the Gowdy T3 model. Phys. Rev. D (2007) 75:084027.

\bibitem{reality} Rendall AD. Unique determination of an inner product by adjointness relations in the algebra of quantum observables. Classical Quantum Gravity  (1993) 10:2261.

\bibitem{reality2} Rendall AD. Adjointness relations as a criterion for choosing an inner product. arXiv:gr-qc/9403001 (1994).

\bibitem{fermilala} Elizaga Navascu\'es B, Mart\'{\i}n-Benito M, Mena Marug\'{a}n GA. Fermions in hybrid loop quantum cosmology. Phys. Rev. D  (2017) 96:044023.

\bibitem{Mikel} Fern\'andez-M\'endez M. (2014) Perturbaciones y Din\'amica Efectiva de Cosmolog\'{\i}a Cu\'antica de Lazos Inhomog\'enea. PhD Thesis. Madrid: Universidad Complutense de Madrid. Available at: https://eprints.ucm.es/28962/1/T35905.pdf.

\bibitem{DEathHall} D'Eath PD, Halliwell JJ. Fermions in quantum cosmology. Phys. Rev. D (1987) 35:1100.

\bibitem{berezin} Berezin FA. The Method of Second Quantization. New York: Academic (1966).

\bibitem{dtorus} Friedrich Th. Zur abh\"angigkeit des Dirac-operators von der spin-struktur. Colloq. Mathematicum (1984) 48:57. 

\bibitem{uniquenessflat} Castell\'o Gomar L, Cortez J, Mart\'{\i}n-de Blas D, Mena Marug\'an GA, Velhinho JM. Uniqueness of the Fock quantization of scalar fields in spatially flat cosmological spacetimes. JCAP (2012) 11:001.

\bibitem{uniquenessscale} Cortez J, Mena Marug\'an GA, Olmedo J, Velhinho JM. Criteria for the determination of time dependent scalings in the Fock quantization of scalar fields with a time dependent mass in ultrastatic spacetimes. Phys. Rev. D (2012) 86:104003.

\bibitem{CGGQMM} Castell\'o Gomar L, Garc\'{\i}a-Quismondo A, Mena Marug\'an GA. Primordial perturbations in the Dapor-Liegener model of hybrid loop quantum cosmology. Phys. Rev. D (2020) 102:083524.

\bibitem{GQMMSP} Garc\'{\i}a-Quismondo A, Mena Marug\'an GA, S\'anchez P\'erez G. The time-dependent mass of cosmological perturbations in loop quantum cosmology: Dapor-Liegener regularization. Classical Quantum Gravity (2020) 37:195003.

\bibitem{Galindo} Galindo A, Pascual P. Quantum Mechanics I. Berlin: Springer-Verlag (1990).

\bibitem{Universe} Elizaga Navascu\'es B, Mart\'{\i}n de Blas D, Mena Marug\'an GA. The vacuum state of primordial fluctuations in hybrid loop quantum cosmology. Universe  (2018) 4:98.

\bibitem{Wang1} Zhu T, Wang A, Kirsten K, Cleaver G, Sheng Q. Universal features of quantum bounce in loop quantum cosmology. Phys. Lett. (2017) B773:196.

\bibitem{Wang2} Zhu T, Wang A, Kirsten K, Cleaver G, Sheng Q.  Pre-inflationary universe in loop quantum cosmology. Phys. Rev. D (2017) 96: 083520.

\bibitem{mass} Elizaga Navascu\'es B, Mart\'{\i}n de Blas D, Mena Marug\'an GA. Time-dependent mass of cosmological perturbations in the hybrid and dressed metric approaches to loop quantum cosmology. Phys. Rev. D (2018) 97:043523.

\bibitem{MBO} Mart\'{\i}n de Blas D, Olmedo J. Primordial power spectra for scalar perturbations in loop quantum cosmology. JCAP  (2016) 06:029.

\bibitem{AAN3} Agullo I, Ashtekar A, Nelson W.  A quantum gravity extension of the inflationary scenario. Phys. Rev. Lett. (2012) 109:251301.

\bibitem{AAN1}  Agullo I, Ashtekar A, Nelson W. Extension of the quantum theory of cosmological perturbations to the Planck era, Phys. Rev. D (2013) 87:043507.

\bibitem{AAN2}  Agullo I, Ashtekar A, Nelson W. The pre-inflationary dynamics of loop quantum cosmology: Confronting quantum gravity with observations. Classical Quantum Gravity (2013) 30:085014.

\bibitem{continuum} Elizaga Navascu\'es B, Mena Marug\'an GA. Perturbations in hybrid loop quantum cosmology: Continuum limit in Fourier space. Phys. Rev. D (2018) 98:103522.

\bibitem{Wang3} Wu Q, Zhu T, Wang A. Non-adiabatic evolution of primordial perturbations and non-Gaussinity in hybrid approach of loop quantum cosmology. Phys. Rev. D (2018) 98:103528.

\bibitem{planck} Ade PAR, {\it et al.} (Planck Collaboration). Planck 2015 results. XIII. Cosmological parameters. A\&A (2016) 594:A13.

\bibitem{planck-inf} Ade PAR, {\it et al.} (Planck Collaboration). Planck 2015 results. XX. Constraints on inflation. A\&A (2016) 594:A20.

\bibitem{linde} Contaldi CR, Peloso M, Kofman L, Linde A. Suppressing the lower multipoles in the CMB anisotropies. JCAP (2003) 07:002.

\bibitem{BD} Bunch TS, Davies P. Quantum field theory in de Sitter space: Renormalization by point splitting. Proc. R. Soc. Lond. A (1978) 360:117.

\bibitem{adiabatic1} Parker L. Quantized fields and particle creation in expanding universes. I. Phys. Rev. (1969) 183:1057.

\bibitem{adiabatic2} L{\"u}ders C, Roberts  JE. Local quasiequivalence and adiabatic vacuum states. Commun. Math. Phys.  (1990) 134:29.

\bibitem{AGvacio1} Ashtekar A, Gupt B. Initial conditions for cosmological perturbations. Classical Quantum Gravity (2017) 34:035004.

\bibitem{AGvacio2} Ashtekar A, Gupt B. Quantum gravity in the sky: Interplay between fundamental theory and observations. Classical Quantum Gravity (2017) 34:014002.

\bibitem{uniquenessThiemann} Elizaga Navascu\'es B, Mena Marug\'an GA, Thiemann T. Hamiltonian diagonalization in hybrid quantum cosmology. Classical Quantum Gravity (2019) 36:185010.

\bibitem{fermidiagonal} Elizaga Navascu\'es B, Mena Marug\'an GA, Prado S. Asymptotic diagonalization of the fermionic Hamiltonian in hybrid loop quantum cosmology. Phys. Rev. D (2019) 99:063535.

\bibitem{fermidesitter} Elizaga Navascu\'es B, Mena Marug\'an GA, Prado S. Unique fermionic vacuum in de Sitter spacetime from hybrid quantum cosmology. Phys. Rev. D (2020) 101:123530.

\bibitem{enmmp} Elizaga Navascu\'es B, Mena Marug\'an GA, Prado S. Non-oscillating power spectra in loop quantum cosmology. Classical Quantum Gravity (2020) 38:035001.

\bibitem{AshLewaDress} Ashtekar A, Kaminski W,  Lewandowski J. Quantum field theory on a cosmological, quantum space-time. Phys. Rev. D (2009) 79:064030.

\bibitem{Bojo0} Bojowald M, Hossain GM, Kagan M, Shankaranarayanan S. Anomaly freedom in perturbative loop quantum gravity. Phys. Rev. D (2008) 78:063547.

\bibitem{Bojo1} Bojowald M, Calcagni G, Tsujikawa S. Observational constraints on loop quantum cosmology. Phys. Rev. Lett. (2011) 107:211302.

\bibitem{CLB} Cailleteau T, Linsefors L, Barrau A. Anomaly-free perturbations with inverse-volume and holonomy corrections in loop quantum cosmology. Classical Quantum Gravity (2014) 31:125011.

\bibitem{Bojo2} Barrau A, Bojowald M, Calcagni G, Grain J, Kagan M. Anomaly-free cosmological perturbations in effective canonical quantum gravity. JCAP (2015) 05:051.

\bibitem{Edward2}  Gerhardt F, Oriti D, Wilson-Ewing E. The separate universe framework in group field theory condensate cosmology. Phys. Rev. D (2018) 98:066011.

\bibitem{OG} Gielen S, Oriti D. Cosmological perturbations from full quantum gravity. Phys. Rev. D (2018) 98:106019.

\bibitem{gielen2} Gielen S. Inhomogeneous universe from group field theory condensate. JCAP (2019) 02:013.

\bibitem{alesci1} Alesci E, Barrau A, Botta G, Martineau K, Stagno G. Phenomenology of quantum reduced loop gravity in the isotropic cosmological sector. Phys. Rev. D (2018) 98:106022.

\bibitem{alesci2} Olmedo J, Alesci E. Power spectrum of primordial perturbations for an emergent universe in quantum reduced loop gravity. JCAP (2019) 04:030.

\bibitem{susanvegan} Schander S, Thiemann T. Quantum cosmological backreactions IV: Constrained quantum cosmological perturbation theory.  arXiv:1909.07271 (2020).

\bibitem{bojogow} Bojowald M, Brahma S. Covariance in models of loop quantum gravity: Gowdy systems. Phys. Rev. D (2015) 92:065002.

\bibitem{gowlmedo} Mart\'{\i}n de Blas D, Olmedo J, Paw{\l}owski T. Loop quantization of the Gowdy model with local rotational symmetry. Phys. Rev. D (2017) 96:106016.

\bibitem{bojocr} Bojowald M. Non-covariance of the dressed-metric approach in loop quantum cosmology. Phys. Rev. D (2020) 102:023532.

\bibitem{bojocr2} Bojowald M. No-go result for covariance in models of loop quantum gravity. Phys. Rev. D (2020) 102:046006.

\bibitem{BM} Bojowald M, Mielczarek J. Some implications of signature-change in cosmological models of loop quantum gravity. JCAP (2015) 08:052.

\bibitem{ebounce} Schander S, Barrau A, Bolliet B, Grain J, Linsefors L, Mielczarek J. Primordial scalar power spectrum from the Euclidean Big Bounce. Phys. Rev. D (2016) 93:023531.

\bibitem{ebounce2} Barrau A, Grain J. Cosmology without time: What to do with a possible signature change from quantum gravitational origin? arXiv:1607.07589 (2016).

\end{thebibliography}
\end{document}